\documentclass[%
 reprint,
 amsmath,amssymb,
 aps,
 prd,
]{revtex4-2}
\usepackage[bottom]{footmisc} 
\usepackage{xspace}
 
 \def\Pmu         {\ensuremath{\mu}\xspace}
 \def\Pnu         {\ensuremath{\nu}\xspace}
 
 \def\Ppi         {\ensuremath{\pi}\xspace}

 \def\Ppsi        {\ensuremath{\psi}\xspace}
 \def\Pphi        {\ensuremath{\phi}\xspace}
 
 \mathchardef\PXi="7104
 \mathchardef\PLambda="7103
 \mathchardef\PPi="7105
 \def\PB      {\ensuremath{B}\xspace}
 \def\PH      {\ensuremath{H}\xspace}
 \def\PJ      {\ensuremath{J}\xspace}
 \def\PK      {\ensuremath{K}\xspace}
 \def\PZ      {\ensuremath{Z}\xspace}
 \def\PD      {\ensuremath{D}\xspace}
 \def\Pb      {\ensuremath{b}\xspace}
 \def\Pc      {\ensuremath{c}\xspace}
 \def\Pd      {\ensuremath{d}\xspace}
 \def\Pe      {\ensuremath{e}\xspace}

 \def\Pq      {\ensuremath{q}\xspace}
 \def\Ps      {\ensuremath{s}\xspace}
 
 \def\Pu      {\ensuremath{u}\xspace}
 \def\thebaroffset{0.18em}
\newcommand{\offsetoverline}[2][\thebaroffset]{\kern #1\overline{\kern -#1 #2}}%

\DeclareRobustCommand{\optbar}[1]{\shortstack{{\miniscule (\rule[.5ex]{1.25em}{.18mm})}
  \\ [-.7ex] $#1$}}

\def\epem       {{\ensuremath{\Pe^+\Pe^-}}\xspace}

\def\mup        {{\ensuremath{\Pmu^+}}\xspace}
\def\mun        {{\ensuremath{\Pmu^-}}\xspace} 

\def\mumu       {{\ensuremath{\Pmu^+\Pmu^-}}\xspace}

\def\ellm       {{\ensuremath{\ell^-}}\xspace}
\def\ellp       {{\ensuremath{\ell^+}}\xspace}
\def\ellell     {\ensuremath{\ell^+ \ell^-}\xspace}

\def\neu        {{\ensuremath{\Pnu}}\xspace}
\def\neub       {{\ensuremath{\overline{\Pnu}}}\xspace}
\def\neul       {{\ensuremath{\neu_\ell}}\xspace}
\def\neulb      {{\ensuremath{\neub_\ell}}\xspace}
\def\neulneulb  {\ensuremath{\neul\neulb}\xspace}

\def\Hz     {{\ensuremath{\PH}}\xspace}
\def\Z      {{\ensuremath{\PZ}}\xspace}

\def\quark     {{\ensuremath{\Pq}}\xspace}
\def\quarkbar  {{\ensuremath{\overline \quark}}\xspace}
\def\qqbar     {{\ensuremath{\quark\quarkbar}}\xspace}
\def\uquark    {{\ensuremath{\Pu}}\xspace}
\def\uquarkbar {{\ensuremath{\overline \uquark}}\xspace}
\def\uubar     {{\ensuremath{\uquark\uquarkbar}}\xspace}
\def\dquark    {{\ensuremath{\Pd}}\xspace}
\def\dquarkbar {{\ensuremath{\overline \dquark}}\xspace}
\def\ddbar     {{\ensuremath{\dquark\dquarkbar}}\xspace}
\def\squark    {{\ensuremath{\Ps}}\xspace}
\def\squarkbar {{\ensuremath{\overline \squark}}\xspace}
\def\ssbar     {{\ensuremath{\squark\squarkbar}}\xspace}
\def\cquark    {{\ensuremath{\Pc}}\xspace}

\def\bquark    {{\ensuremath{\Pb}}\xspace}
\def\bquarkbar {{\ensuremath{\overline \bquark}}\xspace}
\def\bbbar     {{\ensuremath{\bquark\bquarkbar}}\xspace}

\def\pion   {{\ensuremath{\Ppi}}\xspace}

\def\pip    {{\ensuremath{\pion^+}}\xspace}
\def\pim    {{\ensuremath{\pion^-}}\xspace}

\def\kaon    {{\ensuremath{\PK}}\xspace}

\def\KorKbar {\kern \thebaroffset\optbar{\kern -\thebaroffset \PK}{}\xspace}
\def\KS      {{\ensuremath{\kaon^0_{\mathrm{S}}}}\xspace}

\def\Kp      {{\ensuremath{\kaon^+}}\xspace}
\def\Km      {{\ensuremath{\kaon^-}}\xspace}

\def\Kstarz  {{\ensuremath{\kaon^{*0}}}\xspace}
\newcommand{\phiz}{\ensuremath{\Pphi}\xspace}

\def\Dz      {{\ensuremath{\PD^0}}\xspace}

\def\jpsi     {{\ensuremath{{\PJ\mskip -3mu/\mskip -2mu\Ppsi}}}\xspace}

\def\B       {{\ensuremath{\PB}}\xspace}
\def\Bbar    {{\ensuremath{\overline{\PB}}}\xspace}
\def\Bd      {{\ensuremath{\B^0}}\xspace}

\def\BdorBdbar {\kern \thebaroffset\optbar{\kern -\thebaroffset \Bd}\xspace}
\def\Bu      {{\ensuremath{\B^+}}\xspace}

\def\Bs      {{\ensuremath{\B^0_\squark}}\xspace}
\def\Bsb     {{\ensuremath{\Bbar{}^0_\squark}}\xspace}
\def\BsorBsbar {\kern \thebaroffset\optbar{\kern -\thebaroffset \Bs}\xspace}

\def\Lz          {{\ensuremath{\PLambda}}\xspace}

\def\Xires       {{\ensuremath{\PXi}}\xspace}
\def\Lc          {{\ensuremath{\Lz^+_\cquark}}\xspace}

\def\Lb           {{\ensuremath{\Lz^0_\bquark}}\xspace}

\def\Xib          {{\ensuremath{\Xires_\bquark}}\xspace}

\newcommand{\decay}[2]{\mbox{\ensuremath{#1\!\to #2}}\xspace}

\def\BsToKKmm    {\decay{\Bs}{\Kp\Km\mup\mun}}

\def\bsll     {\decay{\bquark}{\squark \ell^+ \ell^-}}
\def\CP                {{\ensuremath{C\!P}}\xspace}

\def\eeZbb   {\decay{\decay{\epem}{\Z}}{\bbbar}}
\def\eeZH   {\decay{\epem}{\Z\Hz}}
\def\eeZHX  {\decay{\epem}{\Z(\to\text{anything})\Hz(\to\text{light quarks})}}

\newcommand{\aunit}[1]{\ensuremath{\text{\,#1}}}

\newcommand{\tev}{\aunit{Te\kern -0.1em V}\xspace}
\newcommand{\gev}{\aunit{Ge\kern -0.1em V}\xspace}
\newcommand{\mev}{\aunit{Me\kern -0.1em V}\xspace}
\newcommand{\kev}{\aunit{ke\kern -0.1em V}\xspace}
\newcommand{\ev}{\aunit{e\kern -0.1em V}\xspace}
 
\newcommand{\mevc}{\ensuremath{\aunit{Me\kern -0.1em V\!/}c}\xspace}
\newcommand{\gevc}{\ensuremath{\aunit{Ge\kern -0.1em V\!/}c}\xspace}
\newcommand{\mevcc}{\ensuremath{\aunit{Me\kern -0.1em V\!/}c^2}\xspace}
\newcommand{\gevcc}{\ensuremath{\aunit{Ge\kern -0.1em V\!/}c^2}\xspace}

\def\ps   {\ensuremath{\aunit{ps}}\xspace}

\def\deriv {\ensuremath{\mathrm{d}}}

\def\evtgen     {\mbox{\textsc{EvtGen}}\xspace}
\def\garfield   {\mbox{\textsc{Garfield}}\xspace}
\def\geant      {\mbox{\textsc{Geant4}}\xspace}
\def\pythia     {\mbox{\textsc{Pythia}}\xspace}
\def\keyhep     {\mbox{\textsc{key4hep}}\xspace}
\def\xgboost    {\mbox{\textsc{xgboost}}\xspace}

\newcommand{\ie}{\mbox{\itshape i.e.}\xspace}

\def\dndx{{\ensuremath{\deriv\!N\!/\!\deriv x}}\xspace}
\def\dedx{{\ensuremath{\deriv\!E\!/\!\deriv x}}\xspace}
\def\qsq       {{\ensuremath{q^2}}\xspace}

\def\idea     {\mbox{IDEA}\xspace}
\def\cld      {\mbox{CLD}\xspace}
\def\fccee    {\mbox{FCC-ee}\xspace}
\usepackage{graphicx}
\usepackage{dcolumn}
\usepackage{bm}
\usepackage{hyperref}
\usepackage{orcidlink} 
\usepackage{pgffor}
\usepackage{booktabs}
\usepackage{xcolor}
\usepackage{multirow}

\begin{document}

\preprint{arxiv:2025.xxxx}
\title{Flavor-physics benchmarks for tracker-based particle identification at the FCC-ee}
\thanks{Contact author: anbeck@mit.edu}
\author{Anja Beck\orcidlink{0000-0003-4872-1213}}
\author{Eluned Smith\orcidlink{0000-0002-9740-0574}}
\affiliation{\it\small Department of Physics and Laboratory for Nuclear Science, MIT, Cambridge, 02139, Massachusetts, USA}

\date{\today}
\begin{abstract}
The correct identification of charged hadrons plays a crucial role in flavor-physics measurements.
The final detector configurations at the proposed Future Circular Collider are yet to be determined and this study aims to contribute to this discussion by benchmarking the particle-identification (PID) performance of the proposed CLD and IDEA detectors using fully simulated events.
At present, neither detector proposal includes dedicated PID systems, relying instead on information from the tracking subsystems.
We estimate the expected level of contamination due to misidentified charged hadrons for \bquark-flavor tagging, rare $\bquark\to\squark$ transitions, and \squark-jet tagging.
The PID information provided by silicon trackers, namely time-of-flight and energy-deposit measurements, leads to significant background suppression with high signal efficiency for the low-momentum hadrons considered for same-side \bquark-flavor tagging.
In order to improve the contamination in rare decays where momenta are in the medium range, only good timing resolution of 30\ps and below can yield an improvement of one order of magnitude below the level achieved by kinematic criteria alone.
Light-quark jet-flavor tagging requires identification of particles with very large momentum,  which is not possible using only time-of-flight or energy-deposit information in silicon.
Access to the number of clusters in a drift-chamber setup, as proposed for the IDEA detector, however, results in strong background suppression in every case.
This suppression can be further improved in some scenarios by time-of-flight resolution of 30--50\ps or better.
The PID quality generally exhibits only a small dependence on the cluster-counting efficiency.
Whether dedicated PID detectors could further enhance flavor-physics sensitivity should be the subject of future study. 
\end{abstract}

\maketitle

\section{Introduction}\label{sec:intro}
The proposed Future Circular Collider (\fccee) is designed to provide electron-positron collisions at different center-of-mass energies in order to determine many parameters of the Standard Model (SM) with unprecedented precision~\cite{FCC:2025lpp}.
In particular, experiments at the \fccee will be able to measure  a range of flavor observables with world-leading precision: the clean environment will result in significantly higher selection efficiency and smaller systematic uncertainty on measurements of heavy-flavor decays compared to current LHC experiments, while still providing access to the full range of \bquark- and \cquark-hadrons inaccessible at Belle~II as well as access to orders of magnitude more data~\cite{Belle-II:2018jsg}.
Also, the flavor reach in the electroweak sector is expected to improve strongly compared to the ATLAS and CMS experiments given the simpler environment and better particle identification (PID).

While the focus and specifications of the detectors is yet to be determined, the composition of high-energy physics detectors mostly follows the same principles.
A staple feature in all collider experiments is the tracking system enabling one to trace the path of charged particles and their subsequent combination to decay topologies.
All modern tracking systems include a silicon vertex tracker sitting close to the interaction region combined with a second tracking subsystem located further away from the interaction point.
The popularity of silicon vertex detectors can be attributed to their excellent spatial resolution, practically instant response time, and the required radiation hardness for elements sitting close to the interaction region of particle beams~\cite{viehhauser2024detectors}.
The hit resolution can be as low as a few micrometers as shown for the novel MAPS technology to be implemented in a future upgrade of the ALICE detector~\cite{Ricci:2023lpm}.
Due to its large material budget and/or cost, however, silicon is not a common choice for large-area detectors far from the interaction region in $ee$ colliders.
More lightweight solutions are, for example, drift chambers or time projection chambers.

Besides tracking, particle-physics experiments also need to deduce the particle type.
This information can be extracted from the signatures that particles leave when traversing different materials.
A calorimeter system is crucial for identifying neutral particles, electrons, muons, and charged hadrons based on the total amount of energy deposited and the spatial extent of the deposition.
Any measurements related to flavor properties usually require a more granular distinction of charged hadrons into protons, kaons, or pions.
Whereas the tracking system provides the momentum of a particle, additional knowledge of its speed allows direct inference of the mass and hence the identity of the particle.
Speed can be measured by clocking the time of flight (ToF) along a known path or via the opening angle of a Cherenkov light cone.
Similarly, the relationship between either the energy deposited through ionization or the number of ionization clusters with a particle speed can enable particle identification.
Depending on the overall design and specialization of a detector, this can be achieved by dedicated particle-identification systems.
It may, however, be convenient and economical to avoid the need for a dedicated subsystem and instead design the tracking system for dual use.
Avoiding a dedicated system may also allow for a reduction in the overall material budget of the detector. 

This paper presents a study of the charged-hadron identification capabilities of the tracking systems as currently envisioned in the \cld~\cite{Bacchetta:2019fmz} and \idea~\cite{IDEAStudyGroup:2025gbt} detector concepts.
The respective tracking detectors and the PID-related quantities obtained from full simulation are presented in Sec.~\ref{sec:detectors}.
Section~\ref{sec:pid} outlines the employed particle-identification algorithm.
Afterward in Sec.~\ref{sec:examples}, this tool is applied to three flavor-physics examples in order to determine the expected improvement in contamination for varying levels of time of flight resolution and cluster-counting efficiency as well as with and without silicon-based \dedx information.
This paper concludes with a discussion and summary in Sec.~\ref{sec:summary}.

\section{Detector concepts and their tracking systems}\label{sec:detectors}
The two most mature detector concepts for the \fccee are \cld and \idea.
As the name suggests, the \textit{CLIC-like detector} (\cld), described in detail in Ref.~\cite{Bacchetta:2019fmz}, was inherited from CLIC studies.
Its tracking system is entirely silicon-based and comprises a pixel vertex tracker as well as two more tracking systems which may use pixels or strips.
Ref.~\cite{IDEAStudyGroup:2025gbt} provides a detailed description of the other investigated concept named the \textit{innovative detector for \epem accelerators} (\idea).
The three tracker subsystems of \idea are a silicon pixel vertex-tracker, a high-transparency drift chamber filled with a mix of 90\% helium and 10\% $\text{C}_4\text{H}_{10}$ gas inspired by the KLOE and MEG2 experiments, and a wrapper using either silicon strips, MAPS, or LGADs.
The results obtained from the \idea studies in this paper can be directly transferred to a third maturing detector concept called \textit{a lepton-lepton collider experiment with granular read-out} (ALLEGRO) as they share the same tracking system~\cite{Pekkanen:2024zbe}.
All studies described in this paper are performed on simulation of the \cld and \idea setups using the \keyhep framework~\cite{Ganis:2021vgv}.

The silicon trackers in all detector concepts can provide particle identification by measuring the speed of a particle as the ratio of flight distance over time of flight.
In this study, the reference points for this measurement are always the first and last recorded hit in any silicon layer.
This choice removes uncertainties invoked for particles originating from displaced vertices and does not rely on assumptions regarding the \fccee clock resolution.
In order to study different values for the time-of-flight precision in the absence of possible improvements due to reconstruction, the true time-of-flight values are smeared using Gaussian distributions.
As a consequence, the term \textit{time-of-flight resolution} in this publication always refers to the magnitude of the $\sigma$ used in the Gaussian to randomly add noise to the total time of flight rather than the timing resolution of an individual pixel.
This choice is motivated by the fact that sophisticated tracking algorithms can improve the final time-of-flight resolution for a track significantly relative to the hit-time resolution; see e.g. Ref.~\cite{Heijhoff:2020mlk}.
The uncertainty imposed on the flight distance by the finite size of each pixel is naturally included in this study as the hit location is taken from the output of full simulation.
Tracking and clustering algorithms can also improve this spatial uncertainty but it is already so small that the impact on the study at hand is negligible.
Making our results dependent on the time of flight resolution rather than the timing capabilities of the silicon chips, the \fccee clock, or reconstruction details also means that the conclusions are portable to different setups as long as the considered absolute time of flight is similar, e.g. using the \fccee clock and a single timing layer just before the calorimeter system.

Another useful quantity for particle identification is the energy deposit per traveled distance, \dedx, in each silicon sensor.
The ability of the detector to measure the energy deposit is assumed to be modeled realistically in the simulation and used without additional smearing.
Inspired by CMS studies, the \dedx value per track is determined as the harmonic mean with grade $-2$ of all recorded hits in silicon layers~\cite{CMS:2025bbk}.

Finally, the drift chamber in the \idea concept also allows for a measurement of the number of primary ionization clusters per traveled distance, \dndx.
Because the number of clusters follows a Poissonian distribution instead of a Landau distribution, \dndx is preferred over \dedx, which can also be measured with a drift chamber.
The number clusters is not yet available in the \idea simulation but instead extrapolated based on \garfield simulations and test-beam studies~\cite{Cuna:2021sho} while the traveled distance is taken from the detector \geant simulation without modification.
The extrapolated value for the number of clusters in each cell is varied according to a Poissonian.
The \dndx value per track is obtained by adding the number of clusters in all cells and dividing by the total distance.
In order to study the impact of different cluster-counting efficiencies later on, each cluster is kept or discarded stochastically with a probability corresponding to the chosen efficiency.

\begin{figure}
    \centering
    \includegraphics[width=.95\linewidth]{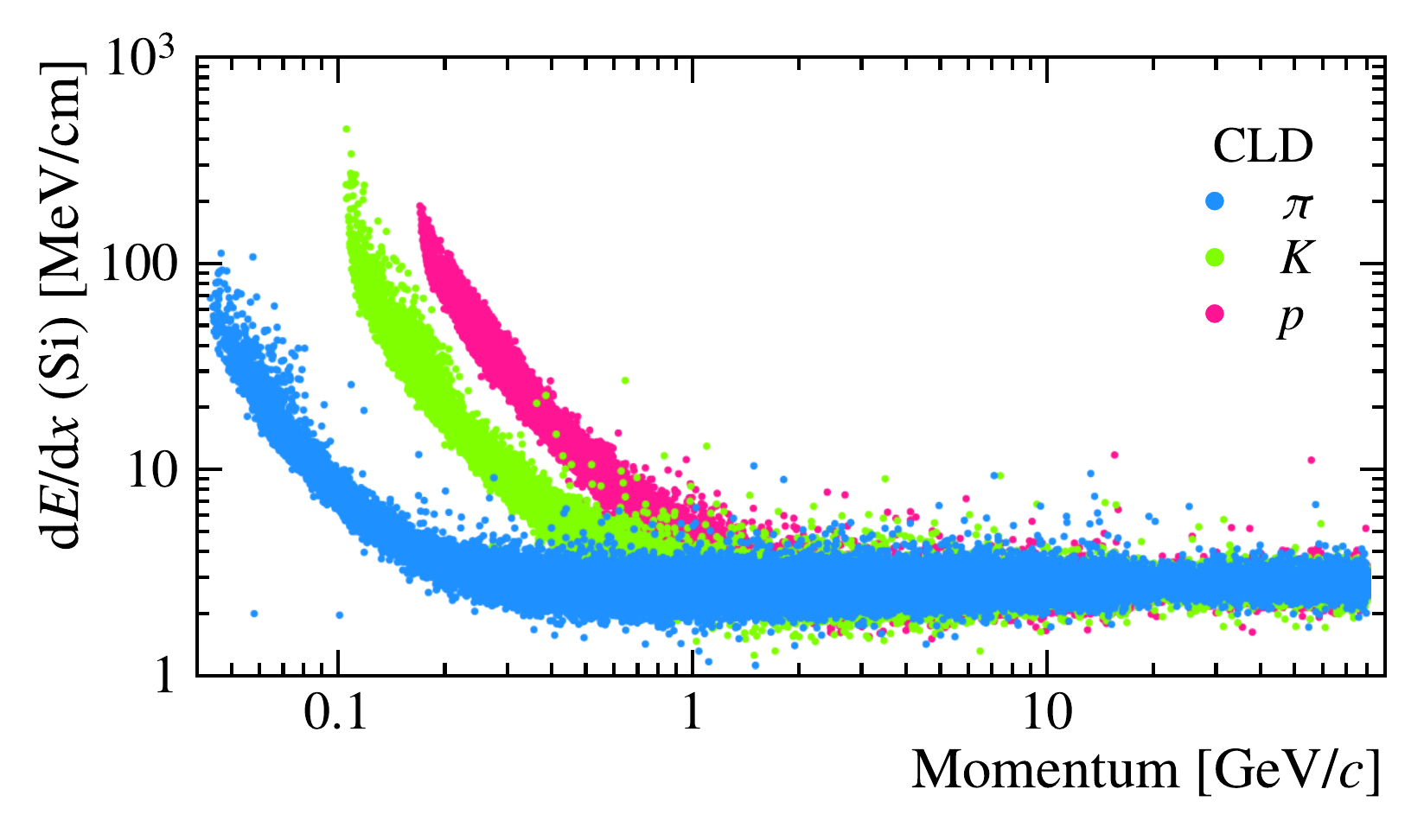}
    \includegraphics[width=.95\linewidth]{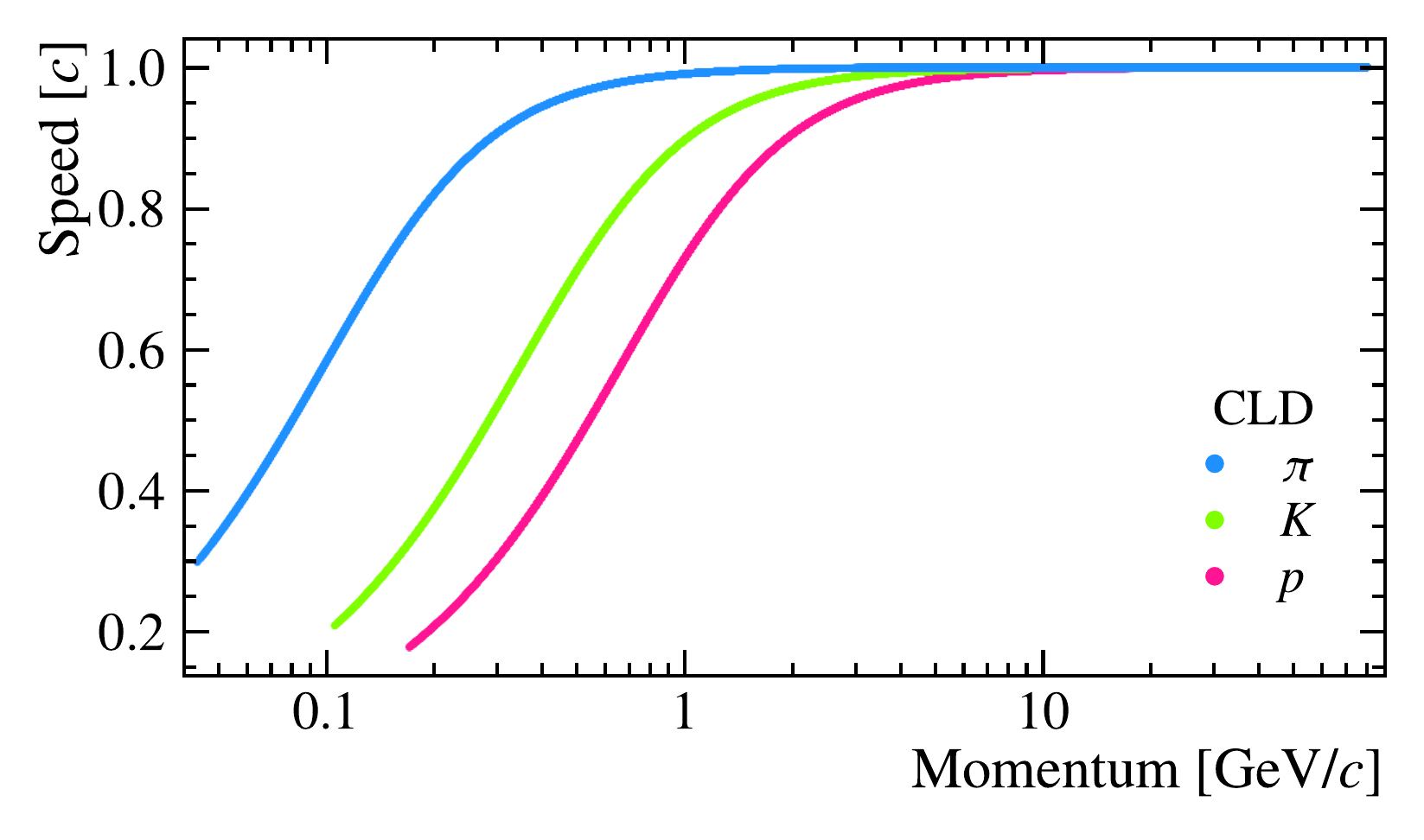}
    \caption{Input quantities for particle identification using \cld tracking detectors. Top: \dedx in the silicon sensors, calculated as the harmonic mean of all hits in silicon trackers. Bottom: true speed measured using the time of flight and flight distance between the first and the last hit in silicon trackers.}
    \label{fig:input:cld}
\end{figure}

\begin{figure}
    \centering
    \includegraphics[width=.95\linewidth]{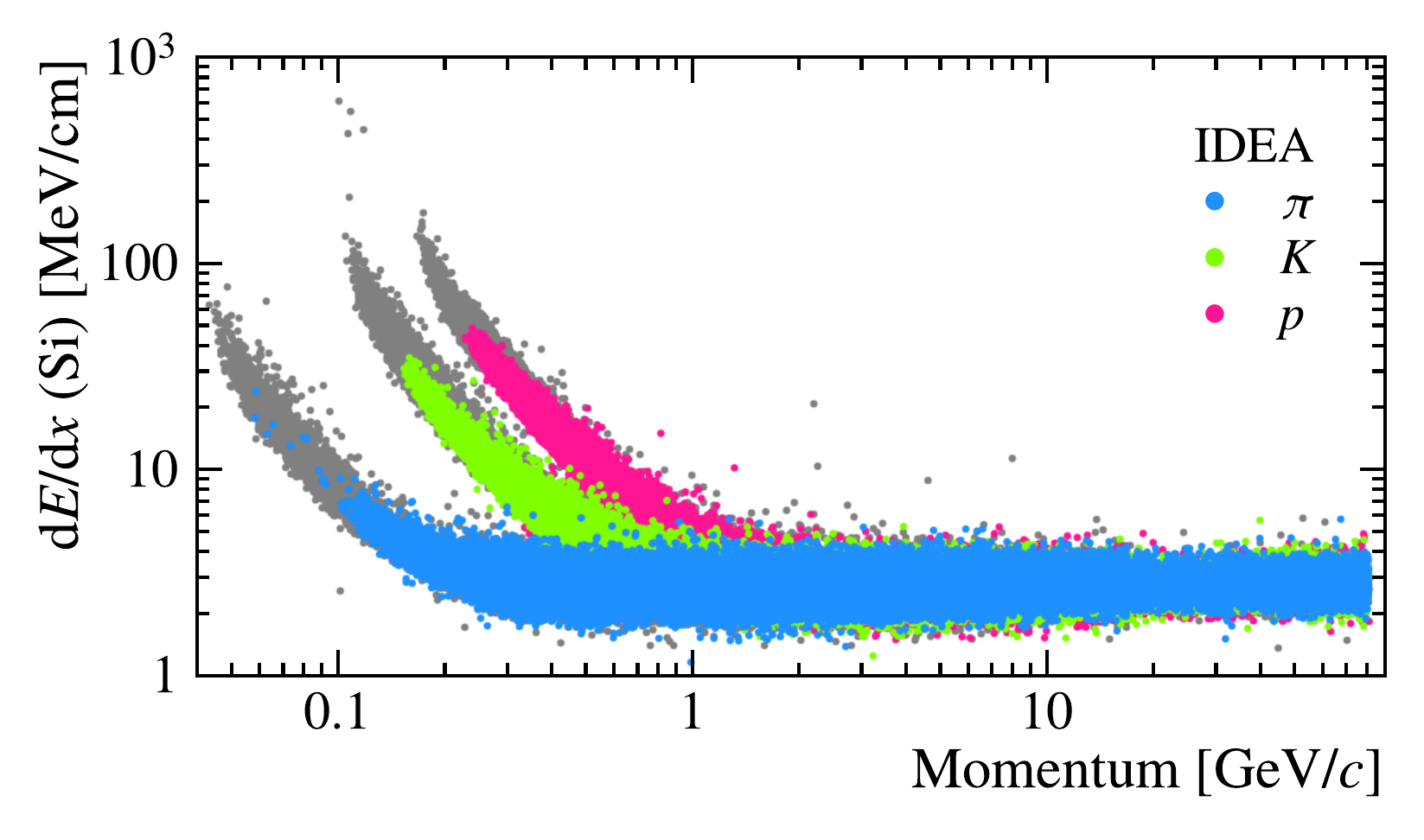}
    \includegraphics[width=.95\linewidth]{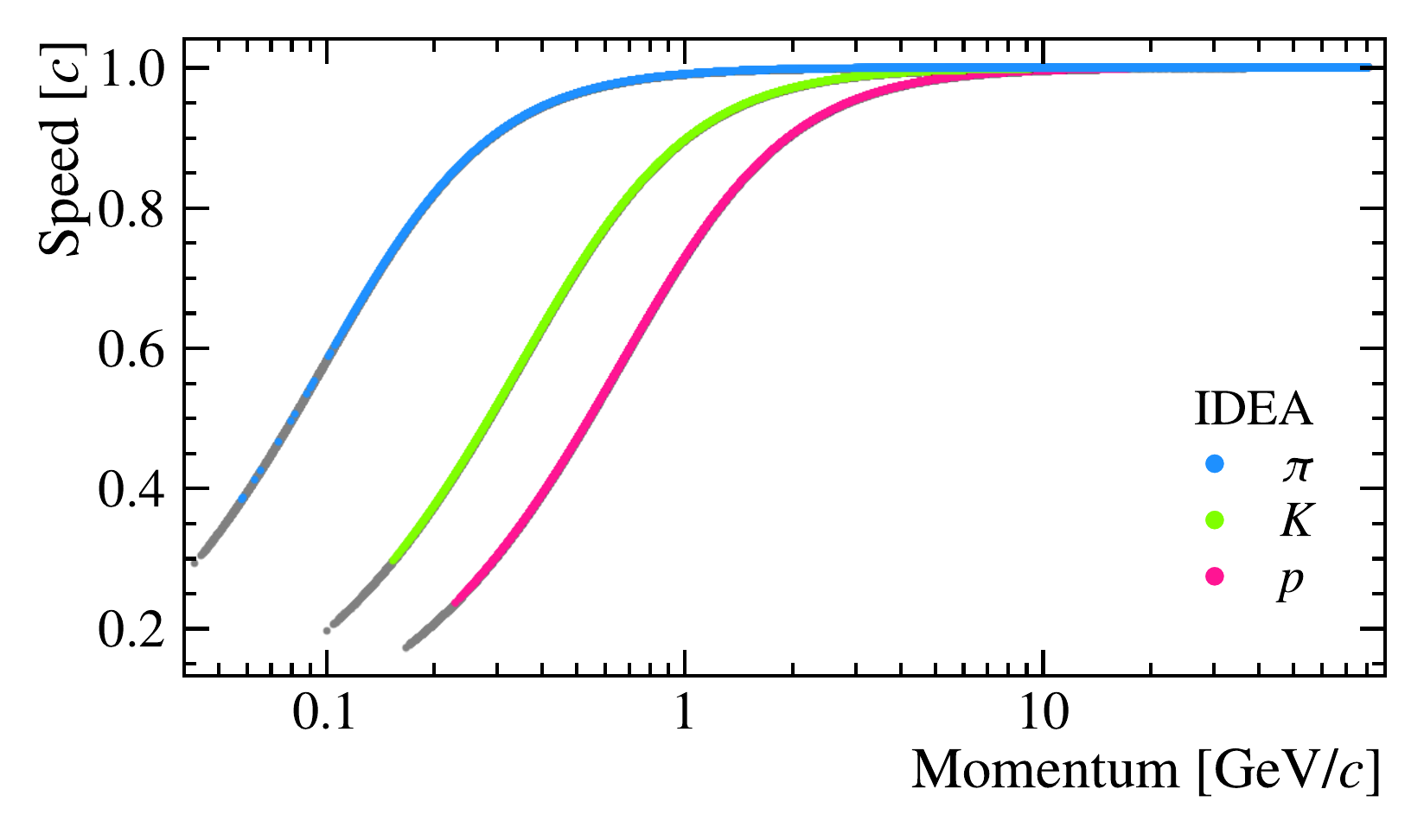}
    \includegraphics[width=.95\linewidth]{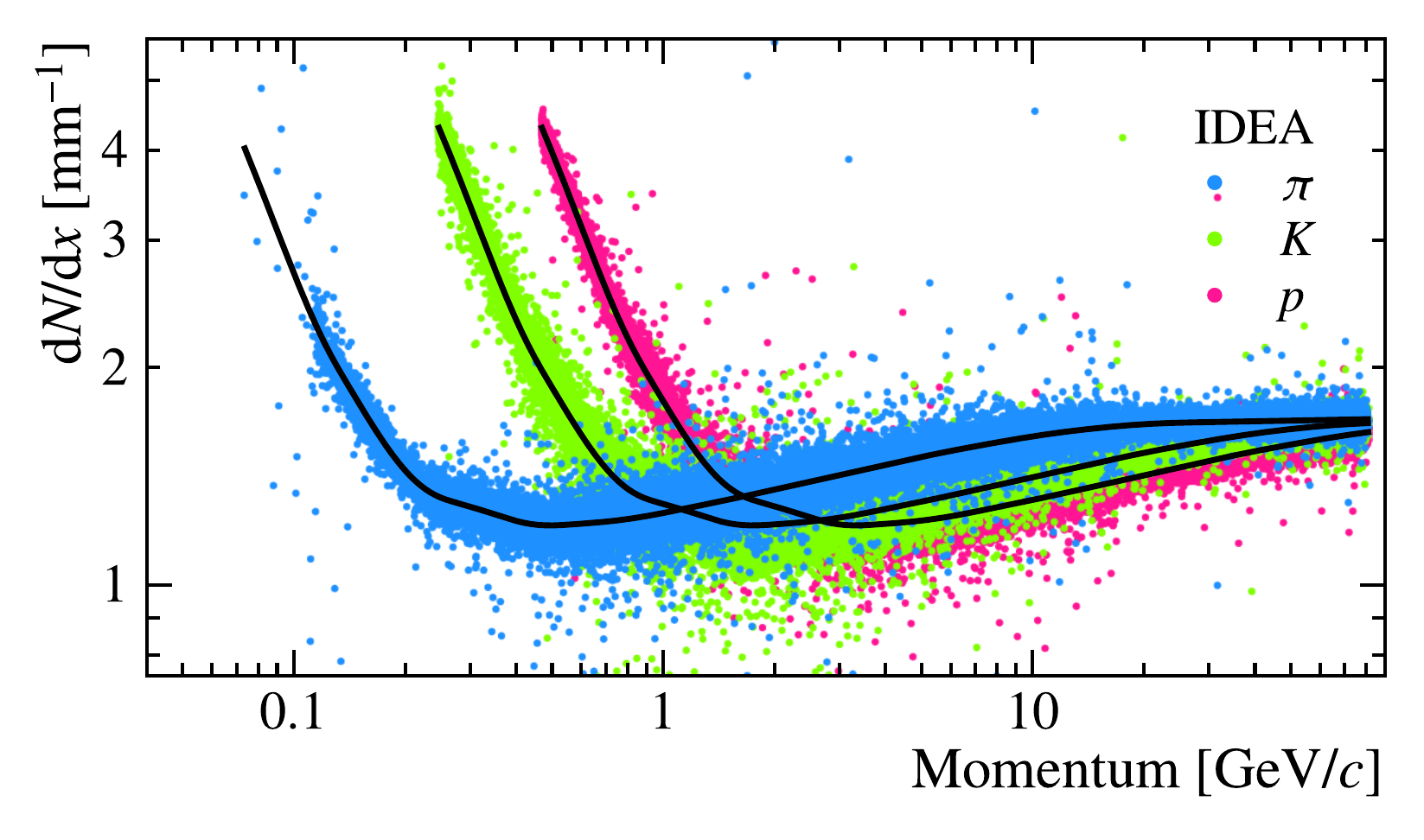}
    \caption{Input quantities for particle identification using \idea tracking detectors. 
    Top: \dedx in the silicon sensors, calculated as the harmonic mean of all hits in silicon trackers.
    Middle: true speed measured using the time of flight and flight-distance between the first and the last hit in silicon trackers.
    Bottom: \dndx interpolated from \garfield simulation (black line) with Poissonian variation at the generation stage assuming perfect efficiency (dots).}
    \label{fig:input:idea}
\end{figure}

Figures~\ref{fig:input:cld} and \ref{fig:input:idea} show the dependence of these particle-identification quantities on the momentum for proton, kaon, and pion tracks in the two detector concepts, assuming perfect time-of-flight resolution.
Similar figures for nonzero time-of-flight resolution are shown in Appendix~\ref{app:tofs}.
The particles used to make these displays are also used to train the classifiers later and are generated isotropically in space using a particle gun.
Because no full tracking algorithm is yet available for the \idea detector, the generated momentum is used for both \idea and \cld studies.
Given that a momentum resolution well below 1\% in the momentum range $p<100\gevcc$ is achievable even for the much more complex events at the LHC~\cite{LHCB-FIGURE-2024-040}, this choice has negligible impact on the results.

Tracks sitting outside the acceptance of the silicon detectors are discarded from the study.
The gray points in the \idea figures correspond to tracks with hits in the silicon trackers resulting in valid speed and \dedx measurements but without hits in the drift chamber.
The two main reasons for the lack of \dndx information are (i)) the available interpolation to calculate the \dndx values is only valid for $\beta\gamma\geq0.5$ and (ii)) the silicon detectors in \idea and \cld have further acceptance reach toward the beam pipe compared to the drift chamber.

Before describing the study, we want to summarize the limitations of the simulation samples to date.
The lack of a reconstruction tool for the \idea detector requires the use of the true momentum and the true time of flight of the particles, where the latter is smeared with a range of possible resolution values.
The spatial resolution resolution, however, is obtained realistically, \ie from the simulated hit location.
Provided that the \fccee environment is very clean and modern reconstruction algorithms are very efficient, these idealizations are likely not far from reality.
Additionally, any inverse correlation between spatial and timing resolution, where improving the latter may require a higher material budget if derived from the first and last hits, is not taken into account.

The \idea simulation does not yet include a realistic drift-chamber response resulting in the need for external simulation and test-beam extrapolations for the number of drift-chamber clusters.
In summary, this study sits at the intersection of fast simulation and full simulation.
All quantities are calculated on a hit-level basis which is only possible in full simulation.
The resolution in the hit location as well as the energy deposit are fully simulated.
Only the time of flight resolution as well as the cluster-counting efficiency are introduced artificially resembling fast simulation.

\section{Particle-identification algorithm}\label{sec:pid}
\begin{figure}
    \centering
    \includegraphics[width=\linewidth]{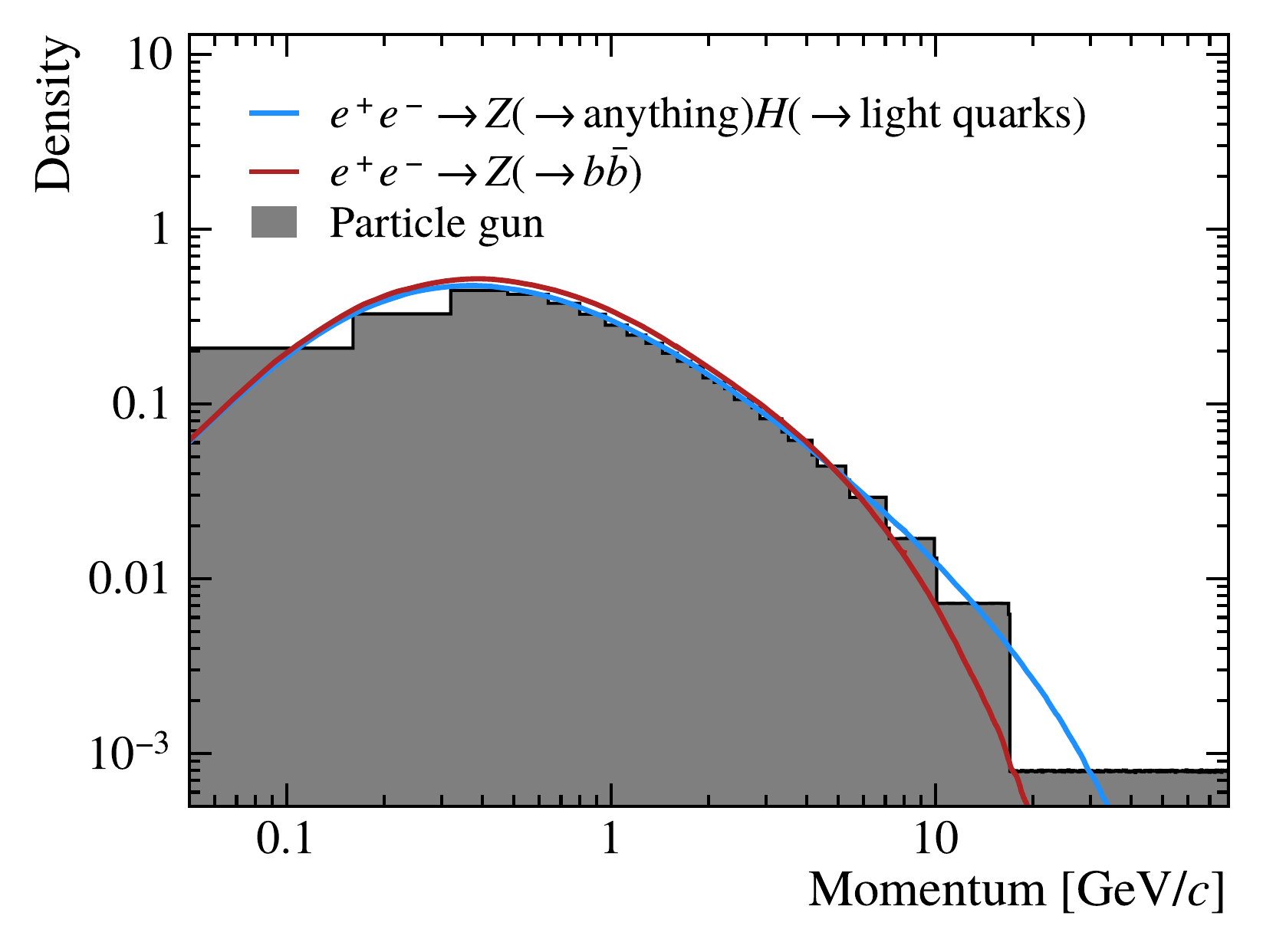}
    \caption{Generator-level distributions of the charged-hadron momenta in the samples used to study specific flavor measurements and the particle-gun samples employed for training the classifiers.}
    \label{fig:momentum}
\end{figure}

We employ boosted decision trees (BDTs) to determine the identity of a particle because they are able to capture high-dimensional correlations while providing high interpretability, effectively producing a nontrivial binning scheme across all input variables.
The training events are individual particles generated isotropically ensuring full coverage of the angular space.
For each charge and species, 100~000 particle-gun events are produced, ensuring sufficient balance between protons, kaons, and pions despite differences in reconstruction efficiency, resulting in a total of 600~000 training particles.
The three flavor-physics examples considered later stem from \eeZbb and \eeZHX events.
Given that particle identification relies heavily on the momentum of a particle, the particle-gun training samples are generated in 20 bins of equal content defined based on the momentum distribution of the \eeZH events considered later.
Within each bin, the momenta are distributed uniformly.
The $\Z\Hz$ events are chosen as they have a broader momentum profile.
Figure~\ref{fig:momentum} illustrates the momentum distribution of the charged hadrons for different event types and the particle-gun samples.
The particle-gun events are propagated through full \geant simulation of both the \idea and \cld detector concepts~\cite{geant4first,geant4second,geant4third,FCC-config_HEP-FCC,CLDConfig_key4hep}.

The particle-identification algorithm consists of three independent BDTs trained using the \xgboost algorithm~\cite{Chen_2016} to select protons, kaons, or pions over the other two species based on the momentum and different configuration of \dndx, silicon-based \dedx, and speed.
For an unseen particle, the species is assigned according to the BDT with the highest output score.
The BDT architecture is limited to 1000 trees with at most four layers of depth.
The learning rate is set to 0.02.
Because the total number of features is very small (two to four depending on the configuration), one very noisy feature can easily lead to overfitting.
Three measures are taken to mitigate this problem as much as possible.
First, the decision whether and how to split a branch relies on a subset of training samples that is randomly chosen at each node and represents 80\% of the total training dataset.
Second, a branch only splits if the associated loss reduction exceeds a value of three and there are at least 15 data points in each child.
Third, the training stops when the validation loss does not change for 40 iterations.
Each BDT is trained on 80\% of the particle-gun samples and the remaining 20\% are kept for validation.
In order to achieve similar magnitudes of the features, the momentum is scaled using $\log(p)$, the speed $v$ is scaled using \mbox{$-\log(1-\tfrac{v}{c})$} and the energy deposit \dedx is scaled using $\log(\dedx)$ before training.

\begin{figure}
    \centering
    \includegraphics[width=\linewidth]{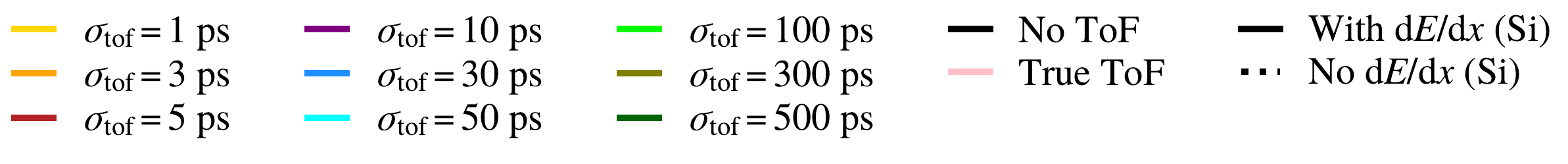}
    \includegraphics[width=0.5\linewidth]{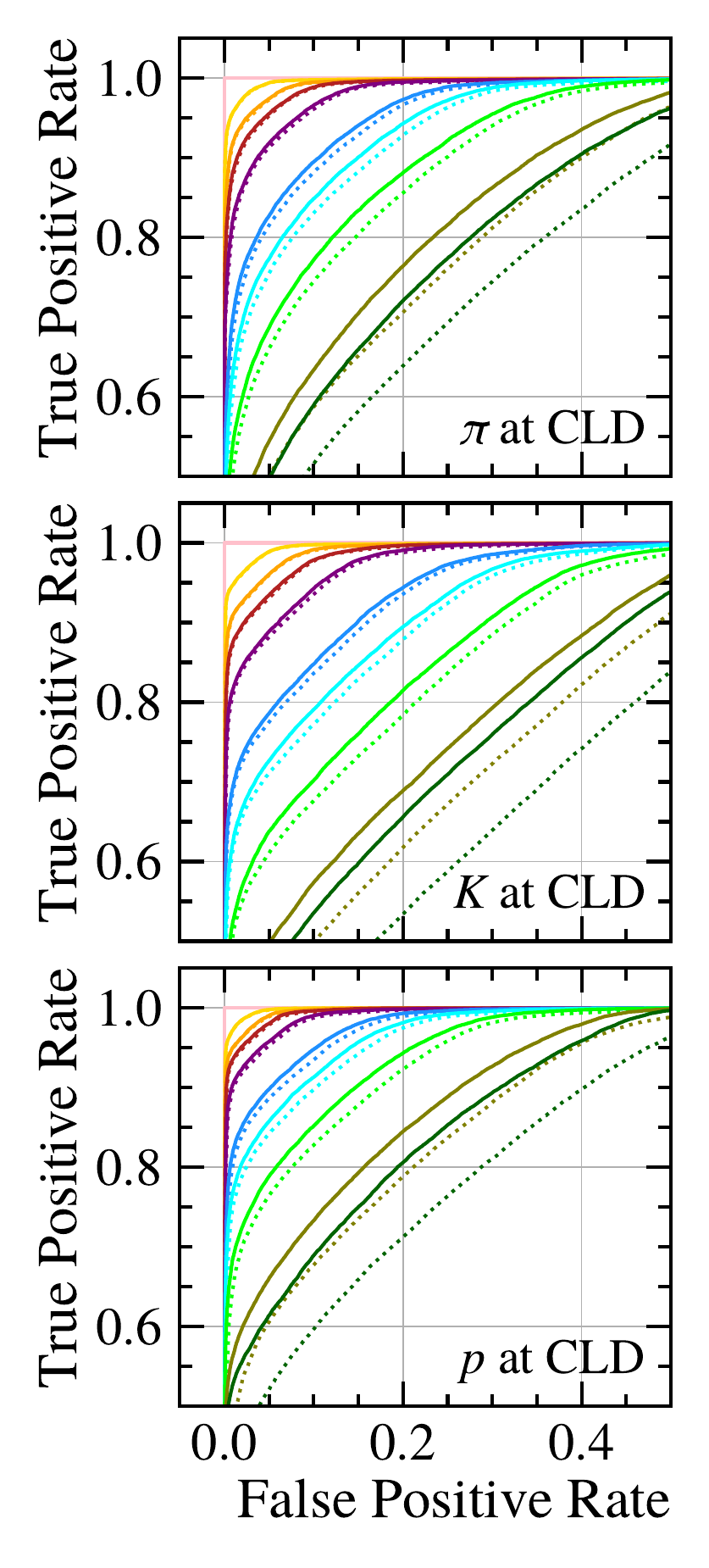}%
    \includegraphics[width=0.5\linewidth]{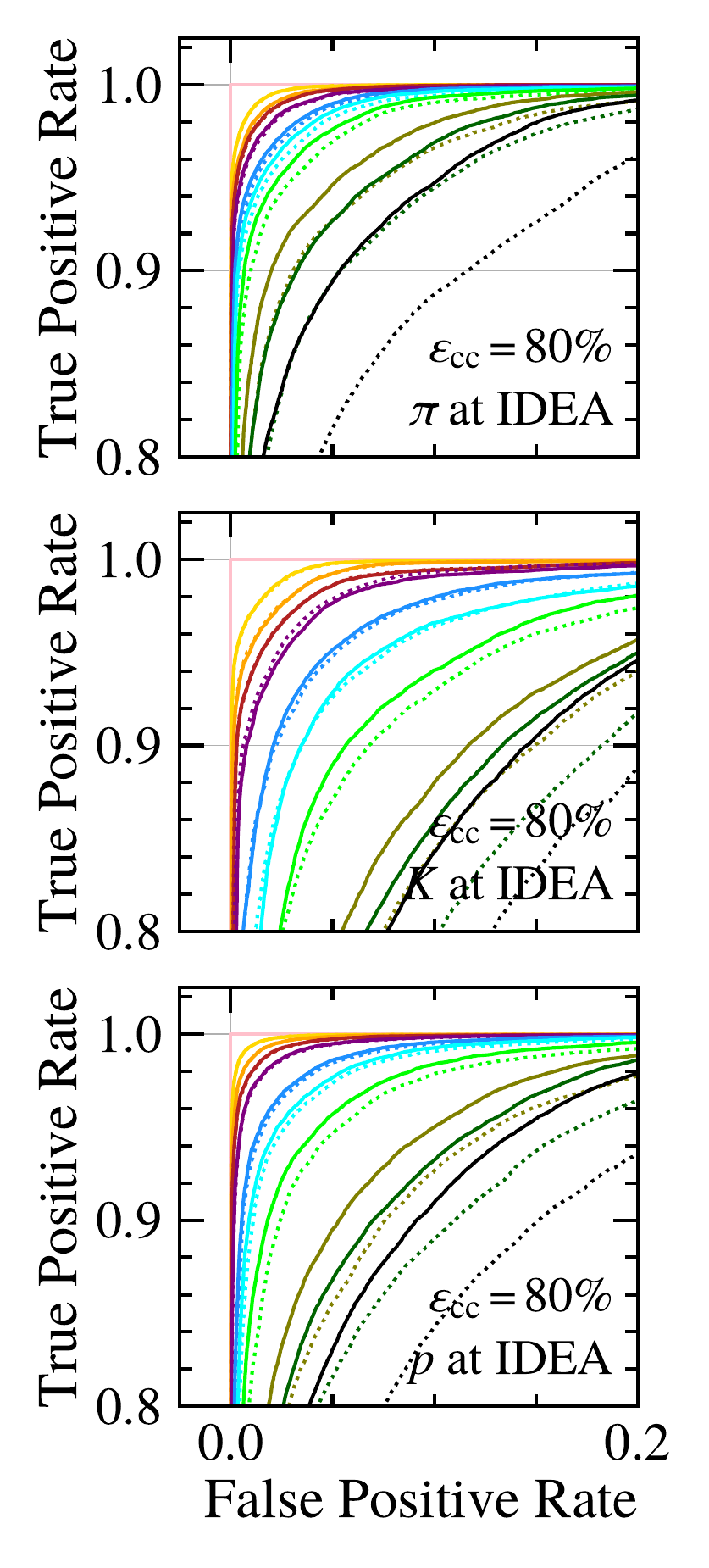}
    \caption{ROC for the (top) pion, (middle) kaon, and (bottom) proton identification at (left) \cld and (right) \idea.
    The colors indicate the time-of-flight (ToF) resolution and the line style indicates whether the silicon-based \dedx information was used.
    A cluster-counting efficiency $\varepsilon_\text{cc}$ of 80\% is assumed for the drift-chamber reconstruction at \idea.
    Note that the axes ranges are different between the \cld and \idea figures.}
    \label{fig:roc}
\end{figure}

The left column of plots in Fig.~\ref{fig:roc} shows the receiver-operator curves (ROCs) for different time of flight resolution (indicated by color) at \cld calculated on the unseen particle-gun validation samples.
The solid lines represent BDTs that use the energy deposit in the silicon trackers.
For each solid line, there is a corresponding dotted line which represents the same setup but without \dedx information.
For large time-of-flight resolutions, the additional \dedx input improves the AUC notably while it has negligible impact for better time-of-flight resolution.
Even the poorest considered time-of-flight resolution of 500\ps leads to AUC values significantly larger than 0.5.

The right column in Fig.~\ref{fig:roc} shows the same set of plots but for \idea with a cluster-counting efficiency of 80\%.
Comparing the solid lines between the \cld (left) and \idea (right) performance reveals that the additional \dndx information improves the classification.
Note that the black lines representing a scenario without time-of-flight measurement reaches a similar AUC as relying on a time-of-flight measurement of 10\ps for pion identification (purple line in the \cld plot).
For all setups, proton and pion identification achieves higher AUCs than kaon identification because in the latter case outliers in both directions lead to misidentification compared to only one direction for protons and pions.

\begin{figure}
    \centering
    \includegraphics[width=\linewidth]{plots/particlegun/legend_auc.pdf}
    \includegraphics[width=0.5\linewidth]{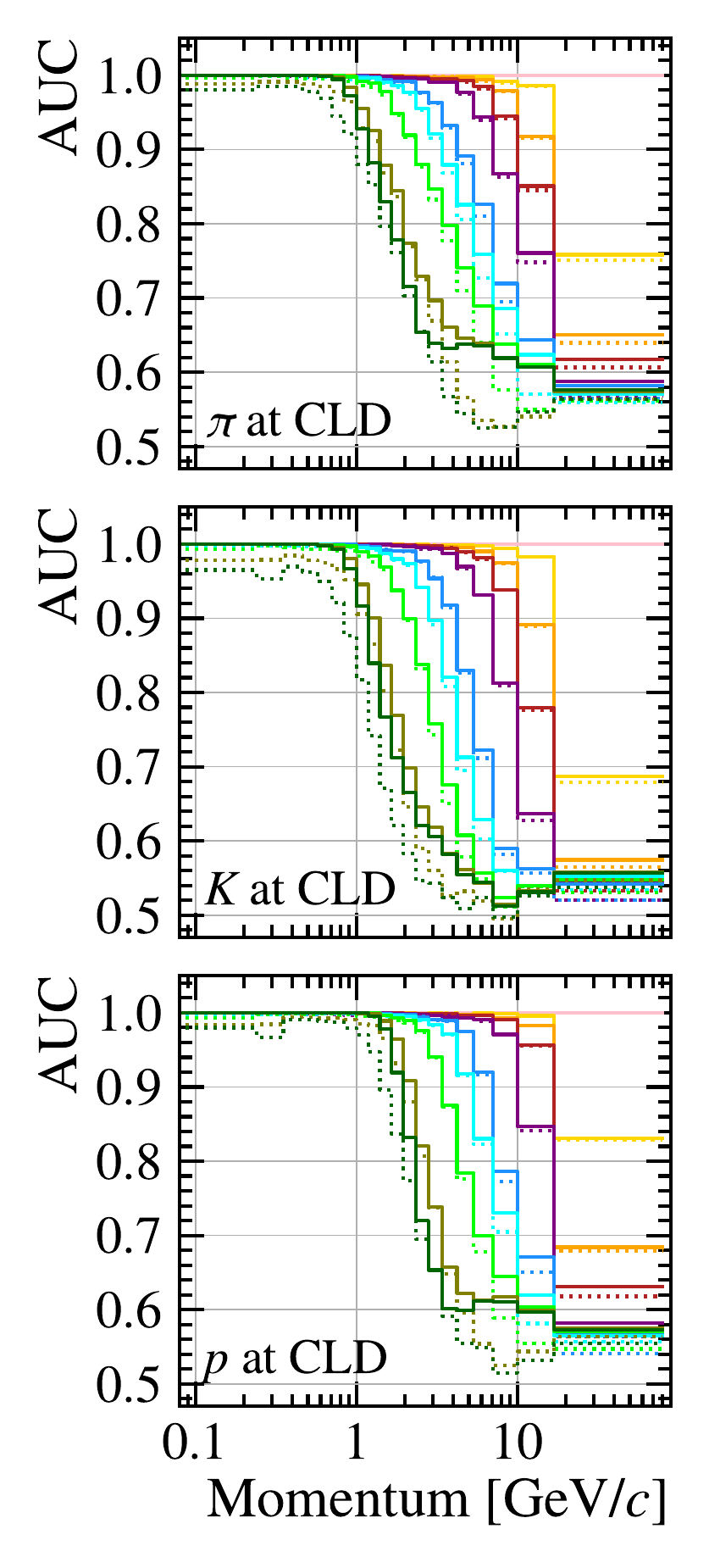}%
    \includegraphics[width=0.5\linewidth]{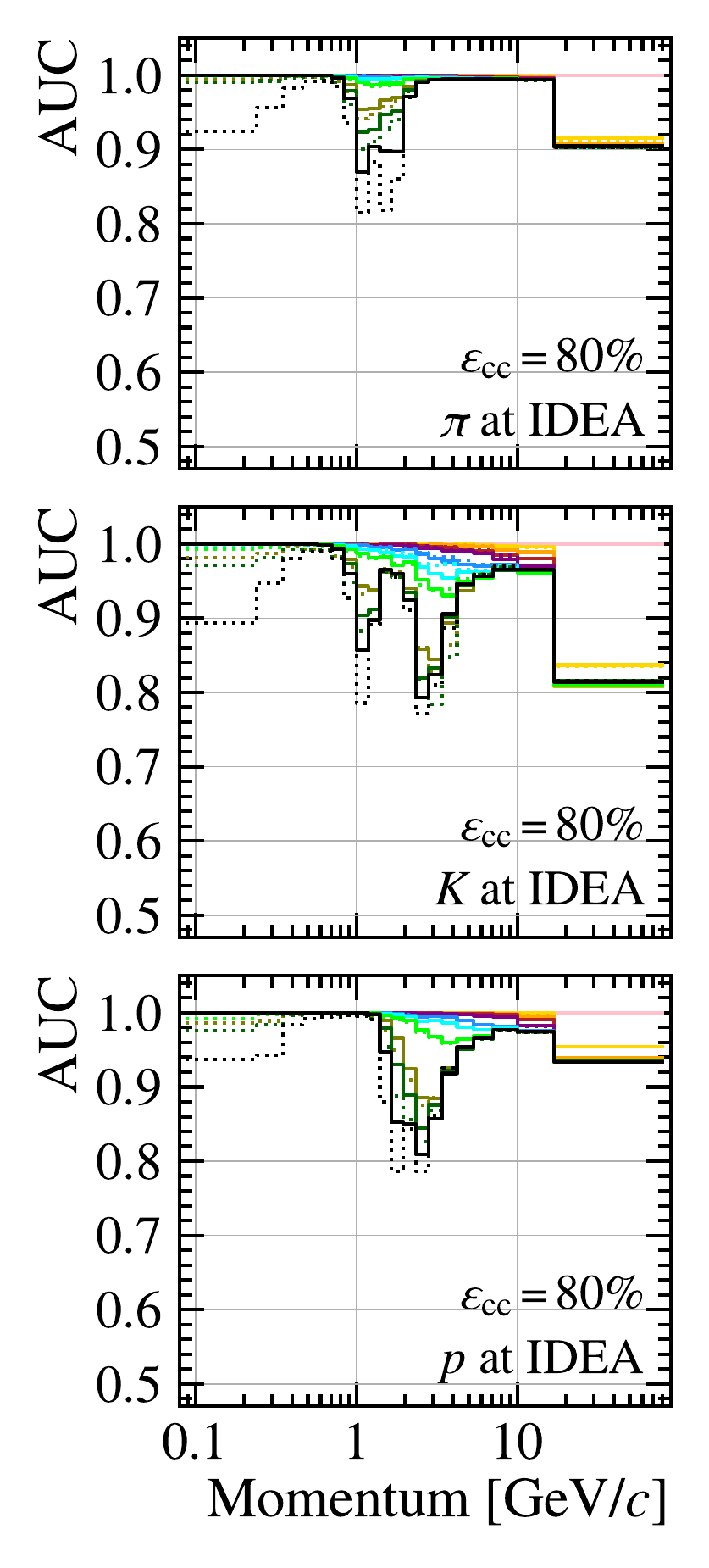}
    \caption{AUC in different momentum ranges for the (top) pion, (middle) kaon, and (bottom) proton identification at (left) \cld and (right) \idea.
    The colors indicate the time of flight (ToF) resolution and the line style indicates whether the silicon-based \dedx information was used.
    A cluster-counting efficiency $\varepsilon_\text{cc}$ of 80\% is assumed for the drift-chamber reconstruction at \idea.}
    \label{fig:roc:bins}
\end{figure}

Figure~\ref{fig:roc:bins} shows the dependence of the AUC on the particle momentum.
While identification in the very low momentum regime is close to perfect in almost any setup, the cluster-counting information at \idea leads to strong particle-identification performance even at higher momenta.
The dips in the \idea AUC scores at medium momentum are a consequence of the crossing between the \dndx curves as shown in the bottom part of Fig.~\ref{fig:input:idea}.
The two intersections of the kaon-pion and kaon-proton curves are clearly visible as two dips in the kaon identification performance.
In the proton (pion) case, the two performance dips merge into one because the distance between the proton-pion and proton-kaon (pion-kaon and pion-proton) curves is smaller.
The theoretical \dedx shapes intersect in the same way which is not visible in Figs.~\ref{fig:input:cld} and \ref{fig:input:idea} due to finite resolution and the lack of density information in the scatter plot.
Despite the dilution, this effect is slightly visible for \cld and poor time of flight resolution.
At 500\ps for example (dark green lines), adding \dedx information (solid line) on top of pure time of flight classification (dotted line) increases the AUC score visibly at low and high momenta while the improvement is only moderate in the 1--3\gevc region.

The dependence on the cluster-counting efficiency is generally rather low as the classifier simply learns the  additional scaling.
Appendix~\ref{app:aucs} shows the information in Figs.~\ref{fig:roc} and \ref{fig:roc:bins} for cluster-counting efficiencies of 50\% and 100\%.

To represent the performance of a particle-identification tool, the fraction of misidentified particles, $f$, is converted into a number of standard deviations, $\sigma$, using
\begin{align}\label{eq:sigmas}
    f_\sigma = \Phi^{-1}(1-f) \ ,
\end{align}
where $\Phi$ is the cumulative distribution function of the normal distribution.
We choose not to explicitly quote a separation power, as the BDT responses are non-Gaussian.
However the separation of the  BDT discriminator distributions for different particle species can be judged by eye from the plots in Appendix~\ref{app:bdtscore} for configurations with a time-of-flight resolution of 30\ps and cluster-counting efficiency of 80\%.

\begin{figure}
    \centering
    \hfill\includegraphics[width=0.9\linewidth]{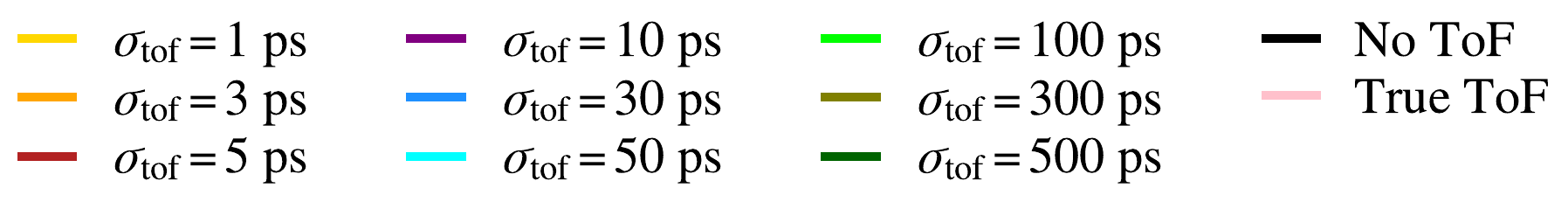}
    \includegraphics[width=\linewidth]{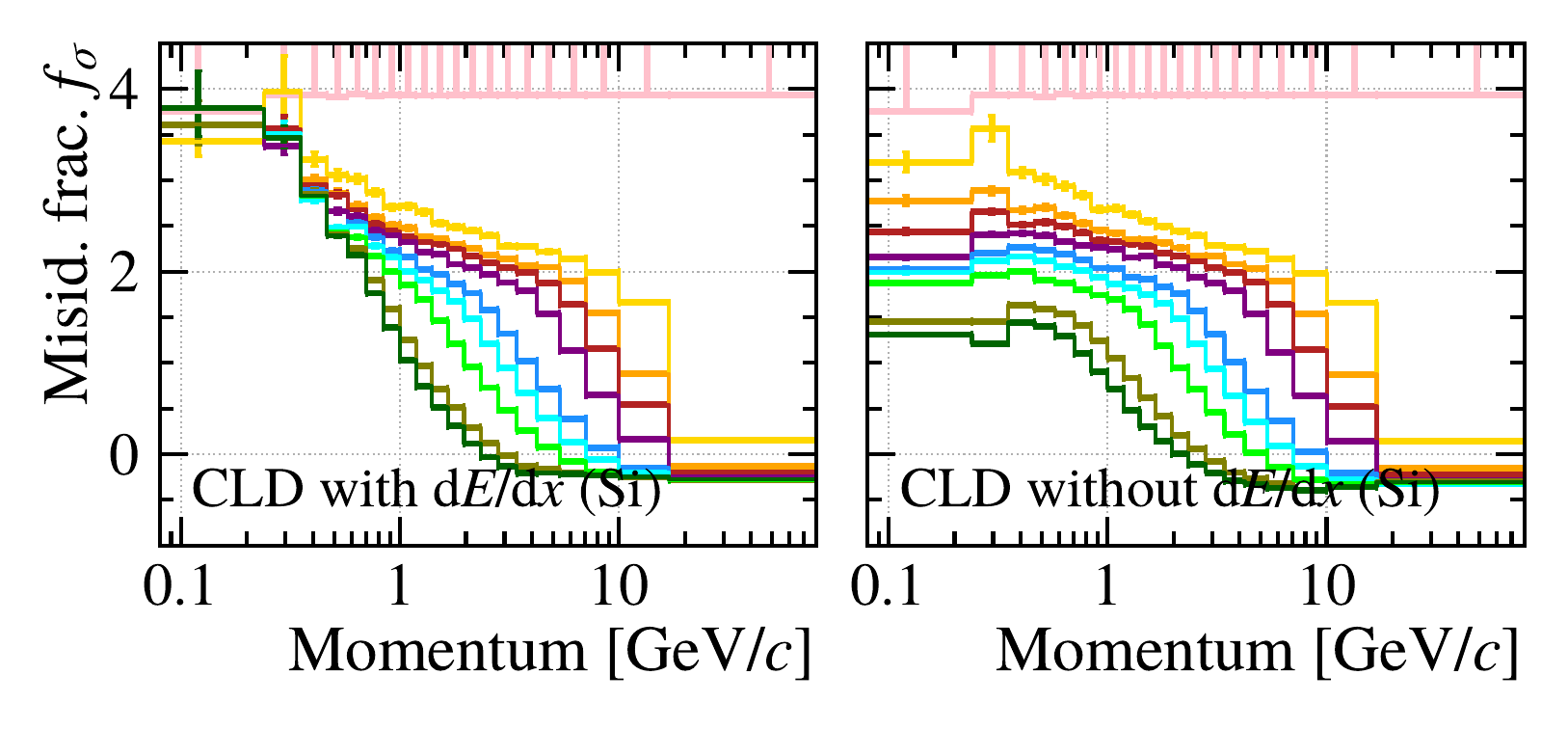}
    \includegraphics[width=\linewidth]{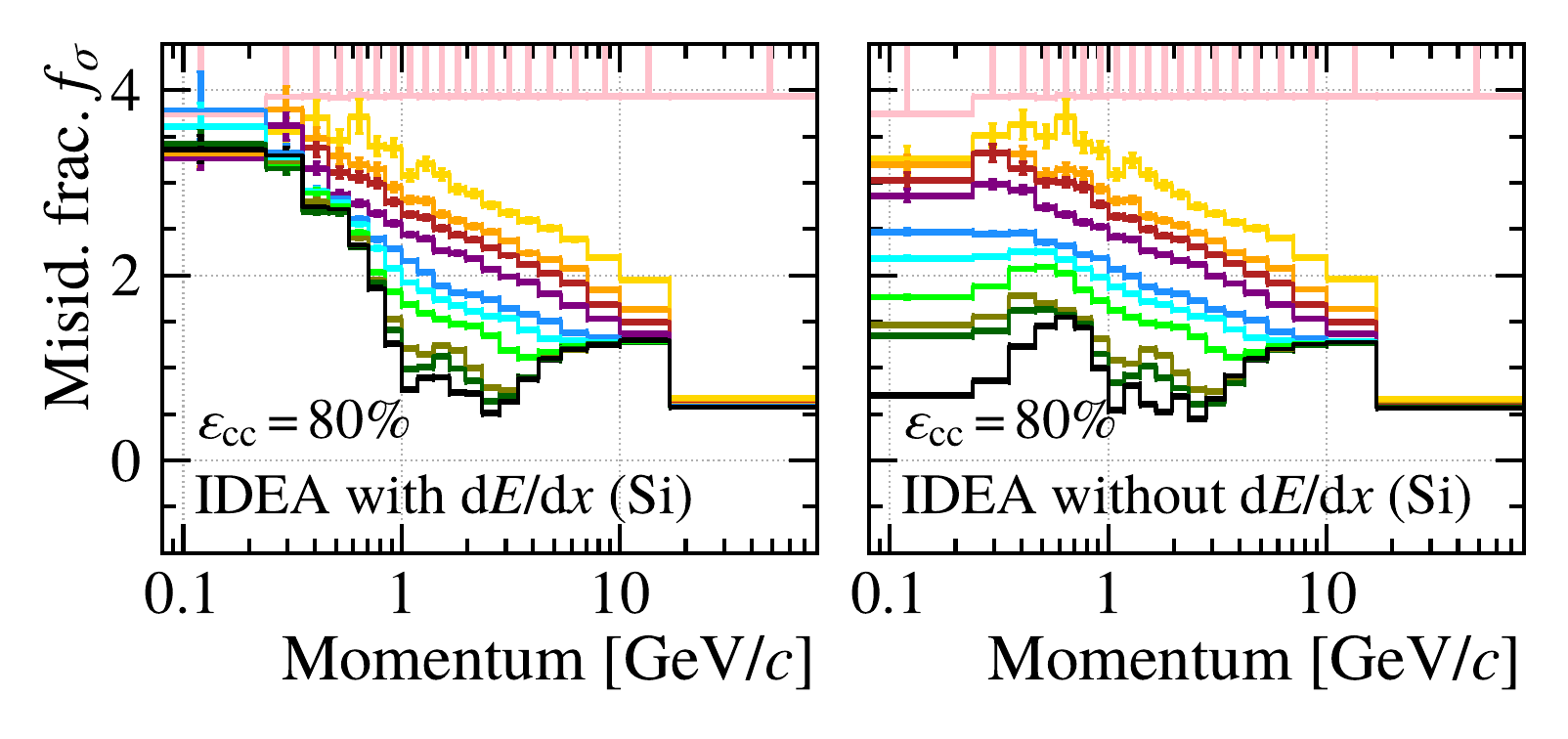}
    \caption{Fraction of misidentified particles $f_\sigma$ expressed as a number of standard deviations using Eq.~\eqref{eq:sigmas} for all particle species combined for different time-of-flight (ToF) resolution and (left) with and (right) without \dedx information at (top) \cld (\ie time-of-flight measurement with or without \dedx measurement) and (bottom) \idea (\ie \dndx measurement with or without time-of-flight and/or \dedx measurement).
    A cluster-counting efficiency $\varepsilon_\text{cc}$ of 80\% is assumed for the \idea.}
    \label{fig:significance:total}
\end{figure}

The top part of Fig.~\ref{fig:significance:total} shows the level of misidentification integrated over the three species for the different configurations tested on \cld (\ie time-of-flight measurement with or without \dedx measurement).
As expected, the misidentification is low at very low momenta where the distance between the speeds for the different particle types is rather large.
Including \dedx information from the silicon tracker (left plot) improves the differentiation at low momentum regardless of time-of-flight resolution.
Note that in case none of the particles in the given bin was misidentified, the lower limit of the 68\% confidence level on $f_\sigma$, sitting just below $4\sigma$ for the available sample size, is shown.

The bottom part of Fig.~\ref{fig:significance:total} shows the same set of plots for the \idea case (\ie \dndx measurement with or without time-of-flight and/or \dedx measurement) with a cluster-counting efficiency of 80\%.
The dependency on the cluster-counting efficiency is limited; corresponding plots for efficiencies of 50\% and 100\% can be found in Appendix~\ref{app:aucs}.
The additional \dndx information generally improves the particle identification significantly, in particular for good to mediocre time-of-flight resolution or very high momenta.
Nevertheless, all configurations suffer a severe deterioration of particle-identification properties at very high momenta caused by the difficult separation in this region as well as the much smaller number of training samples here than in other regions as shown in Fig.~\ref{fig:momentum}.
The constraints put in place to avoid overfitting prevent very fine binning in this region.
Similarly to the \cld case, adding silicon-based \dedx information improves the performance significantly for low momenta.

The performance of configurations without time-of-flight measurement (black) or with very poor time-of-flight resolution (dark green and olive green) drops between 1\gevc and 3\gevc.
This behavior is expected and mirrors the shape of the \dndx curves shown in the bottom plot of Fig.~\ref{fig:input:idea}.
As discussed earlier, this behavior is also reflected in the variation of the AUC score with momentum shown in Fig.~\ref{fig:roc:bins}.
The severity of the drop between 1\gevc and 3\gevc has a slight dependence on the cluster-counting efficiency; see Appendix~\ref{app:aucs}.
In the configurations without \dedx information and with either poor or no time-of-flight information, particle identification is difficult at very low momentum because there are only few training data points but also because the identification relies predominantly or entirely on the \dndx information which is not available for \mbox{$\beta\gamma<0.5$}.

Appendix~\ref{app:significance} contains a figure showing the contamination in units of standard deviation for each combination of particles separately.

Two additional cross-checks are performed.
First, all studies presented in this paper are repeated using a categorical BDT as well as a multilayer perceptron (MLP).
The classification quality criteria obtained using these alternative methods are indistinguishable from the ones presented in this section obtained using three independent BDTs.
Moreover, the results presented hereafter in Sec.~\ref{sec:examples} are also consistent between the three methods.
Second, all studies are repeated with all three methods using particle-gun simulation samples with uniformly generated momentum ranging from 0 to 90\gevc instead of the realistic spectrum shown in Fig.~\ref{fig:momentum}.
While the quality measures are nominally different in this case, e.g. the total AUC scores are much lower because there are many more particles with large momenta that are difficult to classify, the observations remain the same.
The main difference is an increased sensitivity to the noisy variables because, on average, particles with higher momentum have relatively short time of flight and are hence more affected by a poorer time-of-flight resolution.

All shown results in this work are obtained using the combination of three individual BDTs trained on particle-gun simulation samples with a realistic momentum spectrum; see Fig.~\ref{fig:momentum}.

\section{Application to example flavor-physics measurements}\label{sec:examples}
\begin{figure}[h]
    \centering
    \includegraphics[width=\linewidth]{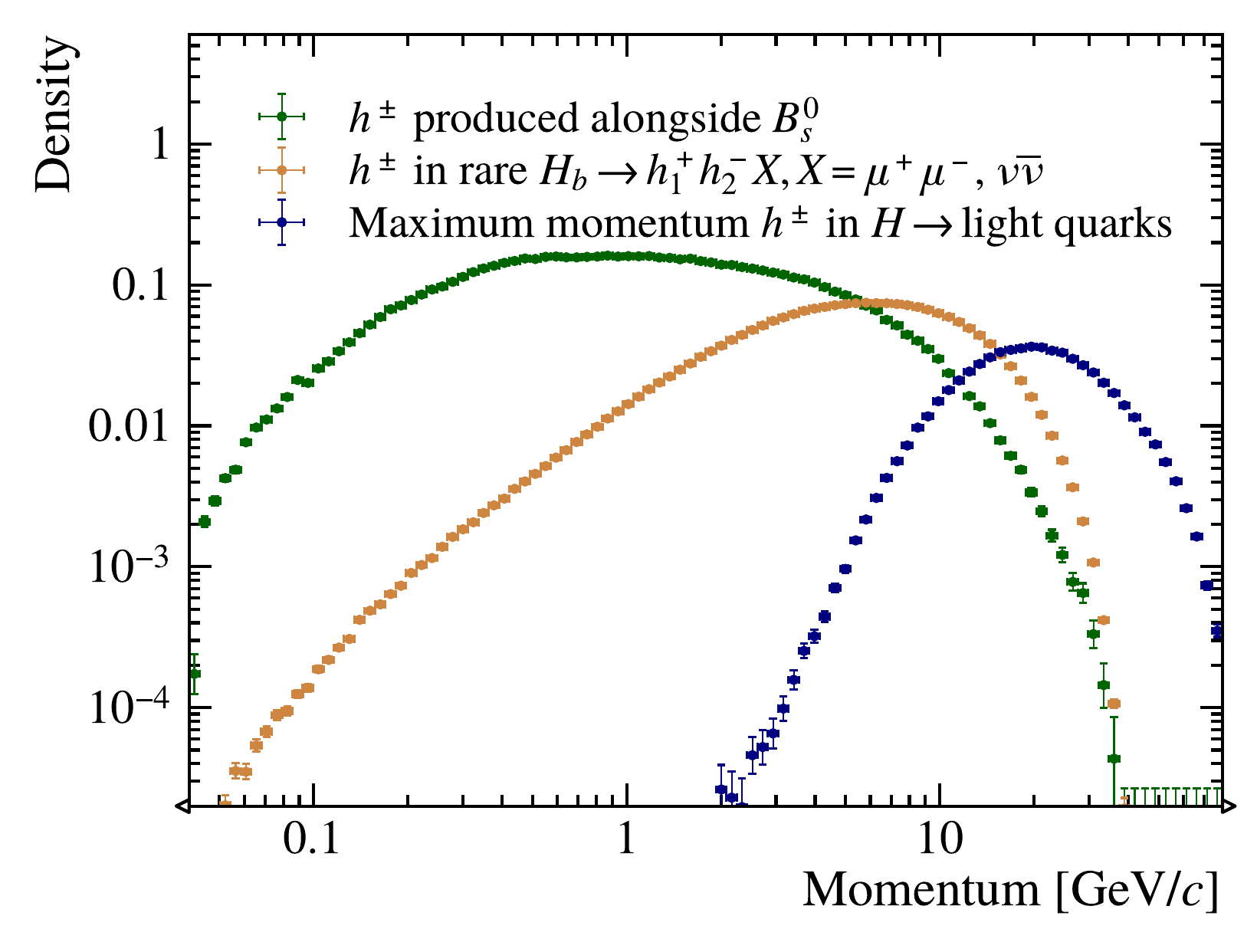}
    \caption{Momentum distributions of the charged hadrons considered in the three example flavor-physics cases.}
    \label{fig:distributions}
\end{figure}

This section illustrates the reduction in contamination achieved through tracker-based particle identification for three example physics cases.
For each example, both \cld and \idea are considered using configurations with different values for the time-of-flight resolution, the cluster-counting efficiency, and with or without silicon-based \dedx information.
The three examples study (A)~the signal and background efficiency for tagging the \bquark-flavor of \Bs mesons based on \textit{same-side tagging} using the net number of charged kaons produced alongside the \Bs meson in the hadronization process, (B)~the contamination due to hadron misidentification in fully visible and the semivisible rare decays, and (C)~the signal and background efficiency when discriminating \squark-jets against jets from other light quarks based on the charged hadron with the highest momentum in the jet.
Figure~\ref{fig:distributions} illustrates how these three examples test vastly different momentum ranges.

\subsection{\bquark-flavor tagging}
Neutral \bquark-mesons can oscillate between their particle and antiparticle states.
Knowledge of the exact meson state at production plays an important role in the flavor sector allowing access to time-dependent and additional \CP-violating amplitudes; see e.g. Ref.~\cite{LHCb-PAPER-2019-013}.
The sensitivity is particularly large for \Bs decays given the small difference between mass eigenstates.

The \fccee running at the \Z pole produces \bbbar pairs flying back to back in the lab frame.
A number of additional, typically lighter, hadrons are produced in the hadronization process.
Because the strong force is flavor conserving, the flavor of the \bquark-quark can be identified by the lighter hadrons produced alongside.
For example, the \bquark-quark might hadronize to a \Bsb meson with the \squarkbar-quark produced via an \ssbar pair from the vacuum.
The additional \squark-quark must hadronize, too, leading to a negatively charged or neutral kaon.
This type of tagging is usually referred to as a \textit{same-side tagger}.

The \bquark-quark might also be identified through first identifying the \bquarkbar-hadron on the other side of the event.
This procedure is typically called an \textit{opposite-side tagger} and relies on identifying the \bquarkbar-hadron based on its decay products.
Due to excellent expected mass resolution at the \fccee, this is likely possible without employing particle-identification tools.
As a consequence, the opposite-side tagger is not investigated here.

Modern \bquark-flavor tagging algorithms employ machine-learning tools to decipher the multidimensional correlations that provide the most efficient tag; see e.g. Refs.~\cite{LHCb:2015olj,LHCb:2016mtq,LHCb:2016yhi,Belle-II:2021zvj}.
Developing such an algorithm for the \fccee environment reaches beyond the scope of this study.
Instead, the investigations focus on the simplest case, tagging \Bs and \Bsb, using the previously discussed proxy, the net number of kaons.
This means that a \bquark-hadron is tagged as \Bs if it is accompanied by a larger number of \Kp than \Km in its hemisphere and vice versa for the \Bsb.
Accounting for the detector acceptance, this classification would catch 45\% of the \Bs and \Bsb mesons if perfect particle identification were possible.

\begin{figure}
    \centering
    \hfill\includegraphics[width=0.85\linewidth]{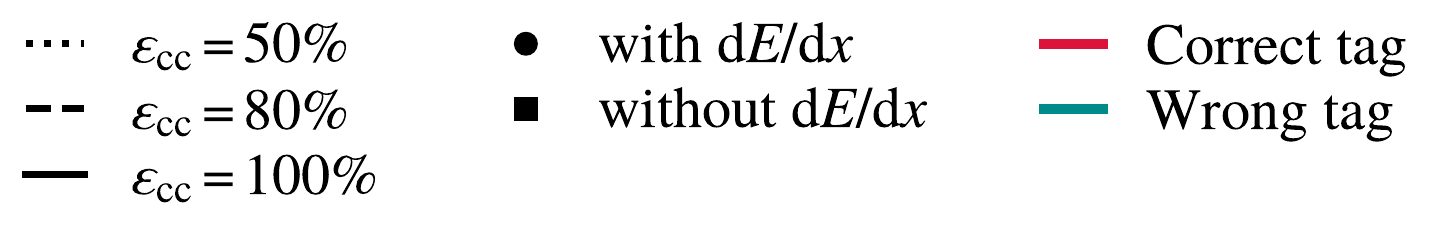}
    \includegraphics[width=\linewidth]{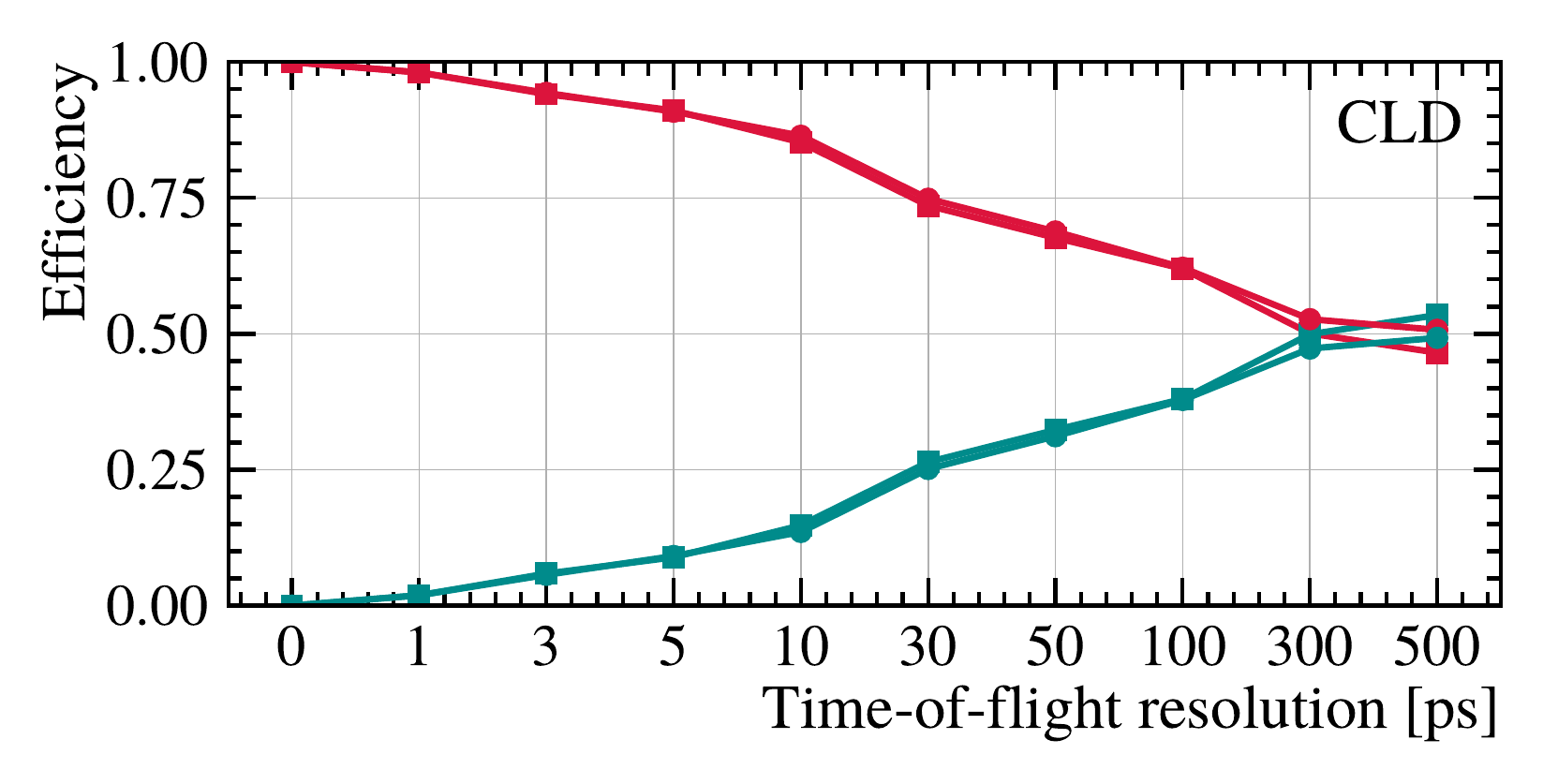}
    \includegraphics[width=\linewidth]{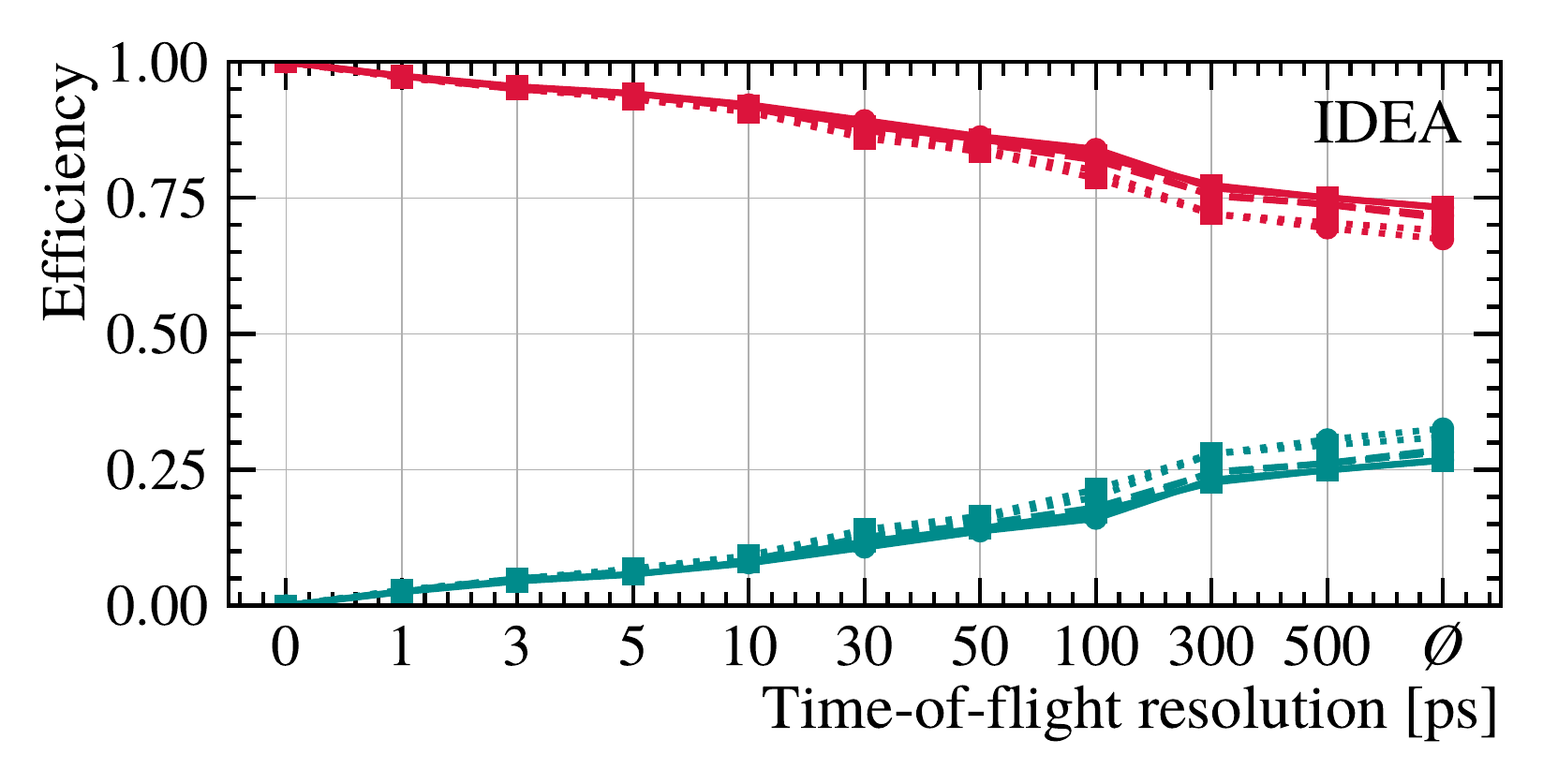}
    \caption{Efficiency of (mis)tagging \Bs mesons by requiring that more \Kp than \Km are produced alongside the \bquarkbar-meson for (top) \cld and (bottom) \idea for different time-of-flight resolution (horizontal axis), different cluster-counting efficiencies $\varepsilon_\text{cc}$ (\idea only, distinguished by line style), and with or without using energy-deposit information from the silicon sensors (marker style).}
    \label{fig:Zbb:contamination}
\end{figure}

Figure~\ref{fig:Zbb:contamination} shows the resulting tagging efficiency to correctly identify a \Bs meson in red and corresponding mistag efficiency in green.
The efficiencies are calculated with respect to the 45\% of \Bs with more \Kp than \Km that can be identified this way in order to show the efficiency improvement through particle identification rather than the efficiency of the algorithm.
As a consequence of the very low momentum of the considered particles (see Fig.~\ref{fig:distributions}), a moderate tagging efficiency can be achieved for time of flight resolution below 50\ps at \cld.
The addition of \dedx information is only slightly beneficial for poor time-of-flight resolution.
While the trends are fairly similar, \idea consistently achieves significantly better performance for time-of-flight resolutions in the most realistic range between 30\ps and 100\ps almost regardless of cluster-counting efficiency.

\subsection{Rare decays}
Precision measurements of electroweak-penguin decays of \bquark-quarks provide sensitivity to effects at energy scales beyond the direct reach of current or future colliders.
Measurements of \bsll transitions in different hadronic systems indicate tensions between measurement and prediction~\cite{CMS:2024atz,Belle:2016fev,ATLAS:2018gqc,LHCb-PAPER-2020-002,LHCb-PAPER-2013-017,LHCb-PAPER-2015-023,LHCb-PAPER-2021-014, LHCb-PAPER-2014-006,LHCb-PAPER-2020-041,LHCb-PAPER-2023-033,LHCb-PAPER-2024-011,CMS:2024syx, BaBar:2012mrf, Belle:2019xld,CMS:2024syx,LHCb-PAPER-2014-024,LHCb-PAPER-2017-013,LHCb-PAPER-2019-040,LHCb-PAPER-2021-004,LHCb-PAPER-2021-038,LHCb-PAPER-2022-045,LHCb-PAPER-2022-046}.
Strong suppression of background is crucial to achieve high precision in rare-decay measurements.
This is particularly important in angular analyses where the contribution from misidentified rare decays with their unique angular distributions can bias the result.
Moreover, this bias is hard to quantify because the background distribution itself is often not well known such that background suppression is the only viable option.

The examples considered in this study are the three hadronic systems \mbox{$\Bs\to \Kp\Km$}, \mbox{$\Bd\to \Km\pip$}, and \mbox{$\Lb\to p\Km$} either with two muons of opposite charge or a neutrino-antineutrino pair.
The \bquark-hadrons are produced using \pythia in \epem collisions at the \Z pole~\cite{Sjostrand:2014zea}.
They are decayed by the \evtgen package forcing a uniform distribution in the dilepton invariant-mass squared~\cite{Lange:2001uf}.
The compositions of the dihadron spectra are inspired by previous analyses at the photon or \jpsi pole~\cite{LHCb:2024blb,LHCb:2024vtc,Belle:2014nuw} with details in Appendix~\ref{app:dihadron}.
The production fractions used to calculate the contamination are \mbox{$f_\Bd=0.407$}, \mbox{$f_\Bs=0.101$}, and \mbox{$f_\Lb=0.075$}~\cite{HFLAV:2019otj}.
The \Lb hadronization fraction is an estimate based on the measured hadronization fraction to any \bquark-baryon of 0.085~\cite{HFLAV:2019otj} and the assumption that 89\% of \bquark-baryons are \Lb baryons, 10\% of \bquark-baryons are neutral or charged $\Xib$ baryons and 1\% are other \bquark-baryons~\cite{Jiang:2018iqa}.
The decay branching fractions are set to be equal.

\subsubsection{Finalstates with neutrinos}
\begin{figure}
    \centering
    \hfill\includegraphics[width=0.85\linewidth]{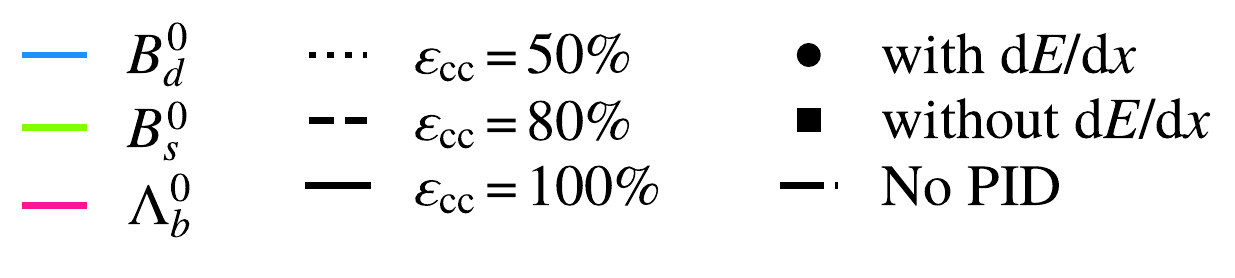}
    \includegraphics[width=\linewidth]{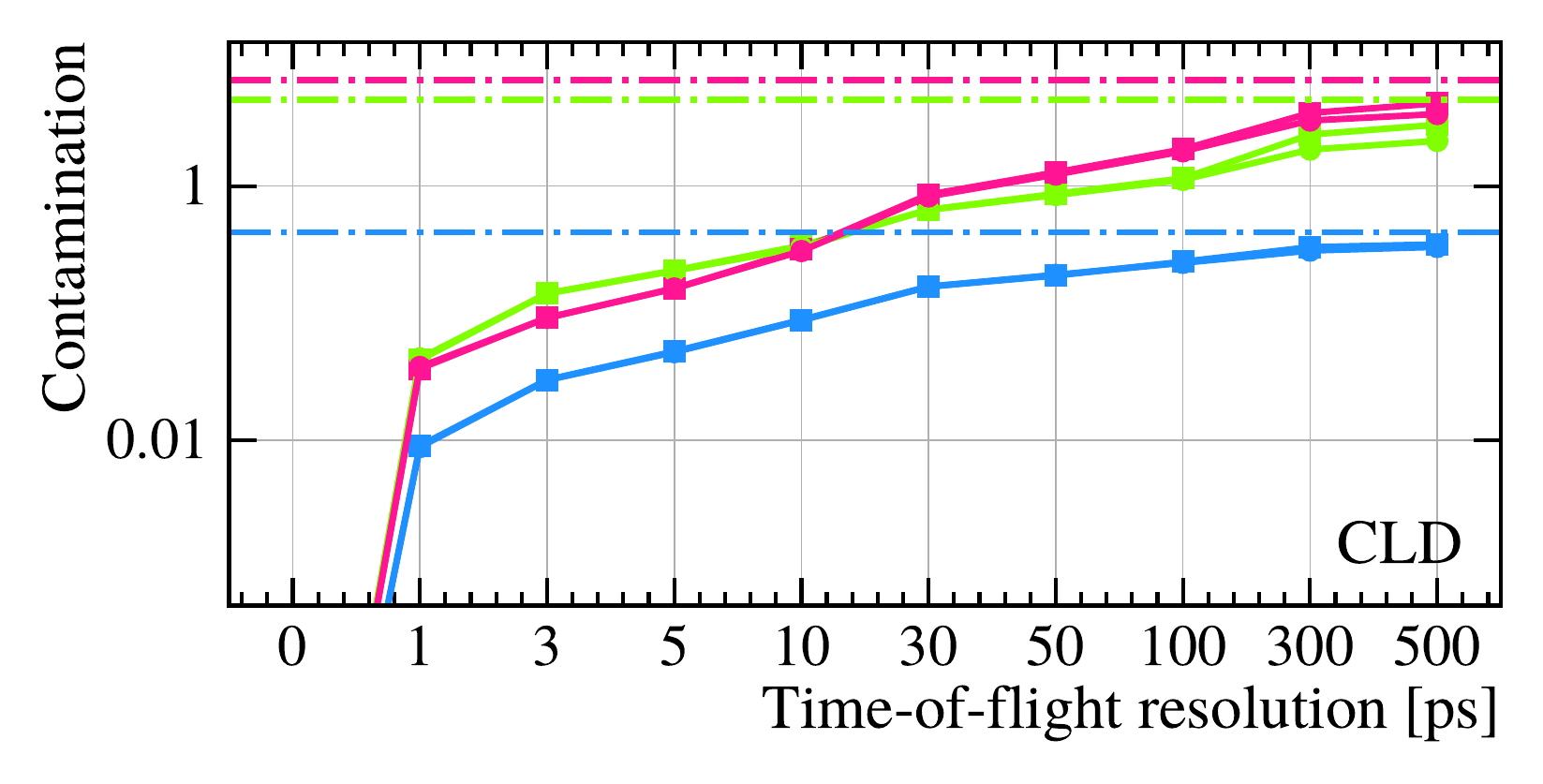}
    \includegraphics[width=\linewidth]{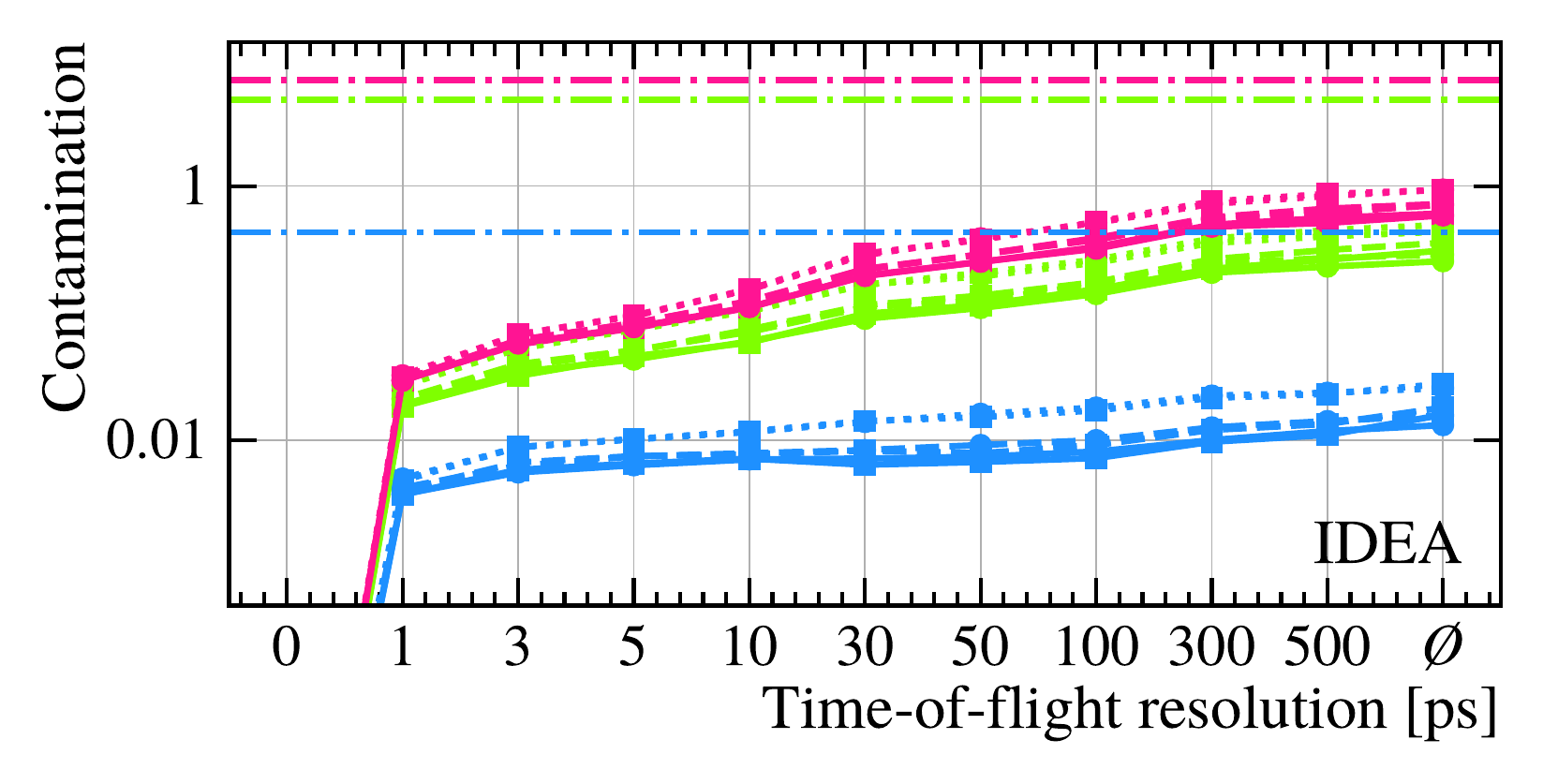}
    \caption{Level of contamination for the three rare decays with semivisible final states (distinguished by color) at (top) \cld and (bottom) \idea for different time-of-flight resolution (horizontal axis), different cluster-counting efficiencies $\varepsilon_\text{cc}$ (\idea only, distinguished by line style), and with or without using energy-deposit information from the silicon sensors (marker style).}
    \label{fig:raredecays:semivisible}
\end{figure}

Figure~\ref{fig:raredecays:semivisible} shows how the level of contamination to each of the three semivisible decays changes for different time of flight resolution at \cld (top) and different time of flight resolution and cluster-counting efficiency at \idea (bottom, including \dndx information) in either case with and without the \dedx measurement.
The different colors represent the different decays where the initial differences in contamination between the decays predominantly stems from very different \bquark-hadron hadronization fractions instead of differences in particle identification.
A reduction in background at \cld by at least an order of magnitude would require time of flight resolution of 30\ps or better depending on the decay.
Adding \dedx information leads to no significant improvement in particle identification.
The rightmost points in the \idea plot (bottom) represent a configuration without time of flight measurement revealing that the \dndx information on its own can reduce the contamination by an order of magnitude with respect to no particle identification (constant, dash-dotted lines).
The cluster-counting efficiency as well as including a \dedx measurement has limited impact.
Additional time of flight information can improve a purely \dndx-based classifier by an order of magnitude for resolutions below 10\ps.

All these observations are a consequence of the momentum spectrum shown in Fig.~\ref{fig:distributions} and agree with the discussions in Sec.~\ref{sec:pid}.
The particles have momenta that are small enough to benefit from moderate time of flight resolution (at least 30--50\ps at \cld).
At the same time, many particles have momenta that are large enough to avoid the region of intersecting \dndx curves such that a \dndx measurement on its own provides strong classification.

Note that searches of these semivisible final states as discussed in Ref.~\cite{Amhis:2023mpj}, would focus on an individual dihadron resonance where additional kinematic selections suppress large amounts of misidentified backgrounds.
Moreover, the presented numbers are purely based on counting whereas real analyses would likely extract the yields by fitting a suitable variable such as the output of a multivariate classifier to achieve additional separation.

\subsubsection{Fully visible final states}
For fully visible final-states, the four-body invariant mass is a helpful discriminator between different \bquark-hadrons.
When considering, for example, \BsToKKmm as the signal decay, the four-body invariant mass of true signal candidates accumulates in a peak at the nominal \Bs mass.
The shape and width of the peak are a consequence of the resolution of the tracking system and reconstruction algorithm.
Previous studies of fully visible muonic rare decays using fast simulation of the \idea detector showed invariant-mass peaks with a full width at half maximum around 10\mevcc for \mbox{$\Bs\to\phi(\to\Kp\Km)\mumu$} decays~\cite{Kwok:2025fza}, 8\mevcc~\cite{Beck:2025bgc} for $\Lb\to\Lz(\to p\pim)\mumu$ decays, and 4-5\mevcc~\cite{DiCanto:2025fpk} for $\Lc\to p\mumu$ decays.
Reference~\cite{FCC:2025lpp} provides similar values for other decays.
Misidentified backgrounds on the other hand appear in very broad smeared-out structures in the signal invariant-mass, which are stretched across several \gevcc in the \fccee setting.
See Appendix~\ref{app:massspread} for an illustration.

The narrow signal peaks and the wide spread of backgrounds allow to efficiently select signal and suppress background by focusing on narrow windows around the signal \bquark-hadrons (\ie the \Bs mass region in the \mbox{$KK\mu\mu$} combination for a \Bs signal) while vetoing narrow windows around the background \bquark-hadrons (\ie the \Lb mass region under the \mbox{$pK\mu\mu$} hypothesis as well as the \Bd mass under the \mbox{$K\pi\mu\mu$} hypothesis).
Because the four-body invariant mass in this study is obtained from the true momenta, different values for the resolution are added artificially by smearing the four-body invariant mass with a Gaussian of predetermined width.
The results shown in the main part of this paper use a Gaussian distribution with \mbox{$\sigma=5\mevcc$} which translates to a conservative full width at half maximum of 11-12\mevcc.
The window for selecting signal and vetoing background peaks is $\pm10$\mevcc around the nominal peak mass.
For this configuration and considering the \Bs decay as signal, the chosen kinematic constraints have an efficiency of 73.8\%, 0.2\%, and 0.1\% for the \Bs,\Bd, and \Lb decay respectively.
Appendix~\ref{app:resolution} contains the results for Gaussian widths of \mbox{$\sigma=3\mevcc$} and \mbox{$\sigma=10\mevcc$} with appropriately scaled selection windows.

\begin{figure}
    \centering
    \hfill\includegraphics[width=0.85\linewidth]{plots/rare_decays/contamination_legend.pdf}
    \includegraphics[width=\linewidth]{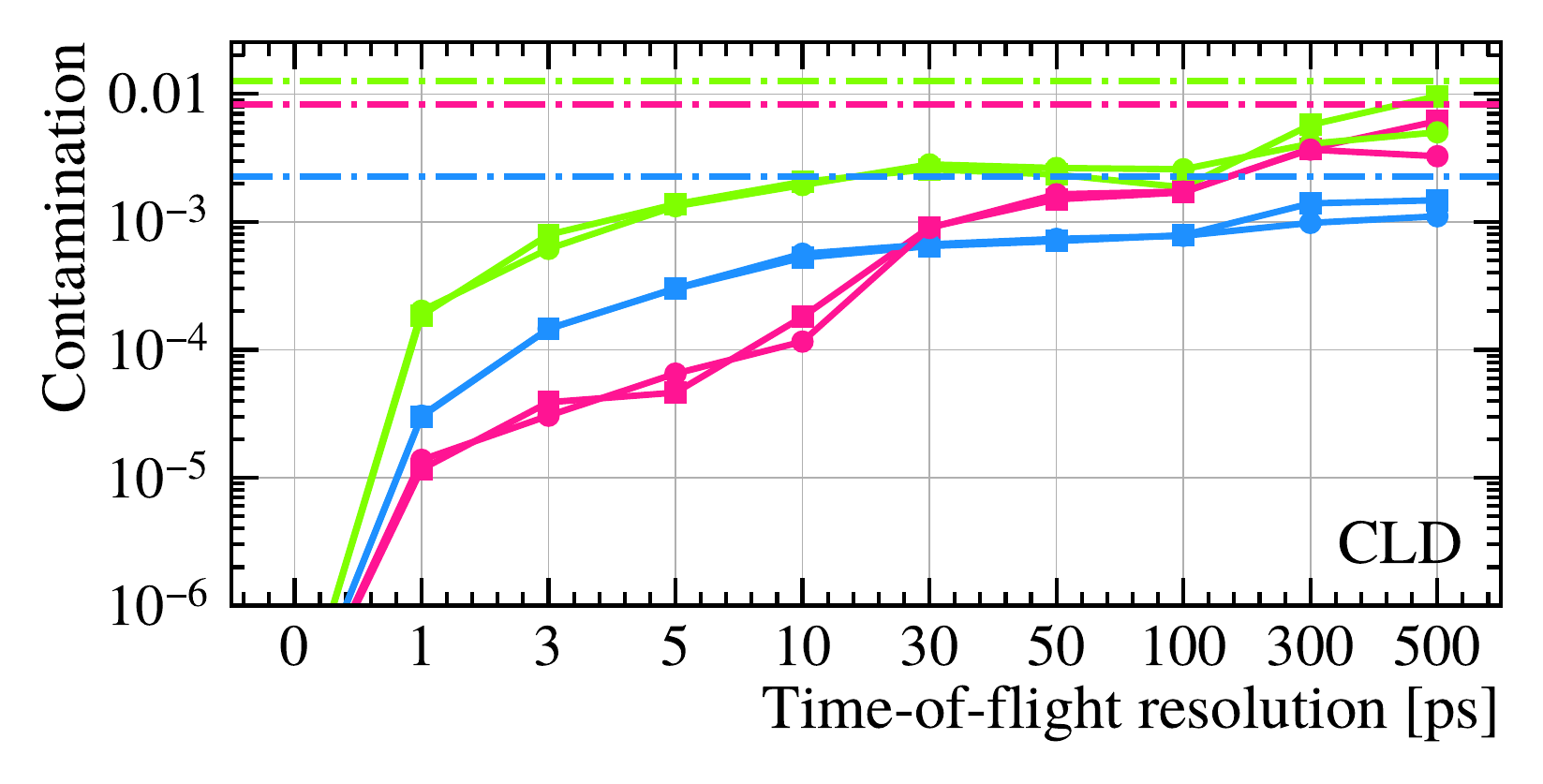}
    \includegraphics[width=\linewidth]{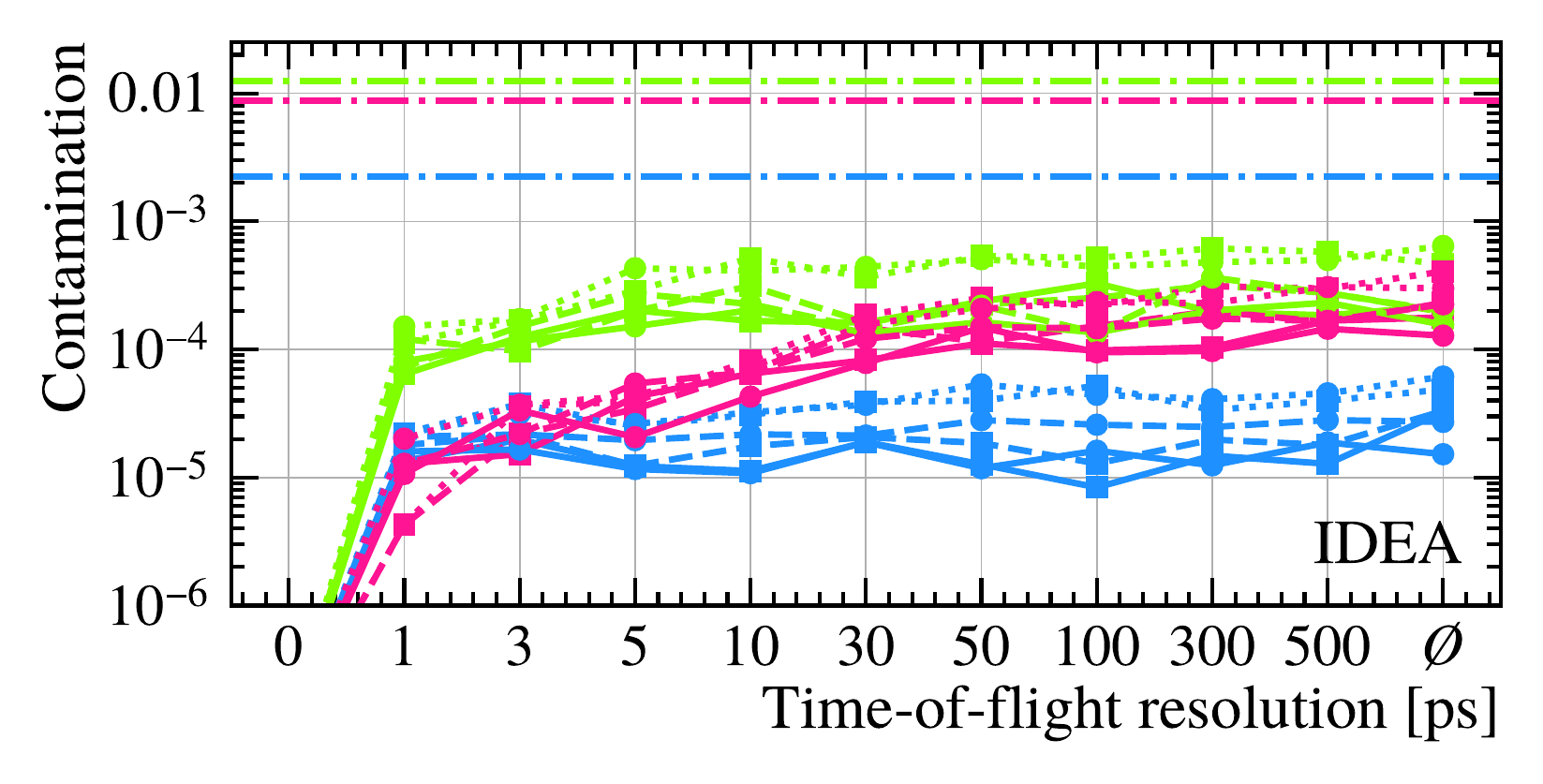}
    \caption{Level of contamination for the three rare decays with fully visible final states (distinguished by color) at (top) \cld and (bottom) \idea for different time-of-flight resolution (horizontal axis), different cluster-counting efficiencies $\varepsilon_\text{cc}$ (\idea only, distinguished by line style), and with or without using energy-deposit information from the silicon sensors (marker style).
    The four-body invariant masses are smeared using a Gaussian of width \mbox{$\sigma=5\mevcc$} with signal window and background vetoes of $\pm10\mevcc$.}
    \label{fig:raredecays:fully-visible}
\end{figure}

Figure~\ref{fig:raredecays:fully-visible} shows the resulting contamination with kinematic selections which suppress backgrounds to below 1\%.
The improvements achieved through particle identification are of similar magnitude as in the semivisible case.
Because the kinematic selections remove different parts of the momentum spectrum for each decay, the trends are slightly different compared to the semivisible example.
Note that the strong fluctuations between the different points at very low contamination values are a consequence of the strong background suppression leaving only a small fraction of the 1~000~000 generated decays per species.

The benefit of particle identification for fully visible rare-decay measurements is limited due to the excellent four-body invariant-mass resolution at all \fccee detectors leading to almost negligible amounts of background.
The practical improvement in contamination, however, might allow for a background-free analysis with negligible associated systematic uncertainty.
Because the contribution of New Physics can vary across the dimuon invariant-mass squared, \qsq, it has been checked that the level of contamination does not vary significantly across \qsq and is dominated by the different extent of the phase space.

\subsection{\squark-jet tagging}
The \fccee provides unique opportunities for measurements involving a Higgs boson decaying to a pair of strange quarks.
We study the impact of particle identification on \squark-jet tagging with simulation samples generated using \pythia assuming \epem collisions at a center-of-mass energy of 240\gev.
Having a collision energy significantly above the $\Z\Hz$ threshold maximises the production cross section~\cite{CEPCStudyGroup:2018ghi} and results in a momentum for each boson of around 52\gevc.
For the sake of this study, it is assumed that the final-state hadrons can be perfectly assigned to the Higgs or \Z boson.
Perfect assignment can be achieved either through anticipated advances in reconstruction tools in combination with the cleanliness of \epem events and a nonzero boost of the \Z and \Hz bosons or by requiring that the \Z boson decays to a dimuon pair, allowing unambiguous assignment of all seen hadrons to the Higgs boson at the cost of significantly reduced sample size.

Other $\Hz\to\qqbar$ and \mbox{$\Hz\to gg$} decays present the most problematic backgrounds for \mbox{$\Hz\to\ssbar$}.
Taggers identifying jets from heavy quarks are very mature to date relying on high-dimensional information including in particular the long lifetime of \bquark- and \cquark-hadrons; see e.g. Refs.~\cite{Abidi:2025mdw}.
Gluon-jet tagging has been explored and often makes use of the significantly increased multiplicity and distinct shape of gluon-jets; see e.g. Refs.~\cite{Bedeschi:2022rnj,Albert:2022mpk}, which study \mbox{$\Hz\to\ssbar$} decays at the \fccee and the International Linear Collider.
This work focuses on the distinction between \squark-jets and \uquark- or \dquark-jets.
A simple but efficient proxy for the jet flavor in this case is the identity of the highest-momentum hadron as it likely stems from the original quarks produced in the Higgs decay.
Based on the simulation produced for these studies, a pion carries most of the momentum in 80\% of the \uquark- or \dquark-jets.
Only slightly above 10\% of the \uquark- or \dquark-jets result in a kaon with highest momentum.
In \squark-jets on the other hand, pions and kaons are similarly likely to obtain the largest momentum at a rate of around 46\%.

\begin{figure}
    \centering
    \hfill\includegraphics[width=0.85\linewidth]{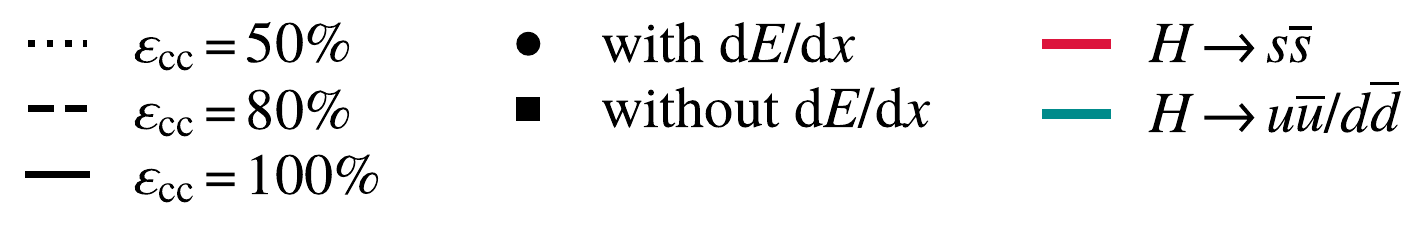}
    \includegraphics[width=\linewidth]{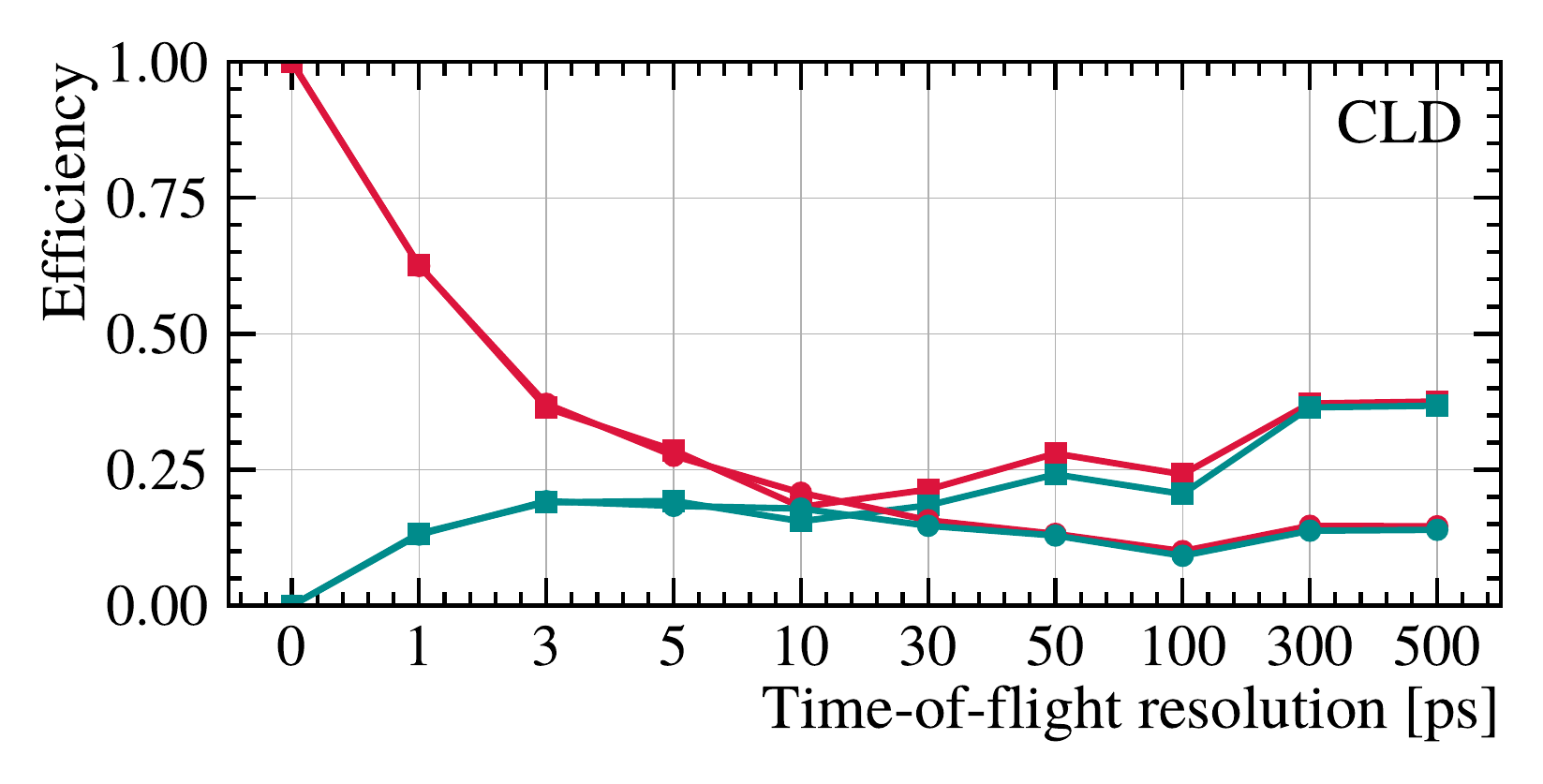}
    \includegraphics[width=\linewidth]{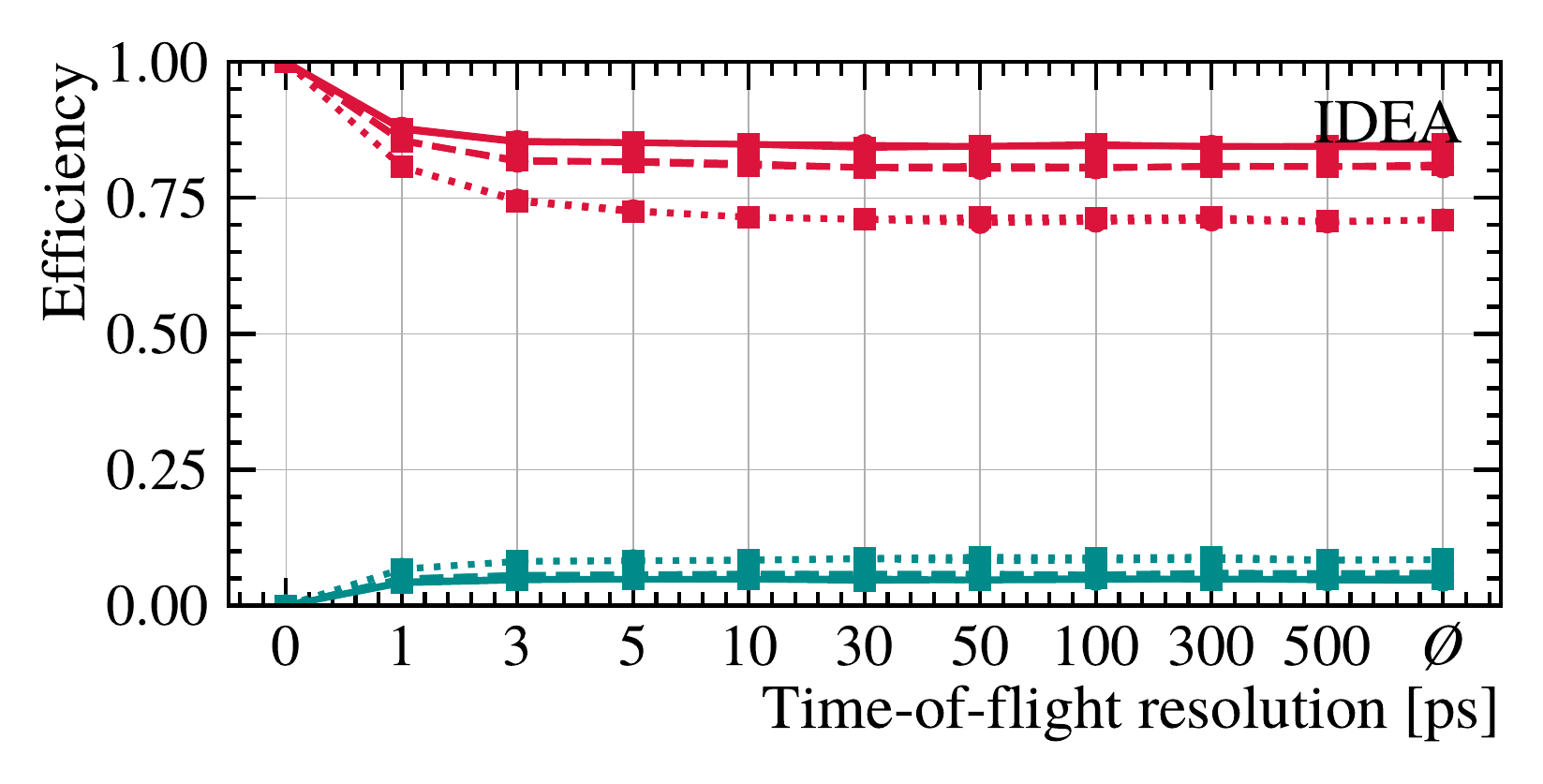}
    \caption{Efficiency of tagging \mbox{$\Hz\to\ssbar$} (in red) and \mbox{$\Hz\to\uubar/\ddbar$} (in green)  based on whether the highest-momentum hadron in the jet is a kaon at (top) \cld and (bottom) \idea for different time-of-flight resolution (horizontal axis), different cluster-counting efficiencies $\varepsilon_\text{cc}$ (\idea only, distinguished by line style), and with or without using energy-deposit information from the silicon sensors (marker style).}
    \label{fig:ZH}
\end{figure}

Figure~\ref{fig:ZH} shows the efficiency of selecting events based on whether the particle with the highest momentum was a kaon.
Note that the efficiency is calculated relative to the 46\% (80\%) of the \mbox{$\Hz\to\ssbar$} (\mbox{$\Hz\to\uubar/\ddbar$}) events where the highest momentum particle was a kaon (pion).
Due to the very high momenta shown in Fig.~\ref{fig:distributions}, \cld cannot easily distinguish the Higgs jets at realistic time-of-flight resolutions.
The \dedx information does not add discrimination power.
The increase in efficiency for signal and background at poor time-of-flight resolution without \dedx information is a reflection of the training instabilities when considering only two variables with high noise in a BDT.
Additional input from cluster counting in the drift chamber at \idea allows strong distinction between the jet flavors with a slight dependence on the cluster-counting efficiency.
The trends described for \idea agree with existing studies of more sophisticated flavor tagging tools performed on fast simulation~\cite{Bedeschi:2022rnj}.

\section{Discussion}\label{sec:summary}
A study of the flavor capabilities of the \cld and \idea detector concepts at the \fccee has been presented.
The particle-identification algorithm used is based on three BDTs predicting a likelihood of a charged hadron to be a pion, kaon, or proton.
The combination of \dndx and moderate time of flight resolution results in a level of misidentification close to $2\sigma$, \ie 2\%, or better for a given particle combination across almost the entire momentum range appearing in \eeZbb and \eeZH events; see Sec.~\ref{sec:pid}.
\dedx information may be useful at very low momentum.

Three example physics studies are presented covering the low, medium, and high momentum regions.
(A)~Same-side flavor tagging of \Bs mesons relies on identification of very low momentum particles.
In this region, good particle identification can be achieved in almost all setups.
In a classifier using only time of flight, a resolution of 30\ps or better is necessary to achieve the same performance as a purely \dndx-based classifier.
(B)~Charged hadrons from rare \bquark-decays sit in an intermediate momentum range.
Here, the particles are still slow enough to benefit significantly from a time-of-flight measurement but require a resolution of at least 30--50\ps for notable improvements.
A classifier relying exclusively on time of flight in this case would need perfect time-of-flight resolution in most cases (\ie 0\ps) to achieve the same performance as a purely \dndx-based classifier.
(C)~Tagging \mbox{$\Hz\to\ssbar$} jets based on the highest momentum hadron in the jet requires particle identification in the momentum range beyond 10\gevc.
A measurement of \dndx in this case leads to solid performance while time of flight only provides useful additional information for vanishing resolution (\ie 0\ps).
It should be noted that the selections used in the three examples are of an illustrative nature and the efficiencies and contaminations optimized for a specific more realistic analysis will differ to an extent.

This and previous studies show that as a consequence of excellent invariant-mass resolution, measurements involving fully reconstructed final states of charged particles do not necessarily require dedicated particle-identification detectors in order to achieve a passable level of background suppression.
In fact, these processes would be useful to calibrate a particle-identification tool.
Decays of narrow \cquark-hadrons, such as \mbox{$\Dz\to\Km\pip/\pip\pim/\Kp\Km$} or \mbox{$\Lc\to p\Km\pip$}, are excellent calibration channels with expected yields between $10^9$ and $10^8$ in a total dataset of $6\cdot10^{12}$ \Z bosons~\cite{ALEPH:1999syy,PDG2024,FCC:2025uan}.
Their decay products span a wide range of momenta due to the fact that the \cquark-hadron can be produced either directly in the primary vertex or from the decay of a heavier hadron.
Any process that cannot be identified based on kinematic criteria requires particle identification.
Such processes include in particular decays with unknown final-state configuration as is the case for inclusive decays, like in \squark-jet and \bquark-flavor tagging, or decays to invisible particles.
The need for particle identification also extends to scenarios with insufficient invariant-mass resolution.
In this context, investigations of decays involving electrons or neutral particles would provide useful further input.

Due to the lack of reconstruction software for the \idea detector in this moment, the studies assume perfect reconstruction of the visible event as well as a uniform magnetic field when calculating the time of flight and flight distance.
Neither assumption should have sizable impact on the results presented in this publication.
In contrast, the accuracy of the simulated \dedx values might not yet reflect a realistic level of noise.
However, the results of this study indicate that the particle identification does not rely heavily on the silicon-based \dedx information in real-world examples.
More sophisticated machine-learning techniques and in particular specific training samples might improve classification performance in some cases without changing the observations made in this paper.
Future investigations may consider realistic simulation of the cluster formation in the drift chamber that cover a potentially more complex cluster-counting efficiency as well as occupancy effects.
Similarly, the inclusion of electrons and muons in the identification procedure and an extension of the classifier to calorimeter and muon chamber information could be of interest.
Finally, the benefits to flavor-physics analyses in the presence of a dedicated particle-identification system should be ascertained in future studies.

The classifiers used for the presented studies are publicly available in Ref.~\cite{tracker-pid}.
Besides the models, the repository also includes a script that provides particle-identification information based on an arbitrary simulation sample for easy use in future studies.

\section*{Acknowledgements}
We thank our colleagues in the FCC organization for providing the software and advice, in particular, Jan Eysermans and Brieuc Francois, as well as the participants in the \textit{MIT and friends} \fccee meeting series for their discussion of this work.
We also acknowledge the US National Science Foundation, whose funding under Award No. 2310073 has helped support this work.

\clearpage
\appendix
\onecolumngrid

\section{Speed spread for different time-of-flight resolutions and cluster-counting efficiencies}\label{app:tofs}

\begin{figure}[h!]
    \centering
    \includegraphics[width=0.33\linewidth]{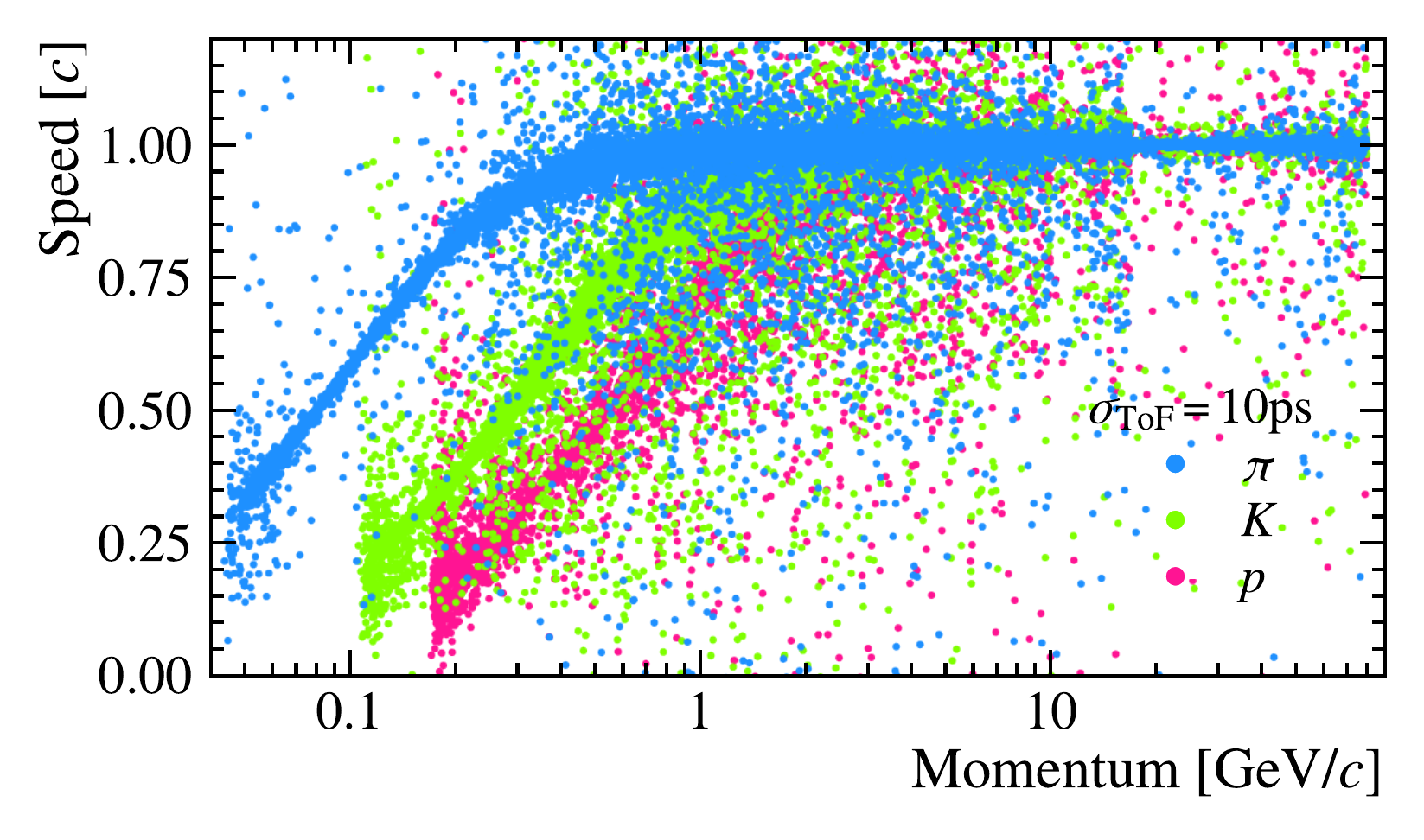}%
    \includegraphics[width=0.33\linewidth]{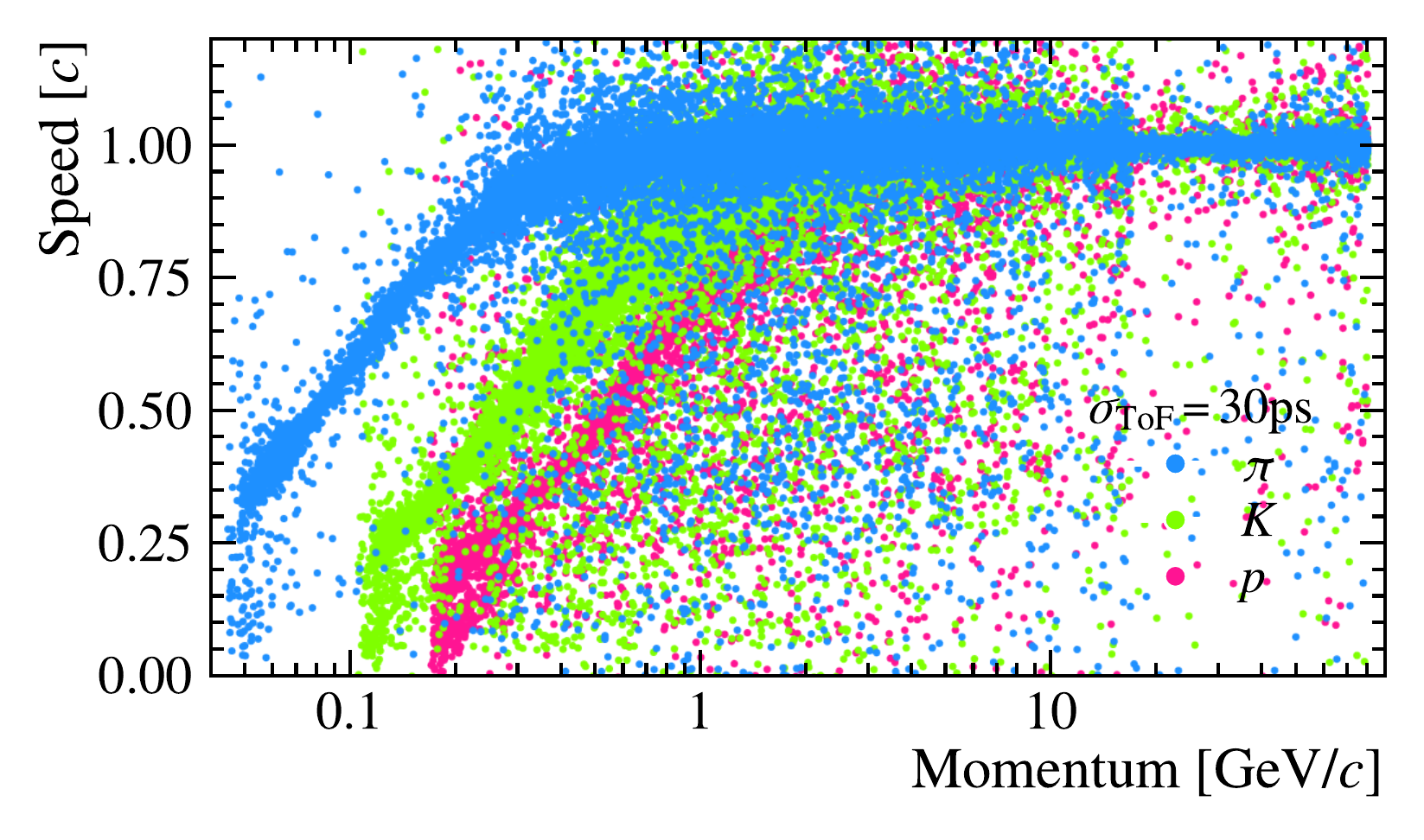}%
    \includegraphics[width=0.33\linewidth]{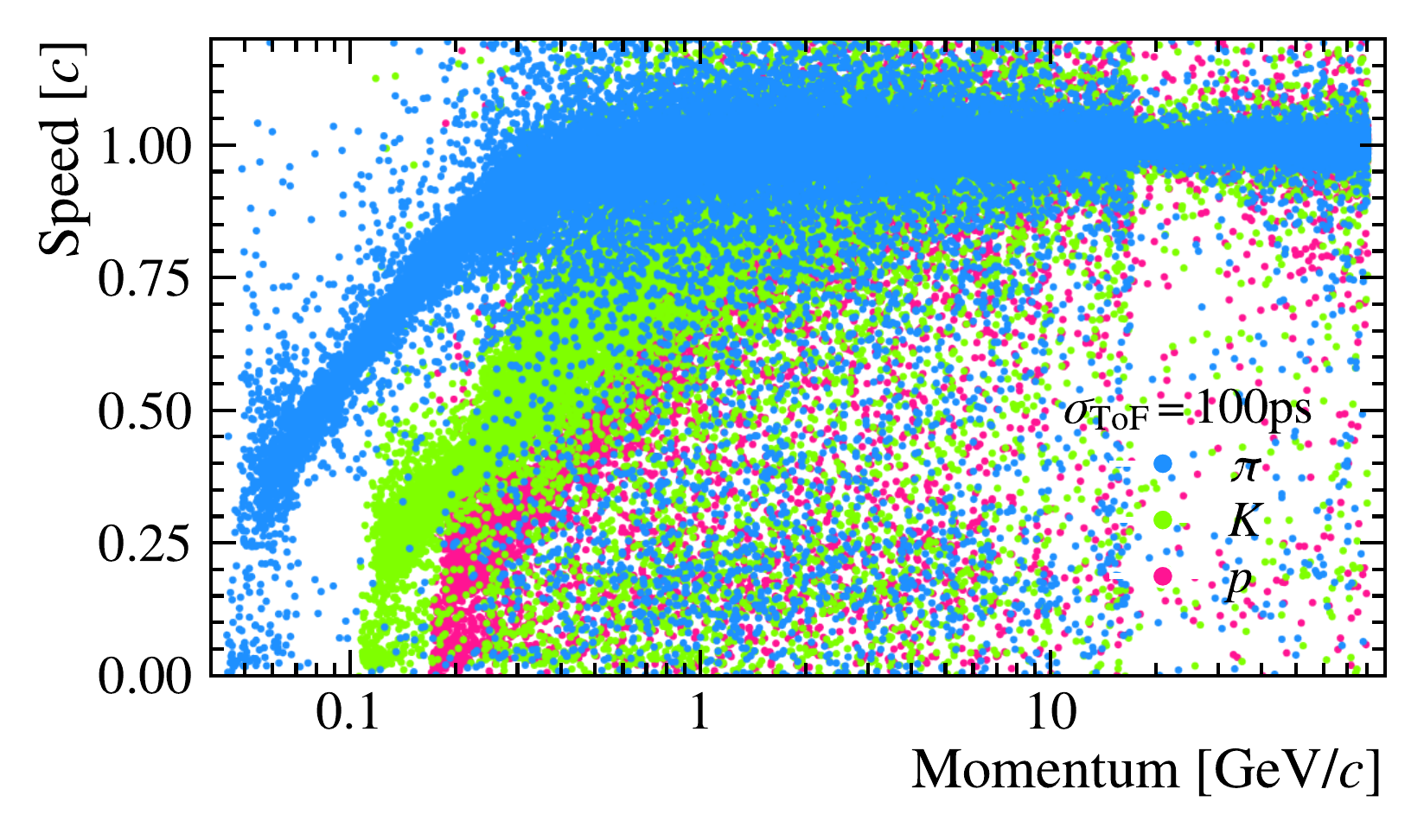}
    \caption{Spread of the speed-momentum curve for different time-of-flight resolutions. The particles shown are produced in bins; see Fig.~\ref{fig:momentum}. The spread in speeds seems to drop abruptly at the bin edges, e.g. between 10 and 11\gevc. This is no physical effect but rather a reflection of the sudden change in sample size.}
    \label{fig:input:tof}
\end{figure}

\section{Performance figures for different cluster-counting efficiencies at \idea}\label{app:aucs}
\twocolumngrid

\begin{figure}[h]
    \centering
    \includegraphics[width=\linewidth]{plots/particlegun/legend_auc.pdf}
    \includegraphics[width=0.5\linewidth]{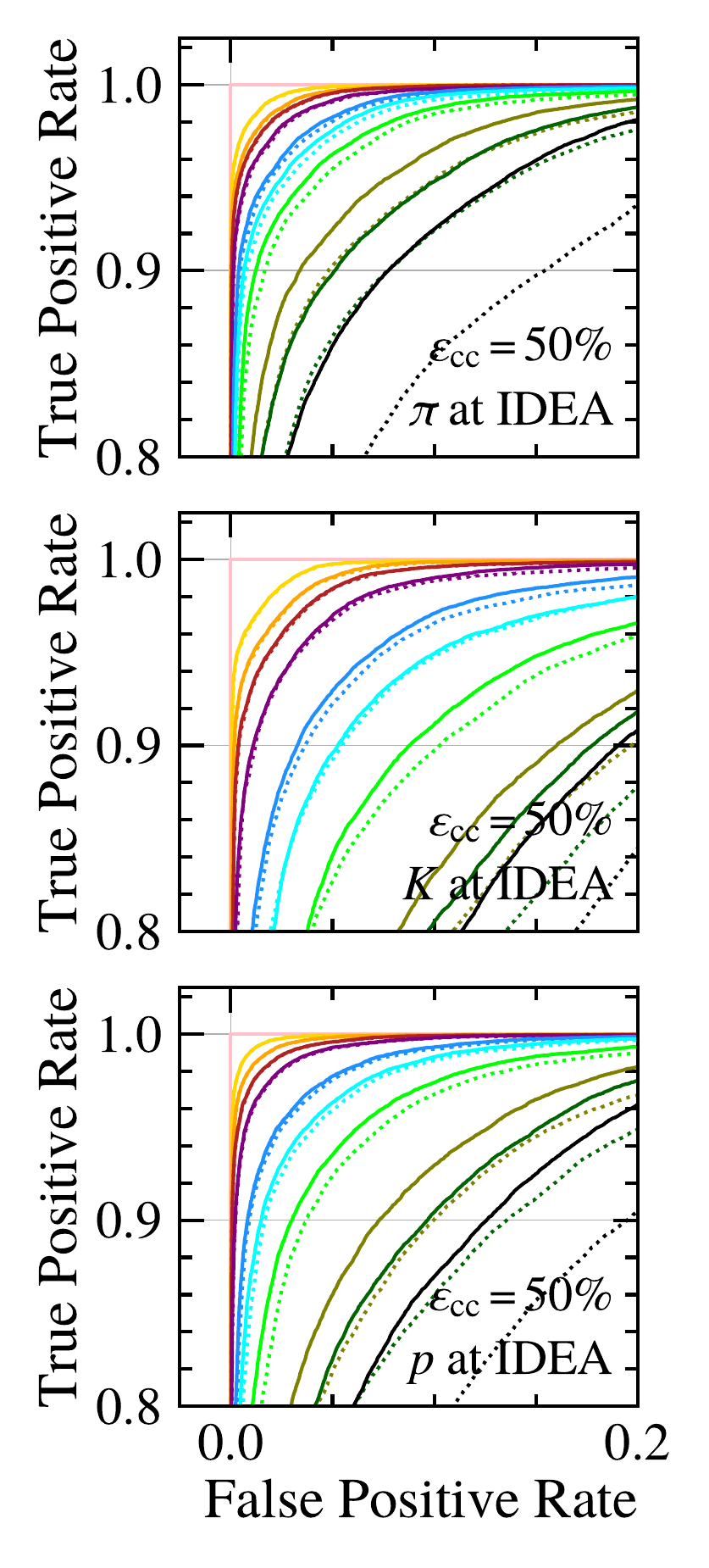}%
    \includegraphics[width=0.5\linewidth]{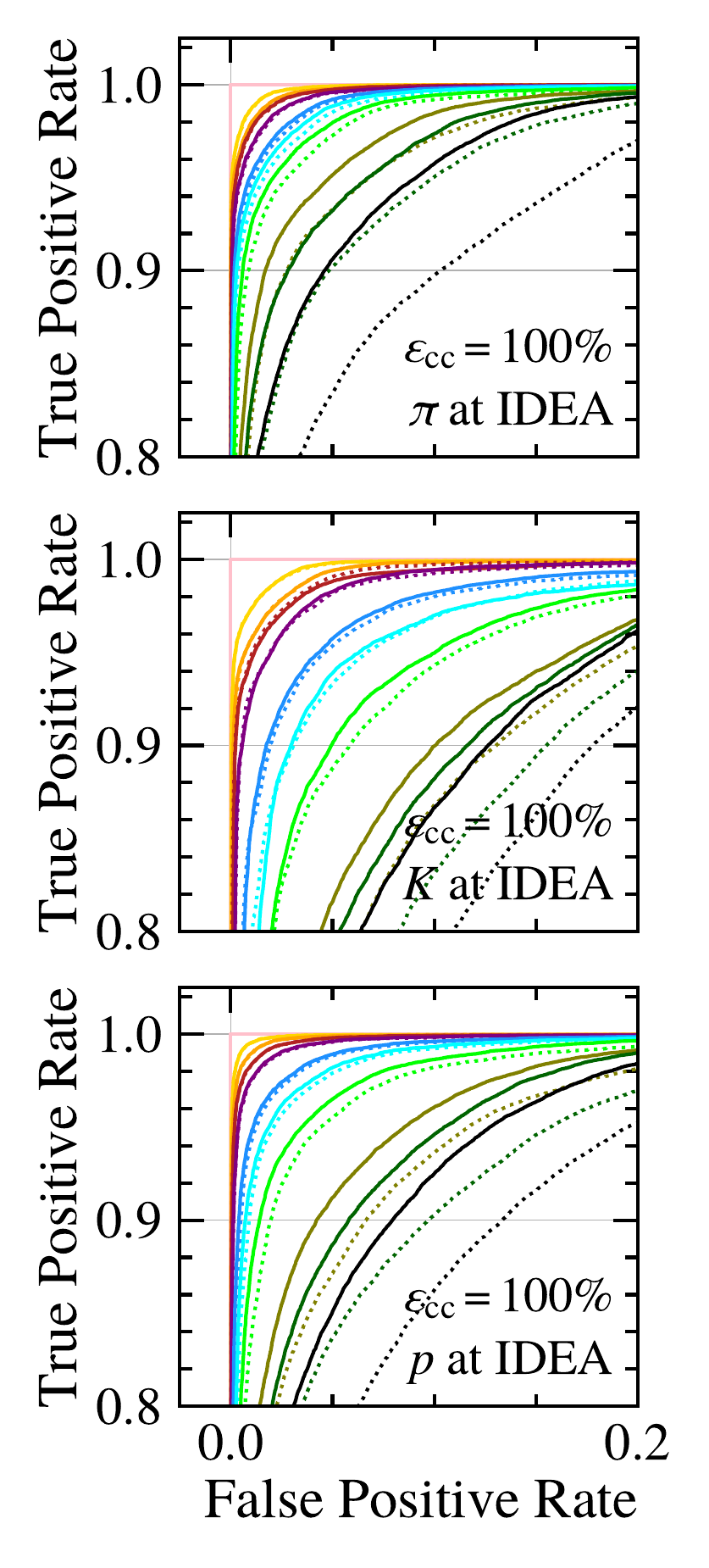}
    \caption{ROC for the (top) pion, (middle) kaon, and (bottom) proton identification at \idea with a cluster-counting efficiency $\varepsilon_\text{cc}$ of (left) 50\% or (right) 100\%.
    The colors indicate the time-of-flight (ToF) resolution and the line style indicates whether the silicon-based \dedx information was used.}
    \label{fig:roc:idea}
\end{figure}
\begin{figure}[h]
    \centering
    \includegraphics[width=\linewidth]{plots/particlegun/legend_auc.pdf}
    \includegraphics[width=0.5\linewidth]{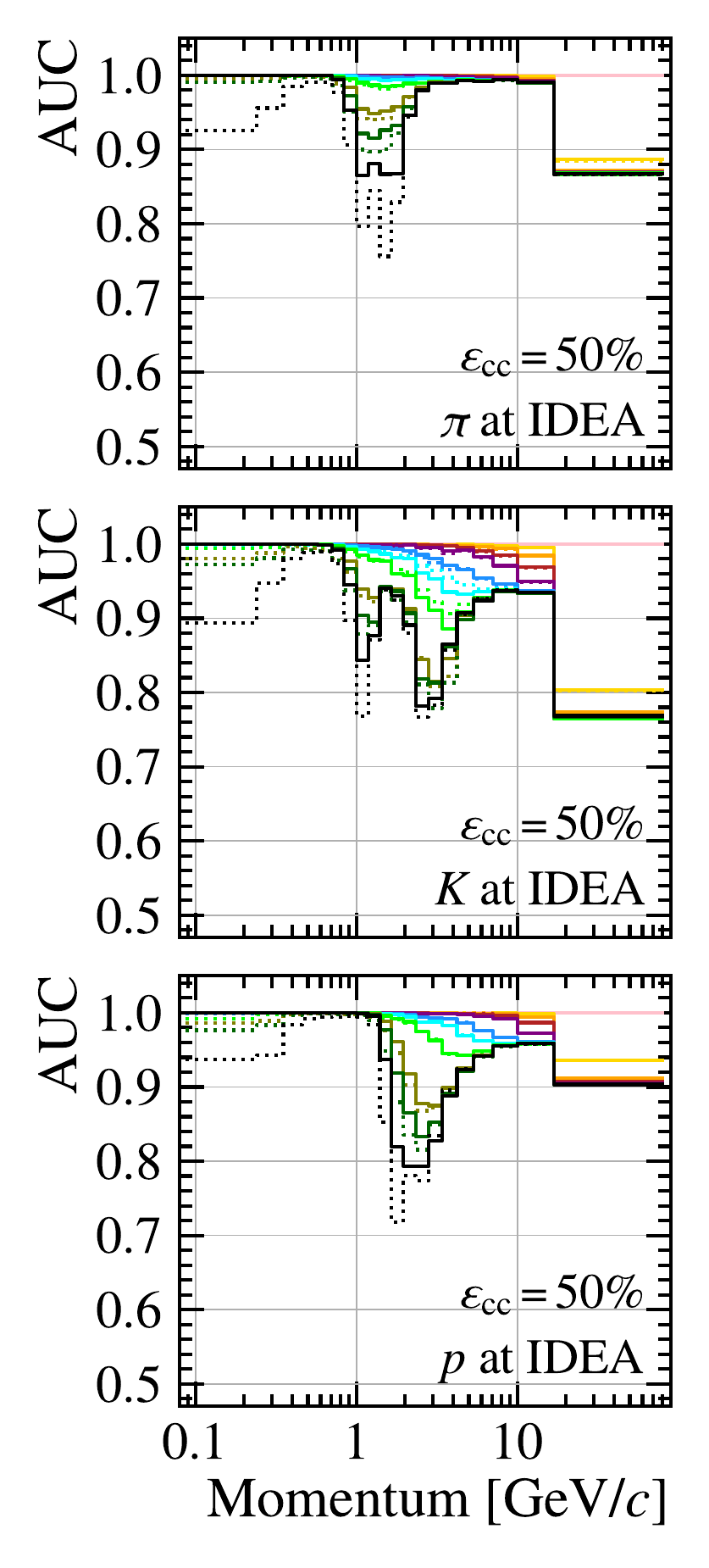}%
    \includegraphics[width=0.5\linewidth]{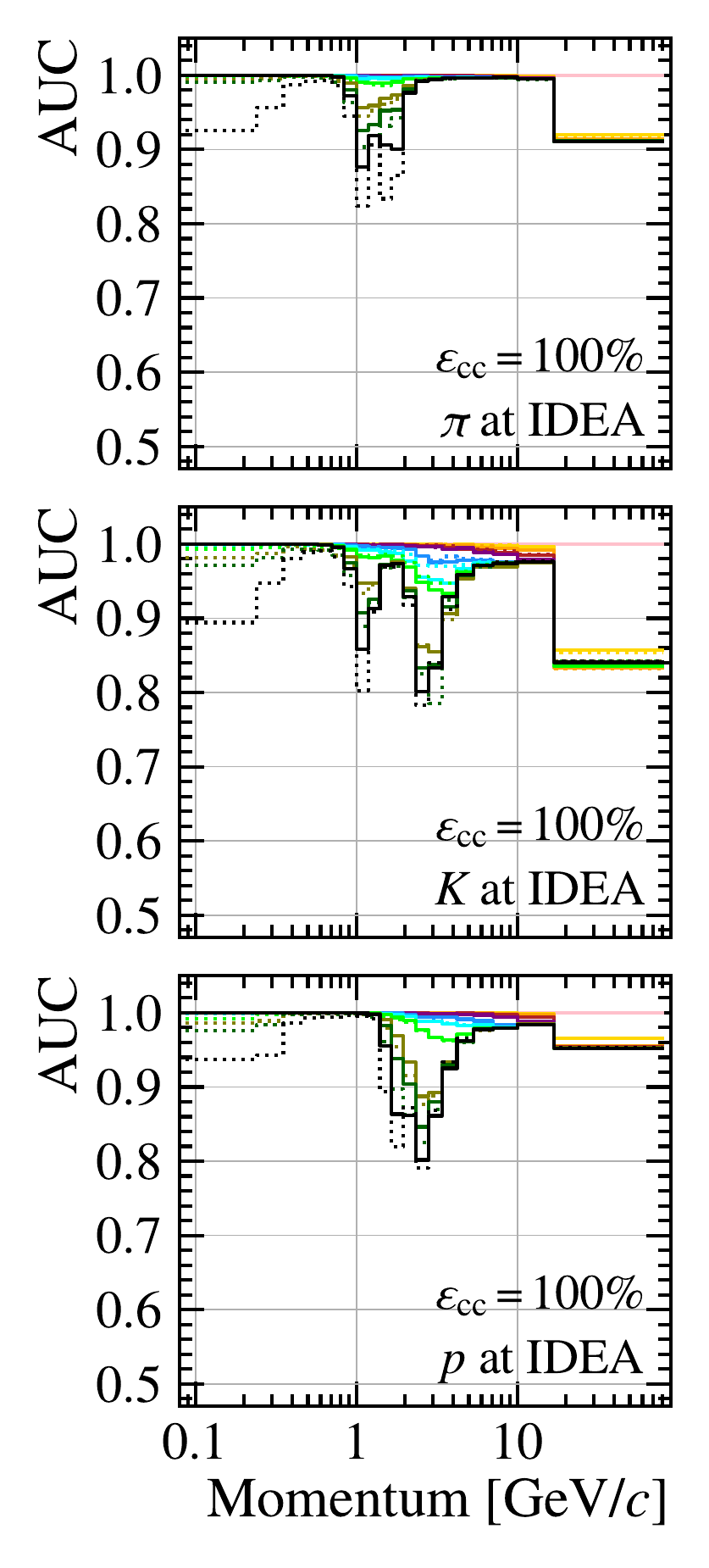}
    \caption{AUC in different momentum ranges for the (top) pion, (middle) kaon, and (bottom) proton identification at \idea with a cluster-counting efficiency $\varepsilon_\text{cc}$ of (left) 50\% or (right) 100\%.
    The colors indicate the time-of-flight (ToF) resolution and the line style indicates whether the silicon-based \dedx information was used.}
    \label{fig:roc:bins:idea}
\end{figure}
\clearpage
\onecolumngrid

\begin{figure}
    \centering
    \includegraphics[width=0.5\linewidth]{plots/particlegun/legend_significance}
    \includegraphics[width=.5\linewidth]{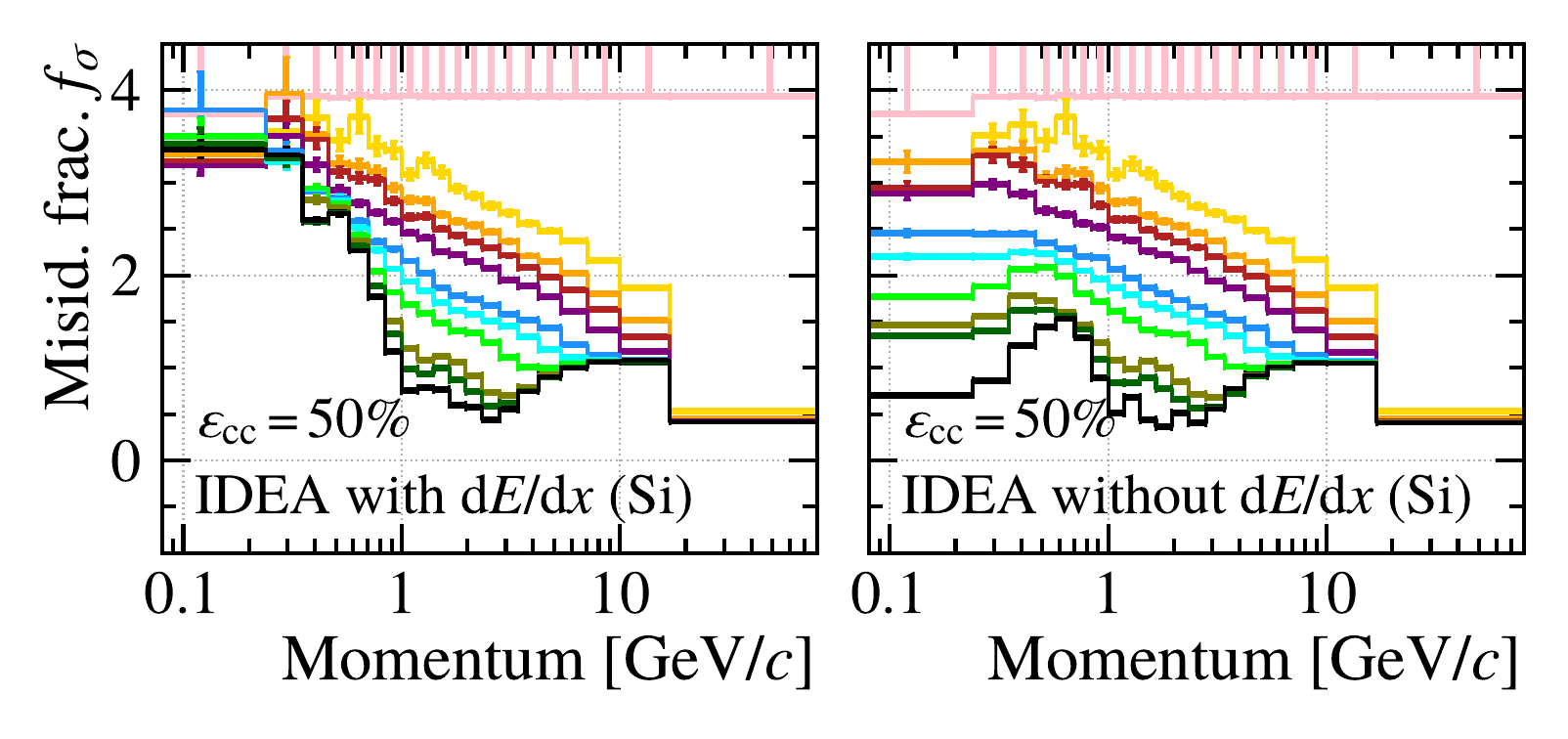}%
    \includegraphics[width=.5\linewidth]{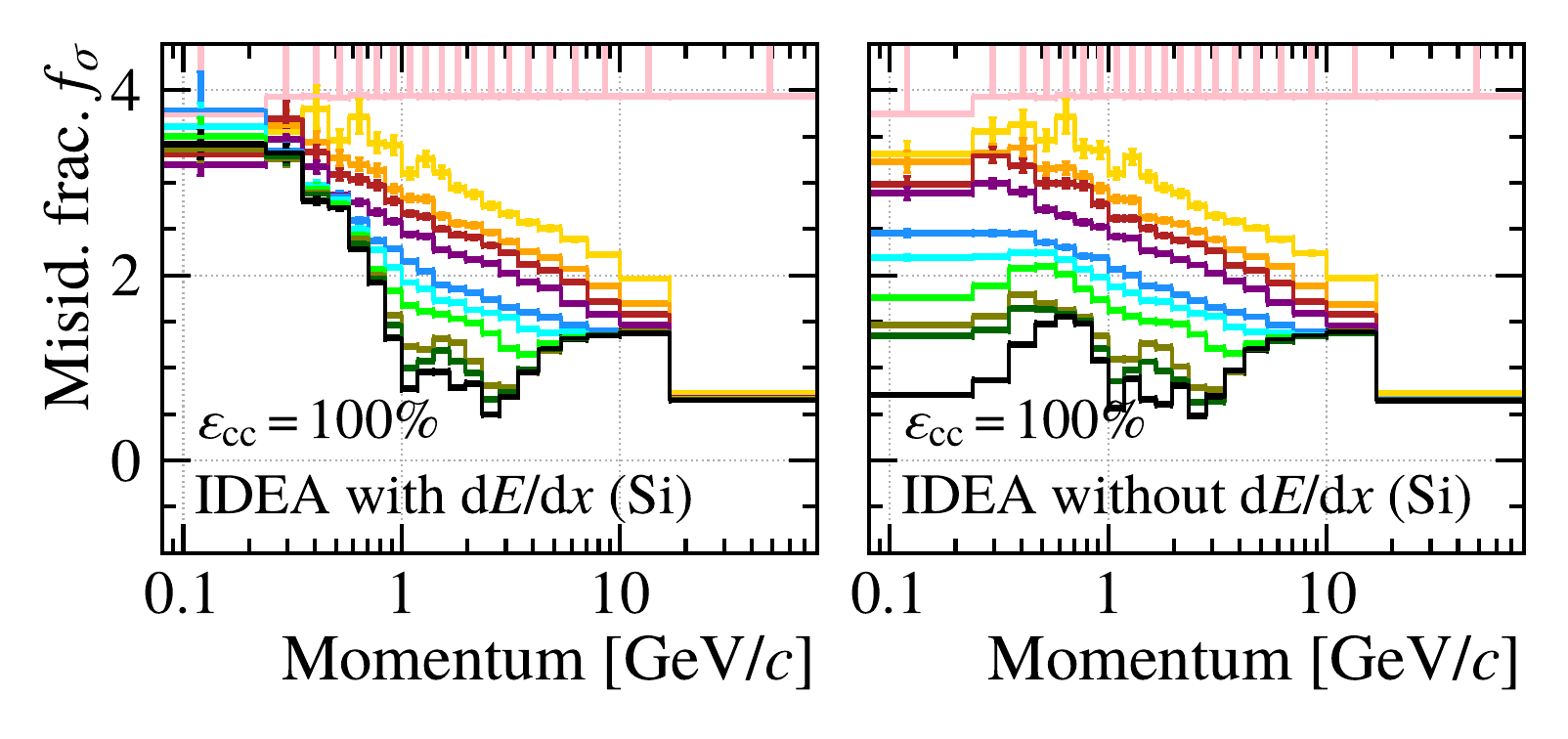}
    \caption{Fraction of misidentified particles $f_\sigma$ expressed as a number of standard deviations using Eq.~\eqref{eq:sigmas} for all particle species combined for different time-of-flight (ToF) resolution and without and with \dedx information at \idea (\ie \dndx measurement with or without time-of-flight and/or \dedx measurement) for a cluster-counting efficiency $\varepsilon_\text{cc}$ of (left pair) 50\% and (right pair) 100\%.}
    \label{fig:significance:idea}
\end{figure}
\twocolumngrid

\clearpage
\onecolumngrid

\section{BDT output scores}\label{app:bdtscore}
\twocolumngrid

\begin{figure}[h]
    \centering
    \includegraphics[width=0.8\linewidth]{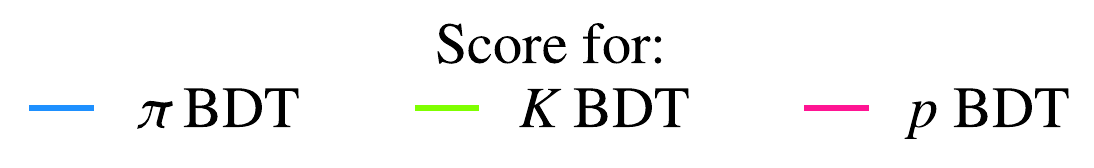}
    \includegraphics[width=0.5\linewidth]{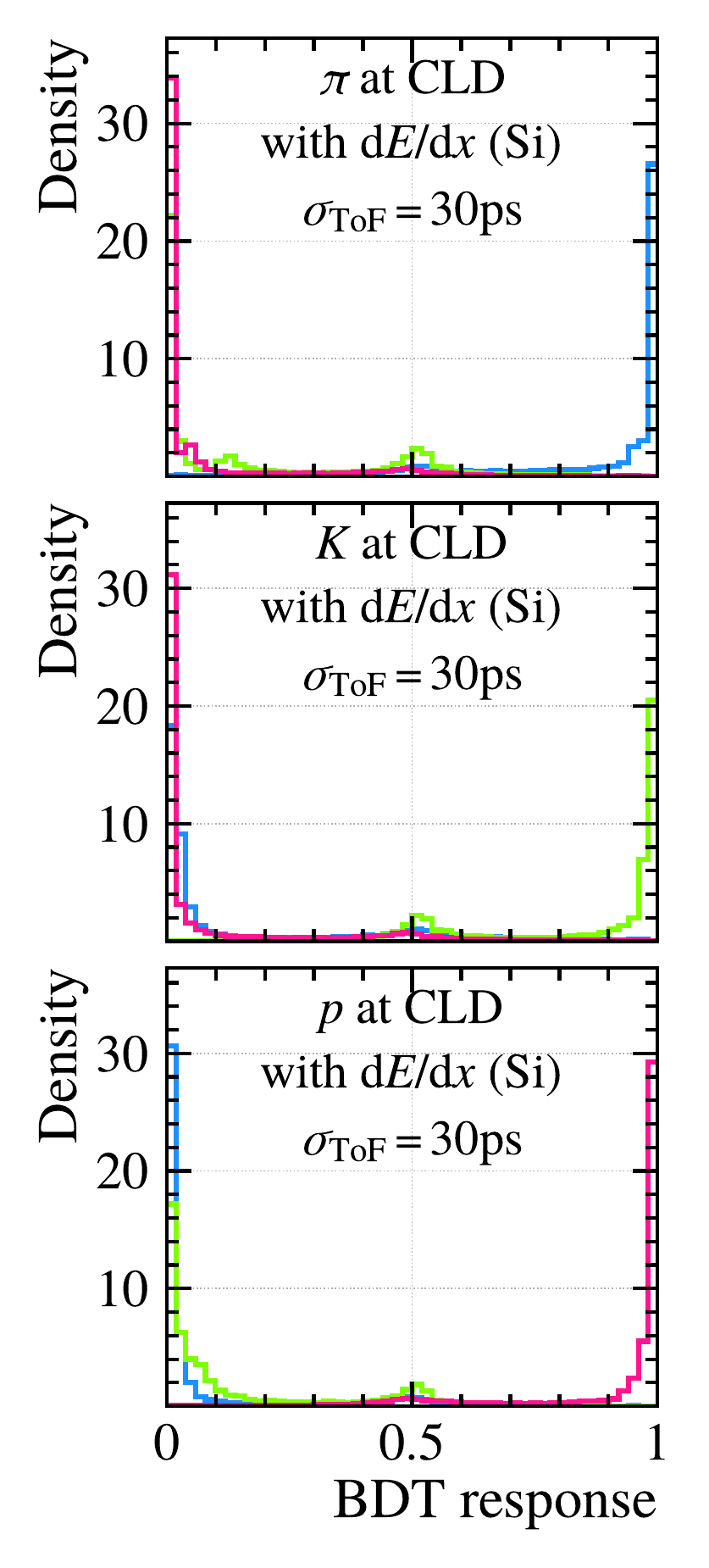}%
    \includegraphics[width=0.5\linewidth]{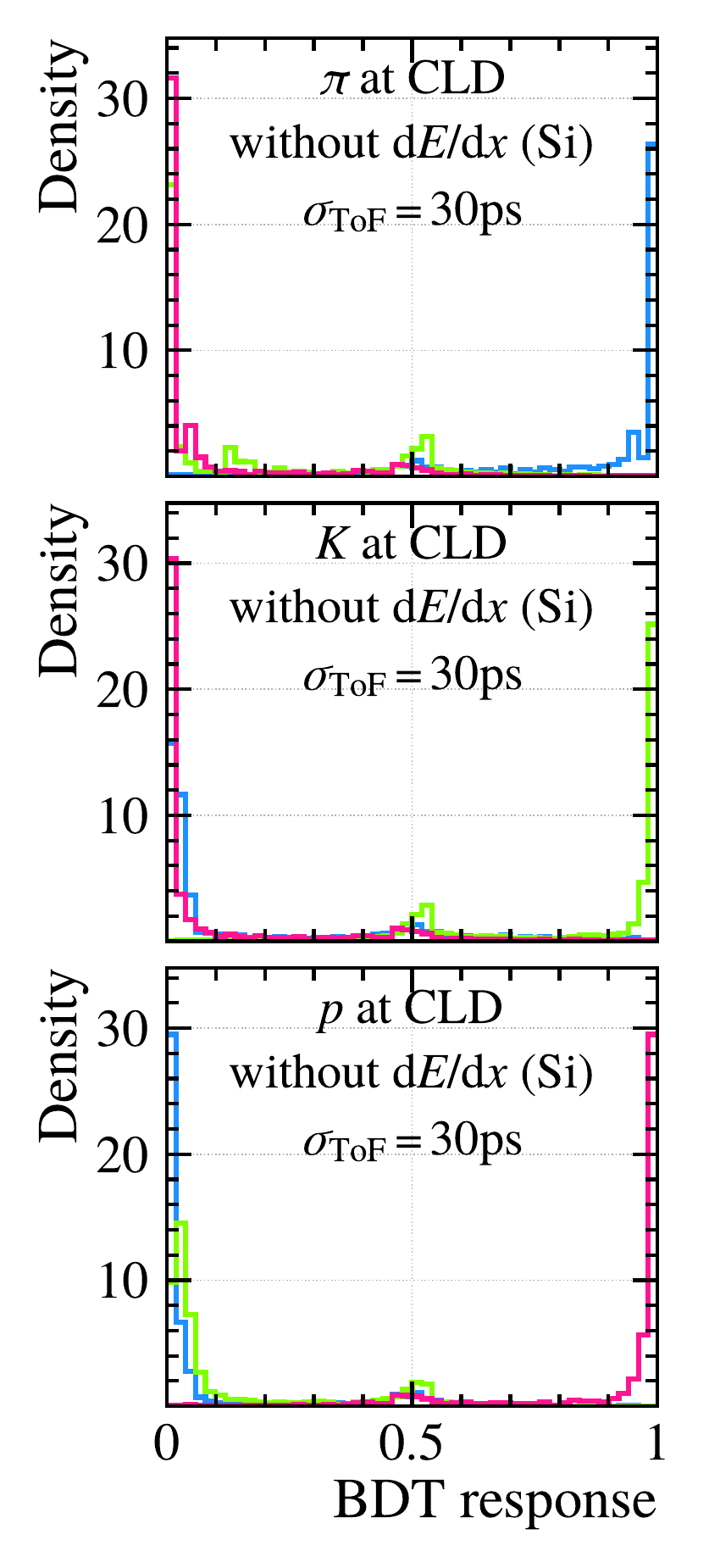}
    \caption{Distribution of the BDT scores for a true (top) pion, (middle) kaon, and (bottom) proton at \cld for time-of-flight resolution of $\sigma_\text{ToF}=30\ps$ (left) with and (right) without silicon-based \dedx measurement at \cld.}
    \label{fig:bdt:scores:cld:30}
\end{figure}

\begin{figure}[h]
    \centering
    \includegraphics[width=0.8\linewidth]{plots/particlegun/legend_bdt_scores.pdf}
    \includegraphics[width=0.5\linewidth]{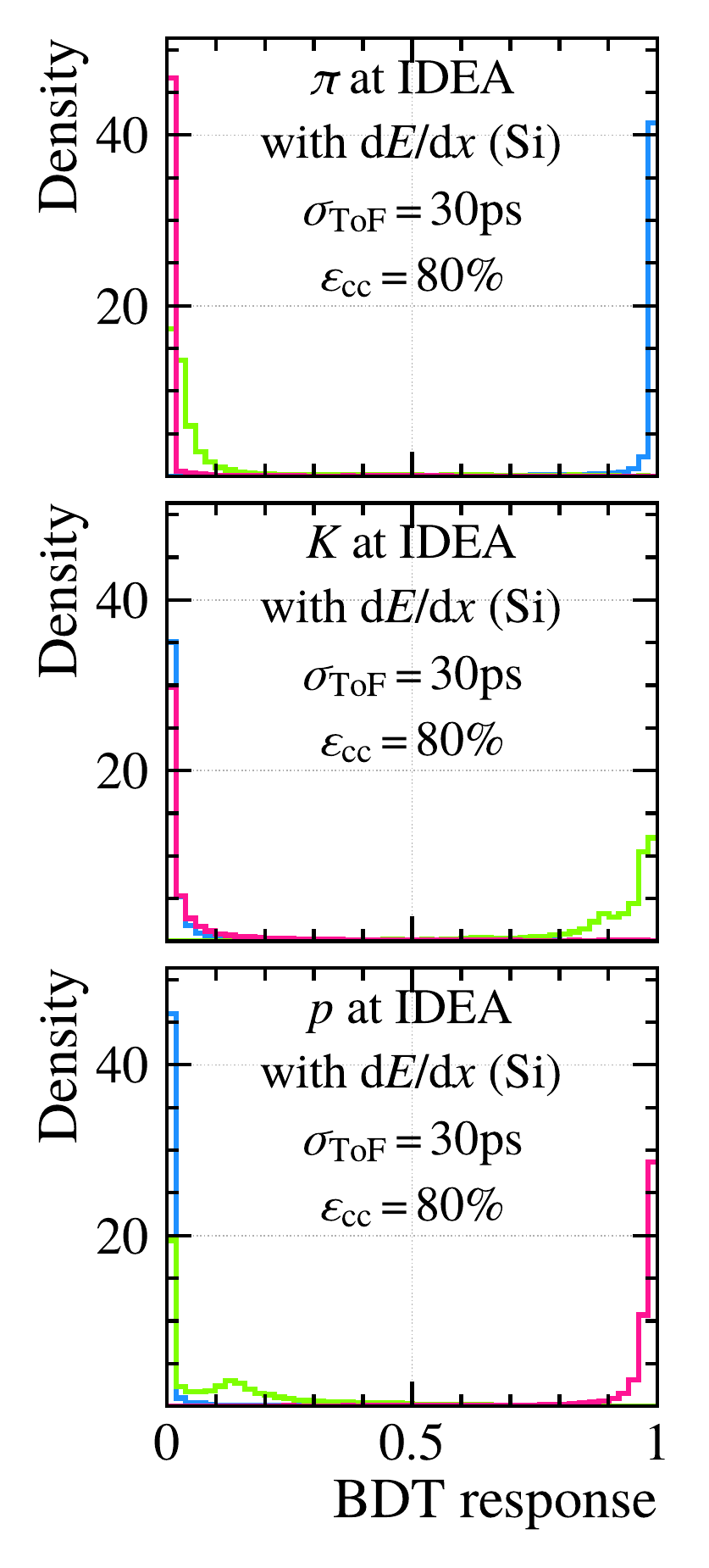}%
    \includegraphics[width=0.5\linewidth]{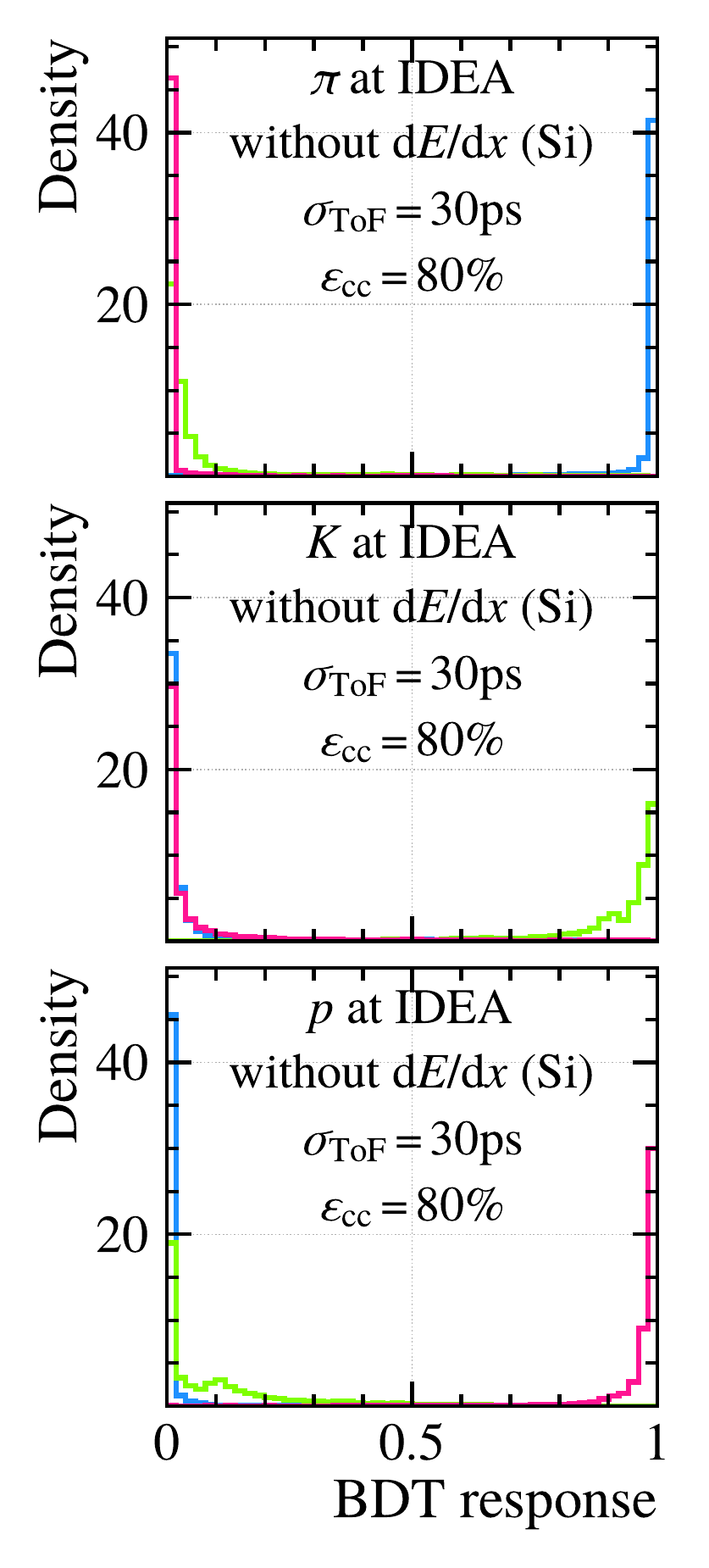}
    \caption{Distribution of the BDT scores for a true (top) pion, (middle) kaon, and (bottom) proton at \idea for time-of-flight resolution of $\sigma_\text{ToF}=30\ps$, cluster-counting efficiency of $\varepsilon_\text{cc}=80\%$, (left) with and (right) without silicon-based \dedx measurement at \idea.}
    \label{fig:bdt:scores:idea:30}
\end{figure}

\clearpage
\onecolumngrid
\section{Per-species significance}\label{app:significance}

\twocolumngrid

\begin{figure}[h]
    \centering
    \includegraphics[width=0.85\linewidth]{plots/particlegun/legend_significance}
    \includegraphics[width=.9\linewidth]{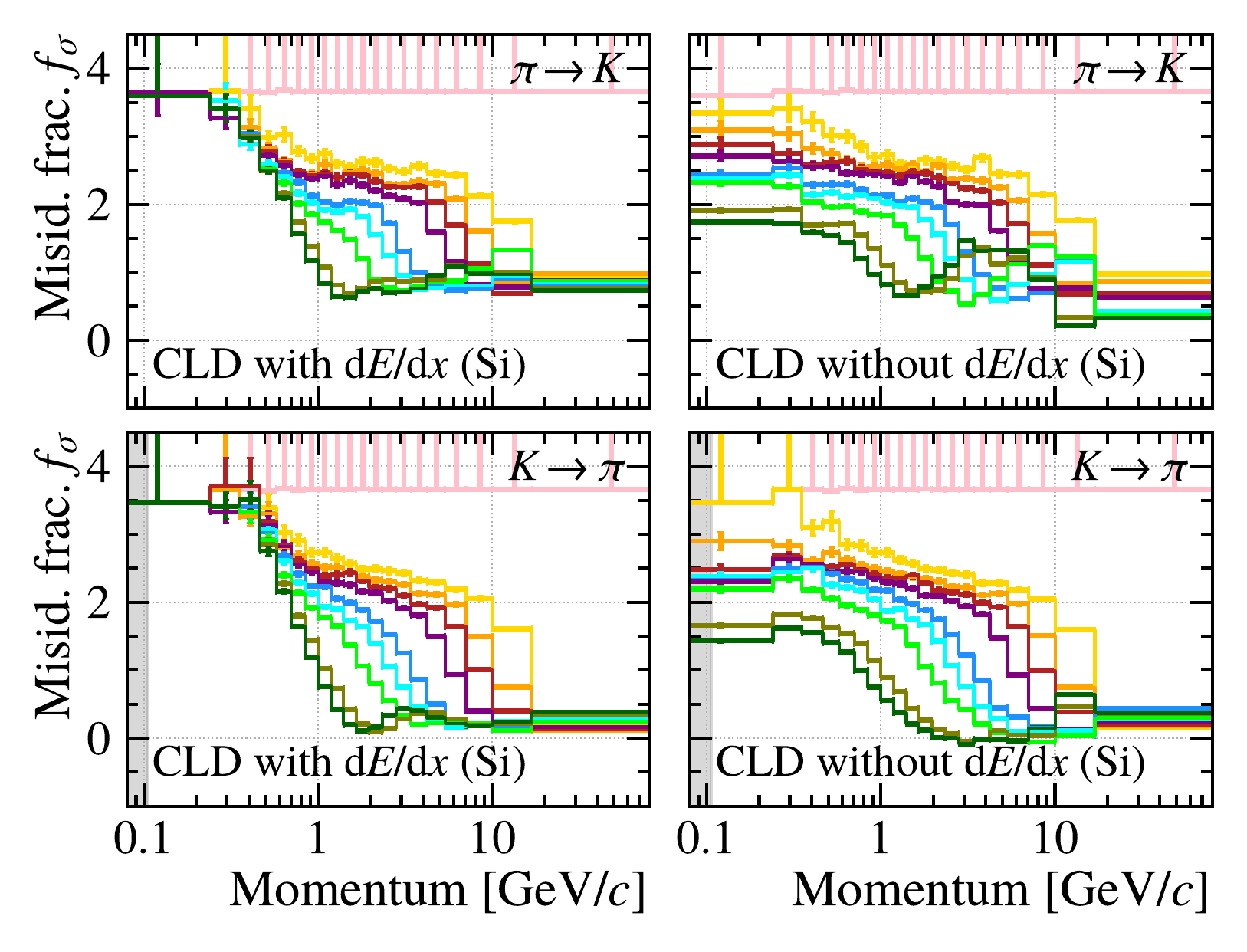}
    \includegraphics[width=.9\linewidth]{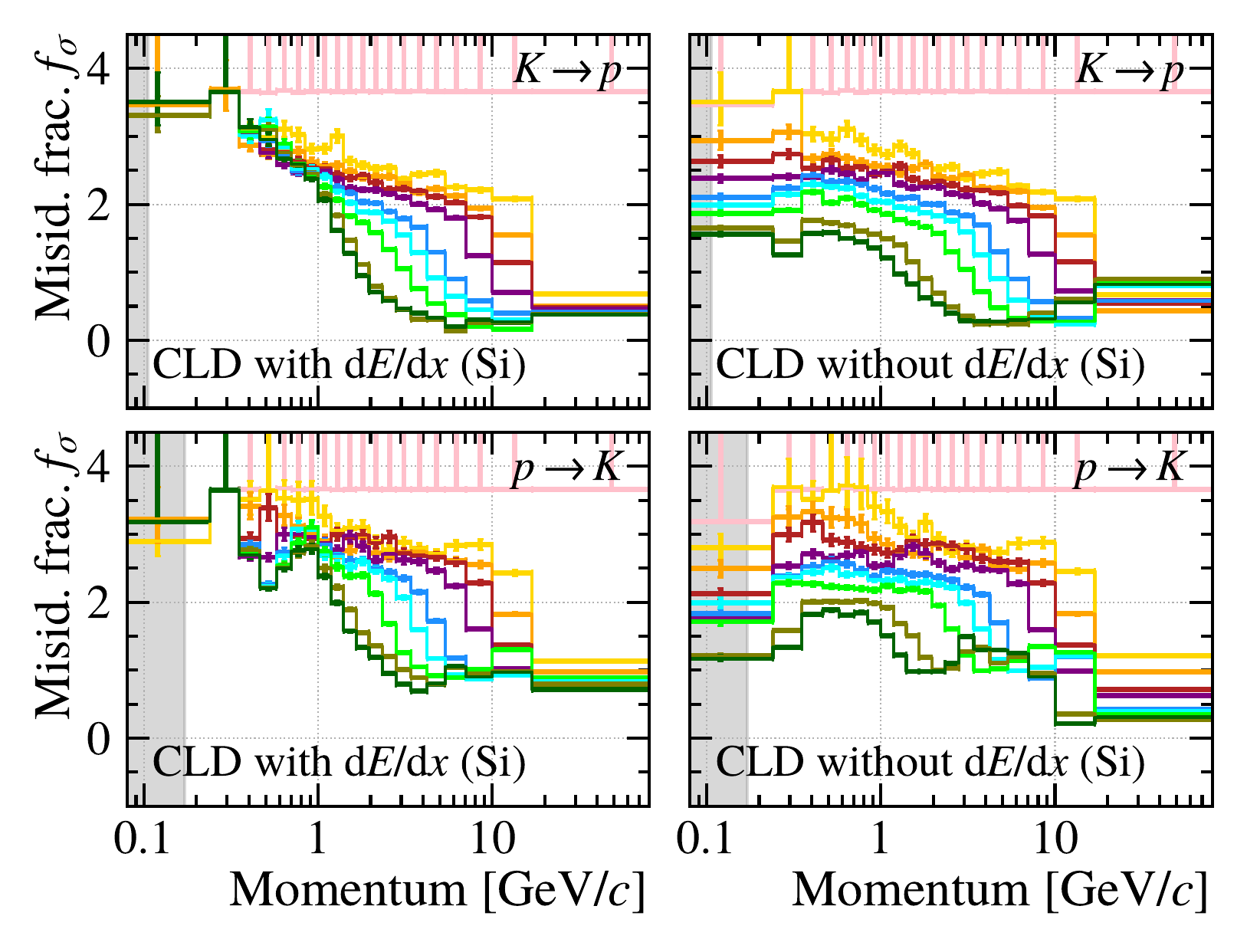}
    \includegraphics[width=.9\linewidth]{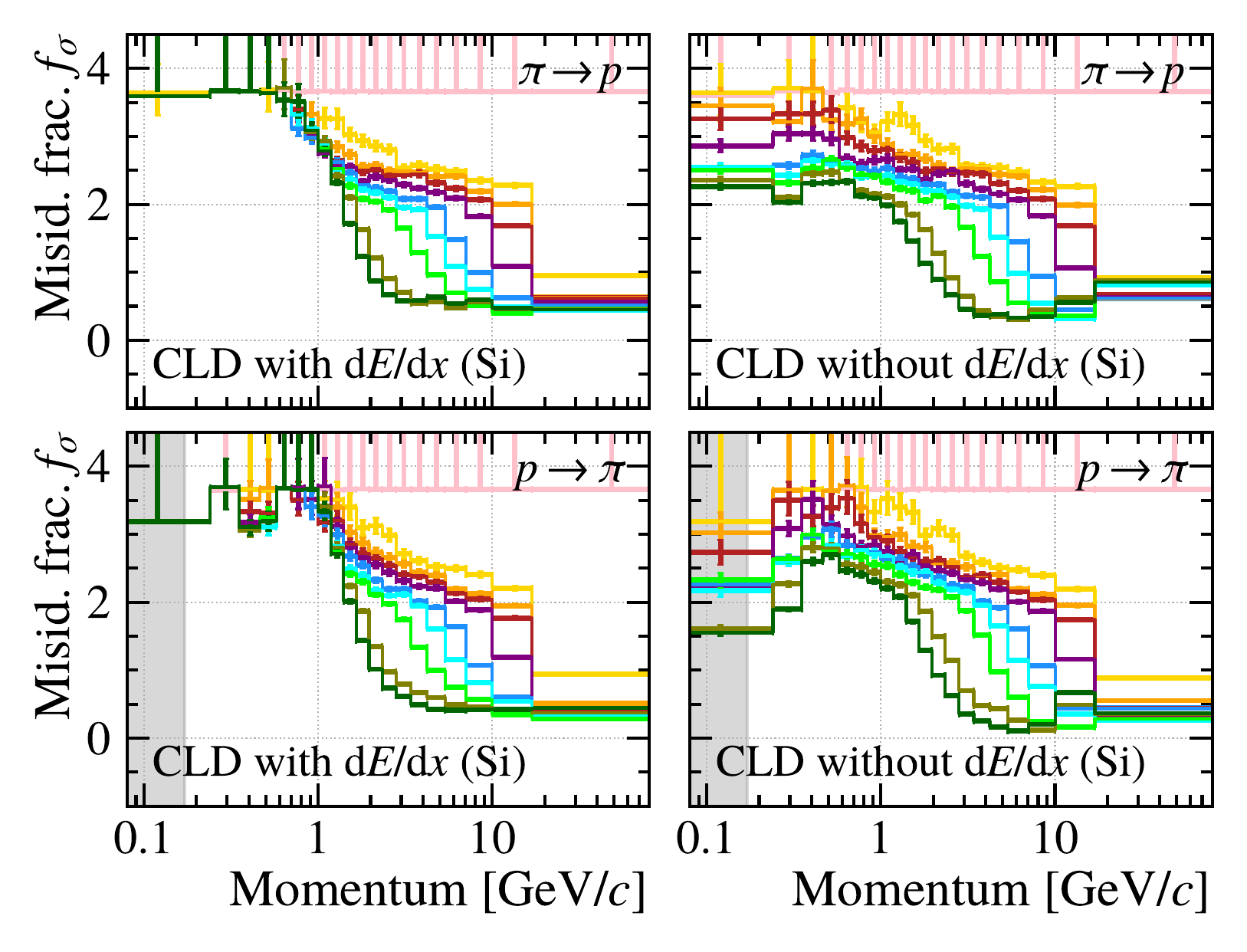}
    \caption{Fraction of misidentified particles $f_\sigma$ expressed as a number of standard deviations using Eq.~\eqref{eq:sigmas} for each combination of hadron species (top, middle, and bottom) for different time-of-flight (ToF) resolution and with (left) and without (right) \dedx information at \cld (\ie time-of-flight measurement with or without \dedx measurement).}
    \label{fig:significance:cld:individual}
\end{figure}
\begin{figure}[h]
    \centering
    \includegraphics[width=0.85\linewidth]{plots/particlegun/legend_significance}
    \includegraphics[width=.9\linewidth]{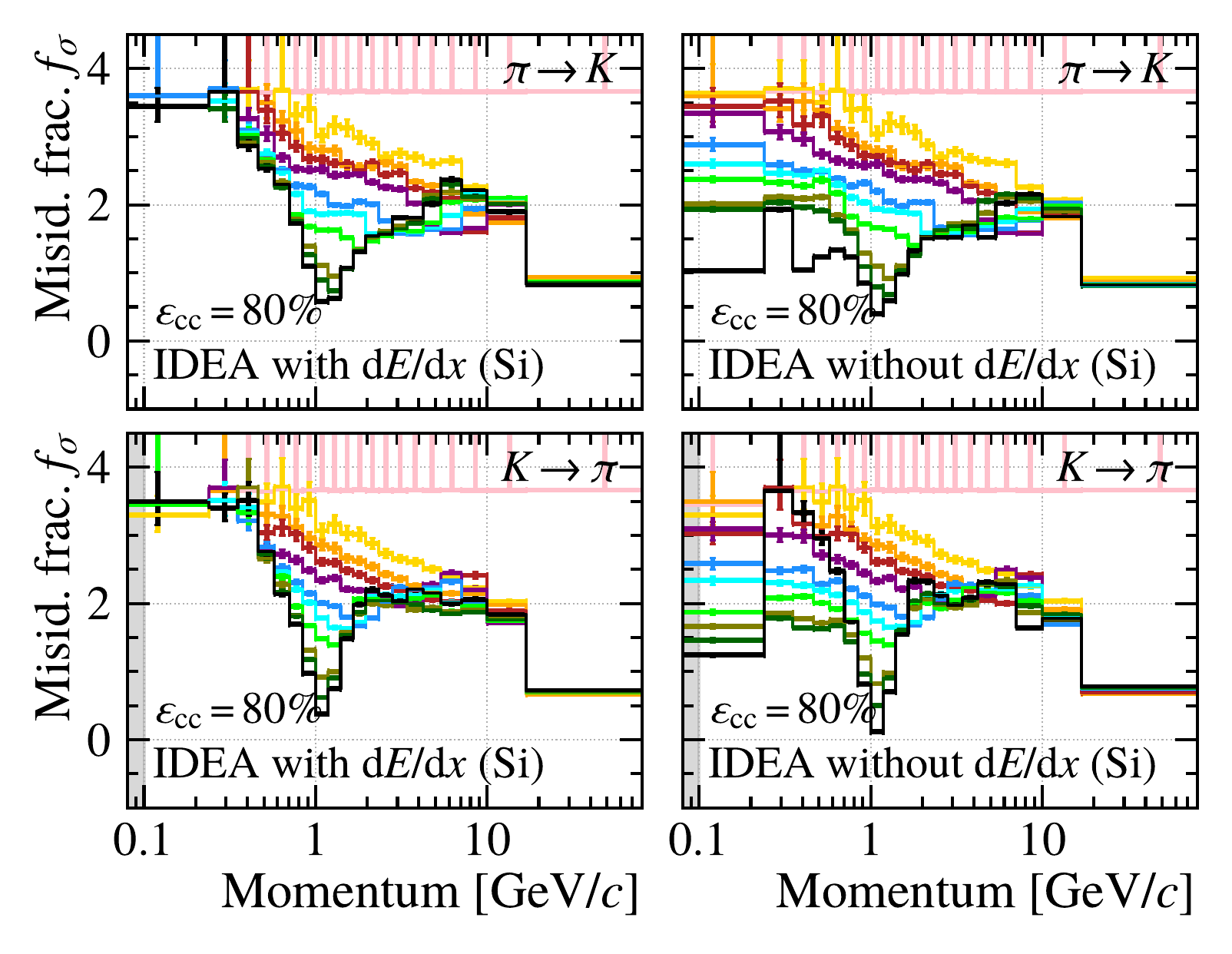}
    \includegraphics[width=.9\linewidth]{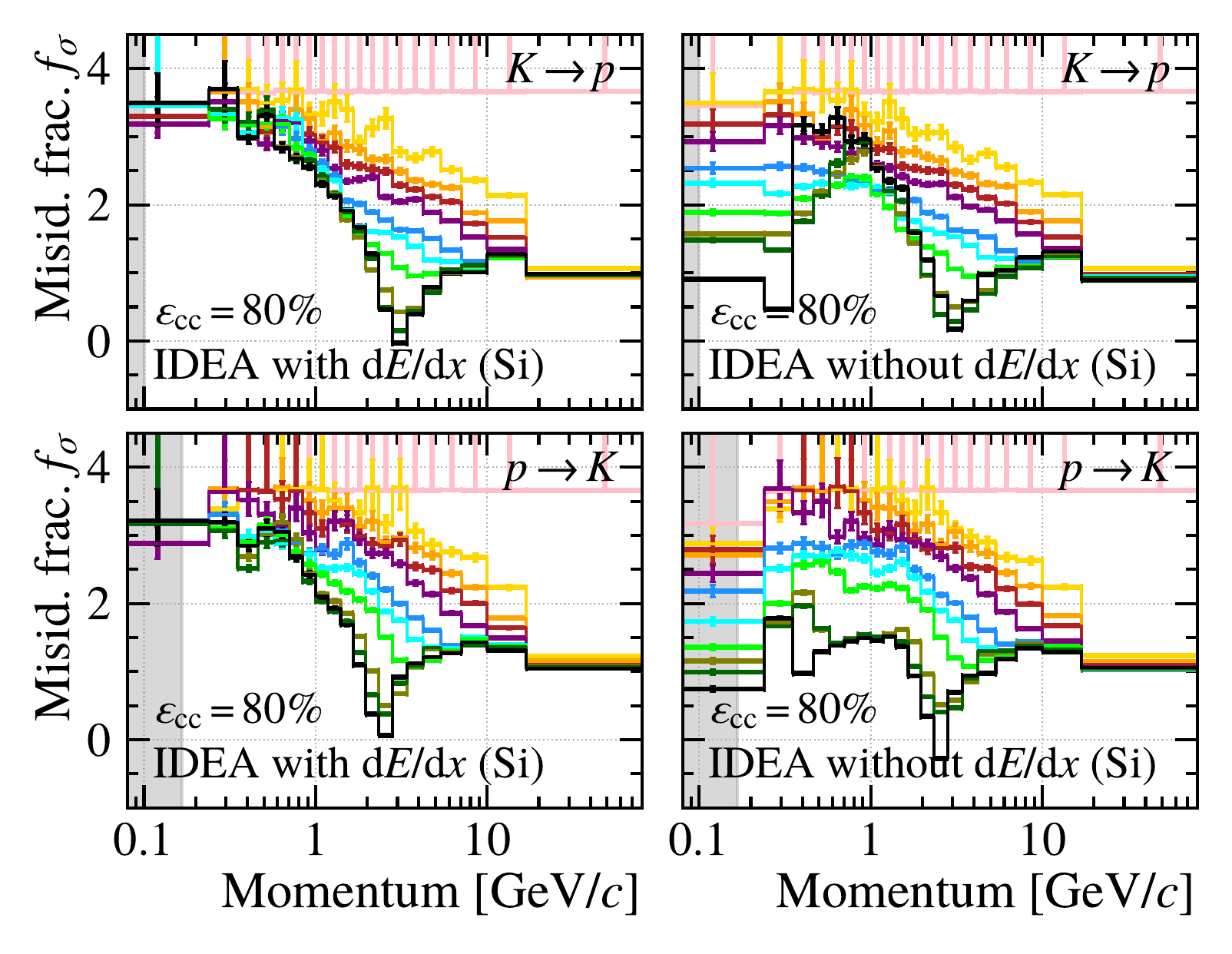}
    \includegraphics[width=.9\linewidth]{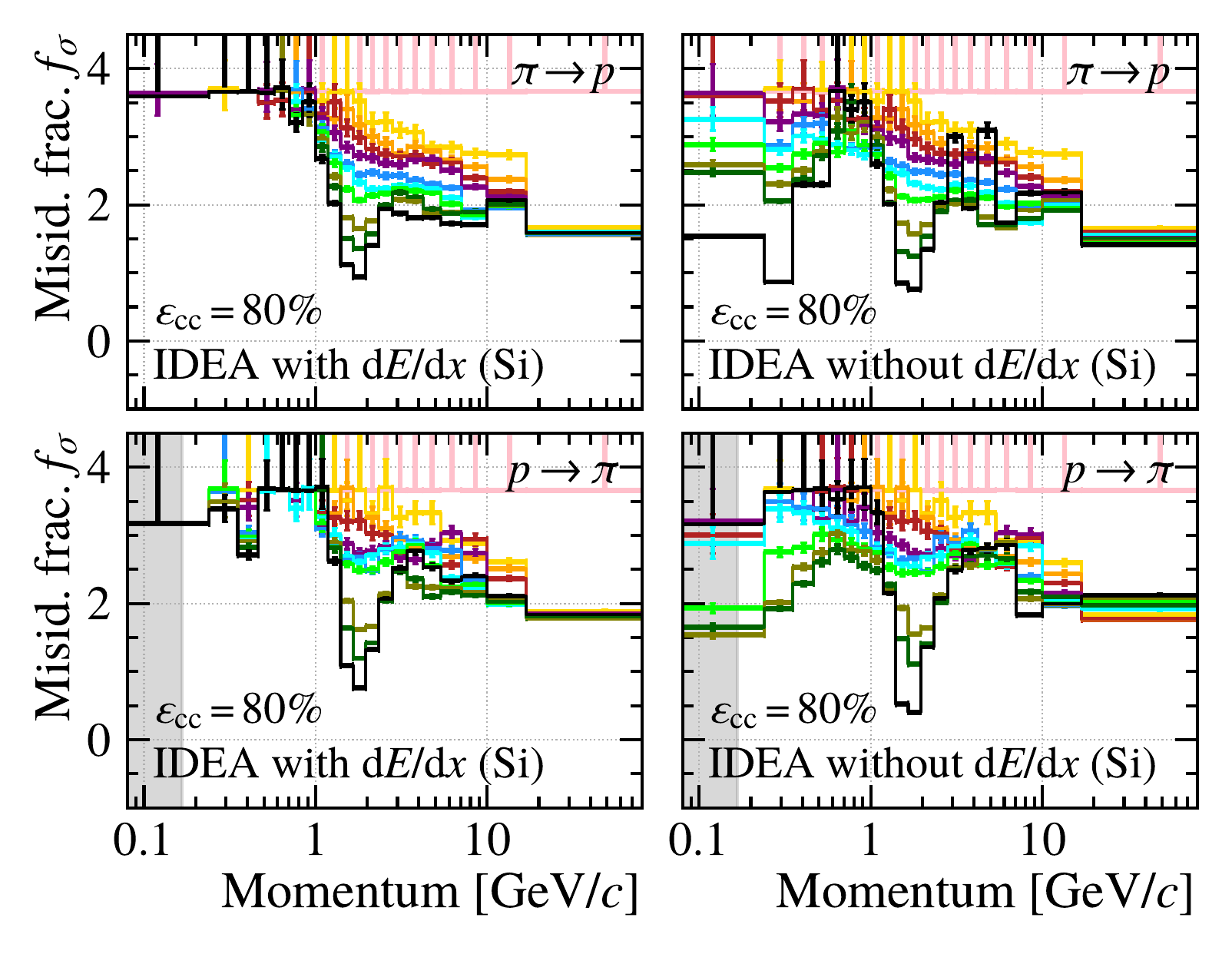}
    \caption{Fraction of misidentified particles $f_\sigma$ expressed as a number of standard deviations  using Eq.~\eqref{eq:sigmas} for each combination of hadron species (top, middle, bottom) for different time-of-flight resolution (ToF) with (left) and without (right) \dedx information at \idea (\ie \dndx measurement with or without time-of-flight and/or \dedx measurement).
    A cluster-counting efficiency $\varepsilon_\text{cc}$ of 80\% is assumed.}
    \label{fig:significance:idea:individual}
\end{figure}
\clearpage

\onecolumngrid

\section{Dihadron spectra}\label{app:dihadron}
\begin{table}[htb!]
    \centering
    \caption{Individual resonances with their relative contribution to the decay rate as simulated in the dihadron spectra of the rare-decay studies.}
    \label{tab:dihadronspectrum}
    \begin{tabular}{c|cc}
    \toprule
        Hadronic system & Resonance & Contribution \\
    \midrule
        \multirow{2}{*}{$\Bd\to\Km\pip$}
         & $\Kstarz(892)$ & 0.850 \\
         & $\Kstarz(1430)$ & 0.150 \\
    \midrule
        \multirow{2}{*}{$\Bs\to\Km\Kp$} & $\phiz(1020)$ & 0.838 \\
        &$f_2^\prime(1525)$ & 0.162 \\
    \midrule
        \multirow{4}{*}{$\Lb\to p\Km$} & $\Lz(1520)$ & 0.198 \\
        & $\Lz(1600)$ & 0.297 \\
        & $\Lz(1800)$ & 0.348 \\
        & $\Lz(1820)$ & 0.158 \\
    \bottomrule
    \end{tabular}
\end{table}
\section{Mass spread in rare decays}\label{app:massspread}
\begin{figure}[h]
    \centering
    \includegraphics[width=.5\linewidth]{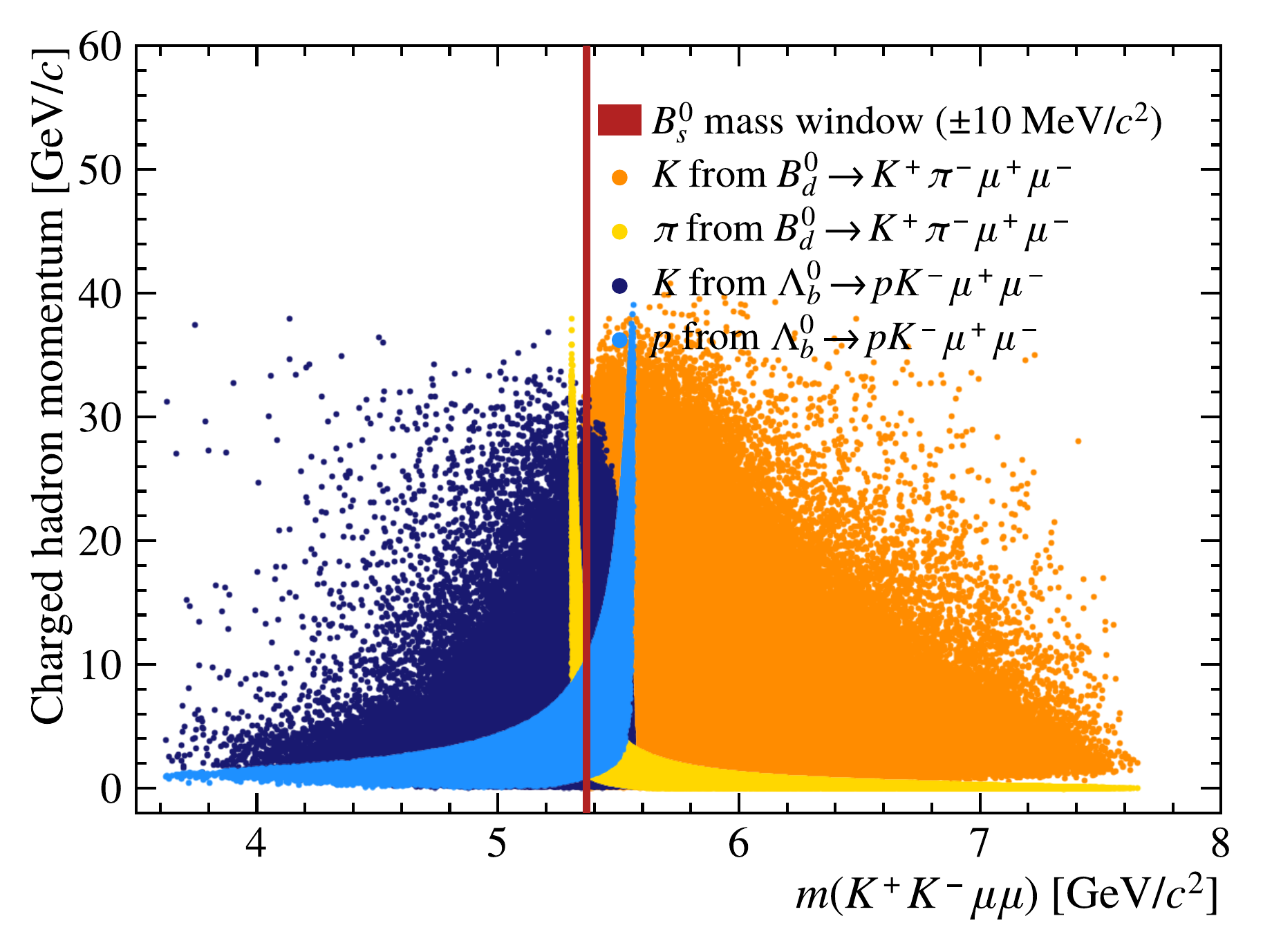}
    \caption{Spread of the four-body invariant  mass for different hadron momenta in the case of correctly identified hadrons (red points, kaons from \Bd and \Lb) and misidentified hadrons (yellow and blue points, $\pi$ from \Bd and $p$ from \Lb).}
    \label{fig:mass-vs-momentum}
\end{figure}

Figure~\ref{fig:mass-vs-momentum} serves to illustrate the wide spread of misidentified rare decays when the signal decay is \BsToKKmm.
The horizontal axis shows the value of the four-body invariant mass reconstructed under the $KK$ hypothesis for the dihadron system.
The vertical axis corresponds to the momentum of either final-state hadron, such that every decay candidate is represented by two points in the figure.
Correctly identified kaons in the \Bd and \Lb decays are shown as orange and dark blue points, respectively.
They accumulate around the \Bs mass but have a large spread.
The yellow and lighter blue points on the other hand, correspond to, respectively, the pion and proton from the \Bd and \Lb decays if they were identified as kaons.
Misidentification of a hadron with high momentum results in a four-body invariant-mass close to the truth; i.e. the yellow (blue) points accumulate at the \Bs (\Lb) mass.
If the pion (proton) from a \Bd (\Lb) decay has very low momentum and is misidentified as a kaon, the system artificially gains (loses) energy leading to increasingly high (low) four-body invariant mass.
In comparison to the background spread, the signal window of $\pm10\gevcc$ is very narrow leading to efficient background suppression.

\newpage

\section{Contamination for different mass resolutions}\label{app:resolution}
\twocolumngrid

\begin{figure}[h]
    \centering
    \includegraphics[width=\linewidth]{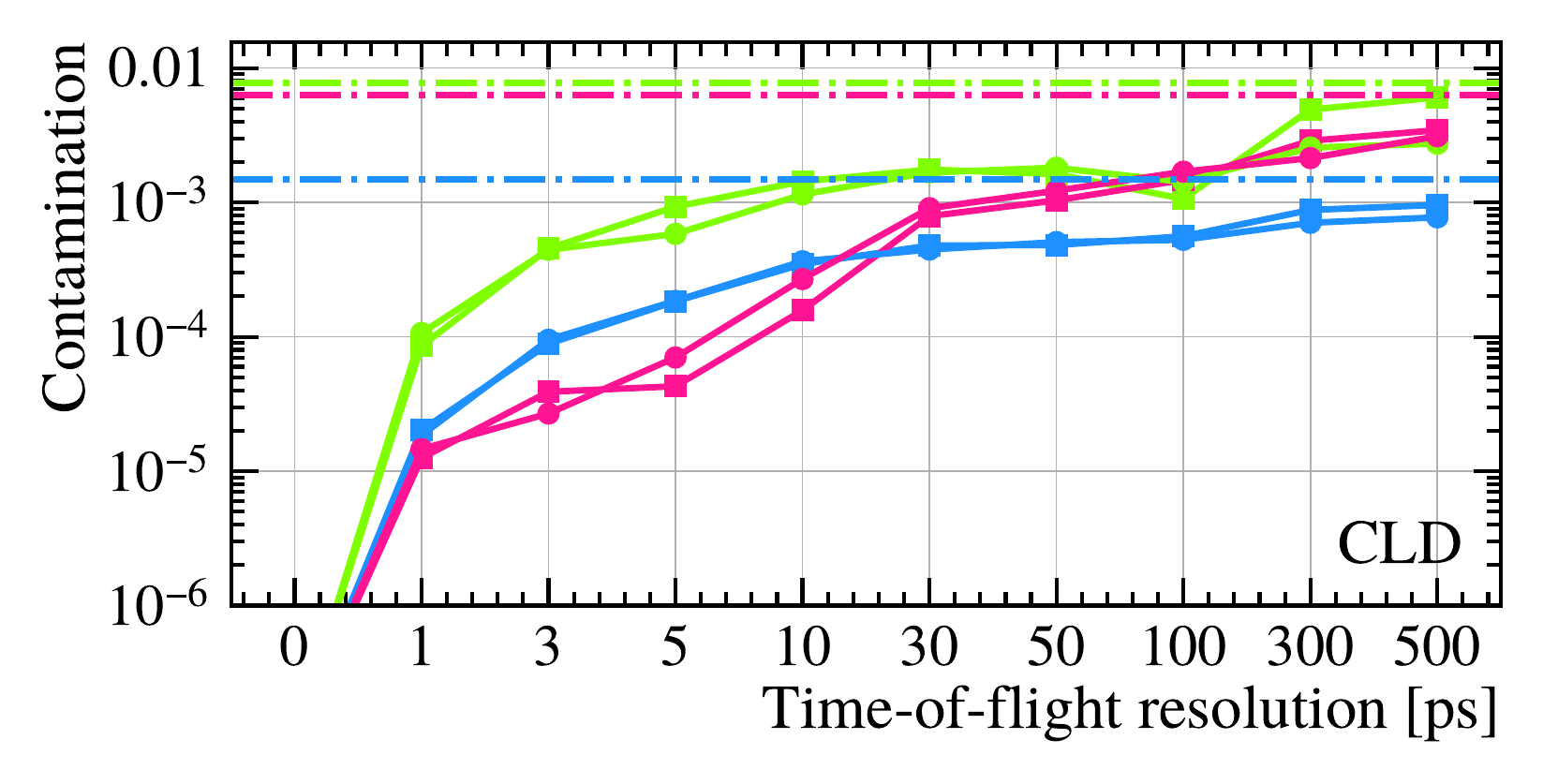}
    \includegraphics[width=\linewidth]{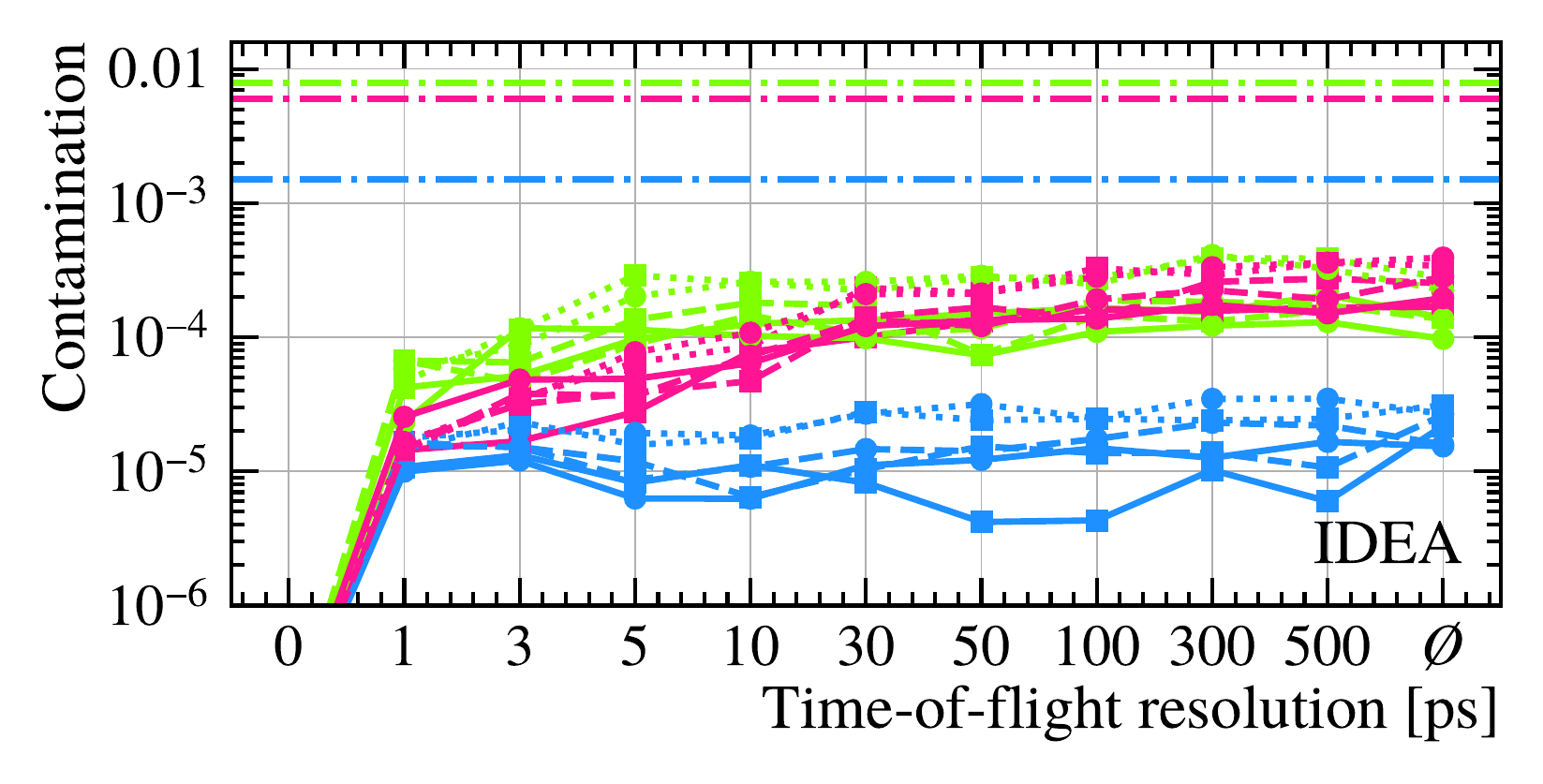}
    \caption{Level of contamination for the three rare decays with fully visible final states (distinguished by color) at (left) \cld and (right) \idea for different time-of-flight resolution (horizontal axis), different cluster-counting efficiencies (\idea only, distinguished by line style), and with or without using energy-deposit information from the silicon sensors (marker style).
    The four-body invariant masses are smeared using a Gaussian of width $\sigma=3\mevcc$ with signal window and background vetoes of $\pm6\mevcc$.}
    \label{fig:raredecays:3}
\end{figure}

\begin{figure}[h]
    \centering
    \includegraphics[width=\linewidth]{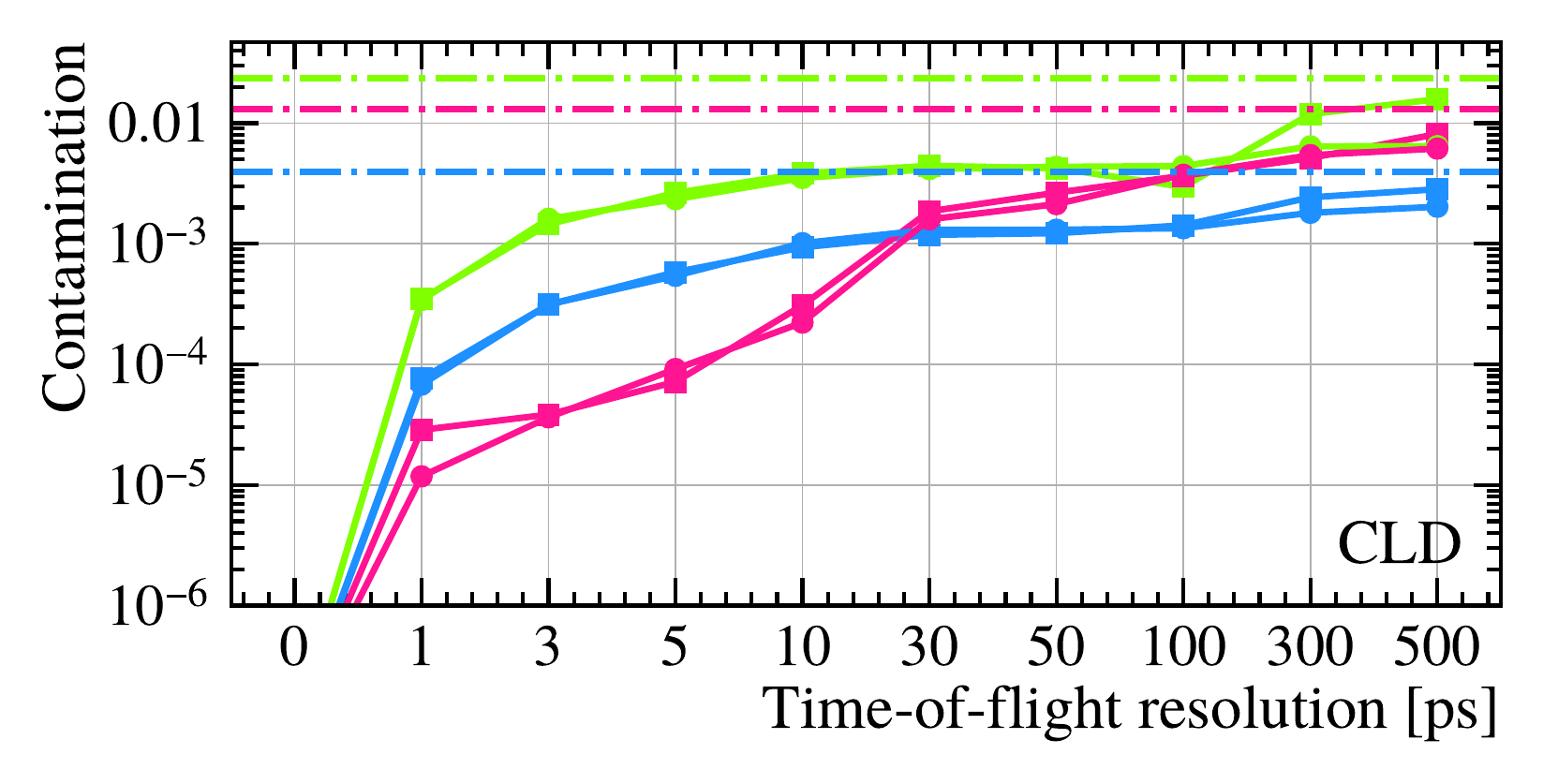}
    \includegraphics[width=\linewidth]{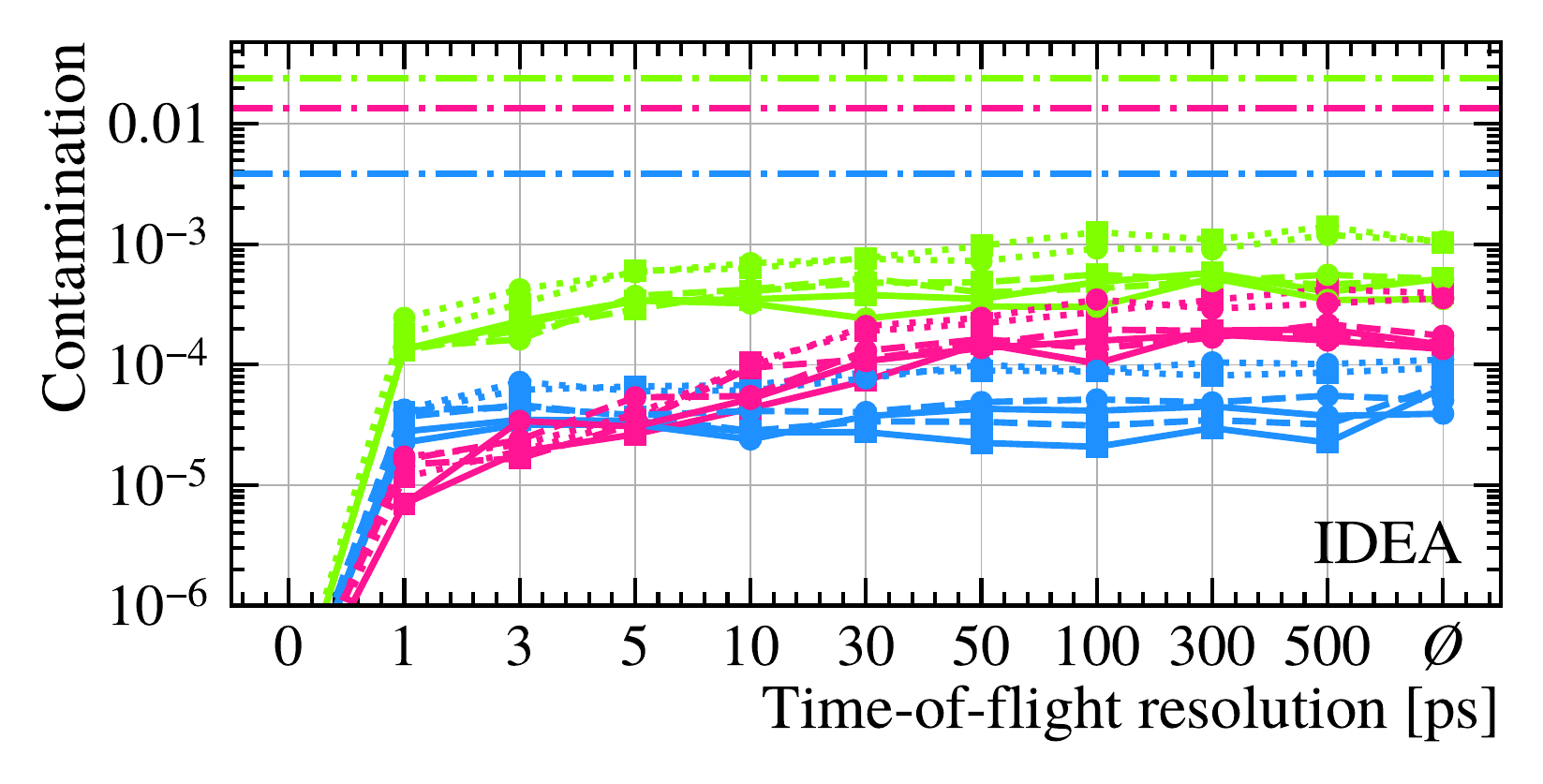}
    \caption{Level of contamination for the three rare decays with fully visible final states (distinguished by color) at (left) \cld and (right) \idea for different time-of-flight resolution (horizontal axis), different cluster-counting efficiencies (\idea only, distinguished by line style), and with or without using energy-deposit information from the silicon sensors (marker style).
    The four-body invariant masses are smeared using a Gaussian of width $\sigma=10\mevcc$ with signal window and background vetoes of $\pm20\mevcc$.}
    \label{fig:raredecays:10}
\end{figure}
\clearpage

\bibliography{main}

\end{document}